\documentclass[useAMS,usenatbib]{mn2e} 

\usepackage{graphicx}
\usepackage{enumerate}
\usepackage[switch]{lineno}
\usepackage{multirow}
\usepackage[fleqn]{amsmath}									
\usepackage{empheq}
\usepackage{hyperref} 
\usepackage{xcolor}

\title[Weighted statistics for irregularly sampled time series]{Weighted statistical parameters for irregularly sampled time series}
\author[L. Rimoldini]{Lorenzo Rimoldini$^{1,2}$\thanks{E-mail: lorenzo@rimoldini.info}\\
$^1$Observatoire astronomique de l'Universit\'e de Gen\`eve, ch. des Maillettes 51, CH-1290 Versoix, Switzerland\\
$^2$ISDC Data Centre for Astrophysics, Universit\'e de Gen\`eve, ch. d'Ecogia 16, CH-1290 Versoix, Switzerland}

\begin{document}

\date{Accepted 0000 Month 00. Received 0000 Month 00; in original form 2013 April 28}

\pagerange{1--19} \pubyear{2013}

\maketitle

\label{firstpage}

\begin{abstract}
Unevenly spaced time series are common in astronomy because of the day-night cycle, weather conditions, dependence on the source position in the sky, allocated telescope time, corrupt measurements, for example, or be inherent to the scanning law of satellites like {\it Hipparcos} and the forthcoming {\it Gaia}. \textcolor{black}{Irregular sampling often causes clumps of measurements and gaps with no data which can severely disrupt the values of estimators. This paper aims at improving the accuracy of common statistical parameters when linear interpolation (in time or phase) can be considered an acceptable approximation of a deterministic signal. A pragmatic solution is formulated in terms of a simple} weighting scheme, adapting to the sampling density and noise level, \textcolor{black}{applicable to large data volumes at minimal computational cost. Tests on} time series from the {\it Hipparcos} periodic catalogue led to significant improvements in the overall accuracy and precision of the estimators with respect to the unweighted counterparts and those weighted by inverse-squared uncertainties. Automated classification procedures employing statistical parameters weighted by the suggested scheme confirmed the benefits of the improved input attributes. The classification of eclipsing binaries, Mira, RR Lyrae, Delta Cephei and Alpha$^2$ Canum Venaticorum stars employing exclusively weighted descriptive statistics achieved an overall accuracy of 92 per cent, about 6~per~cent higher than with unweighted estimators. 
\end{abstract}

\section{Introduction \label{sec:intro}}

\textcolor{black}{Unevenly sampled astronomical time series are common in both ground- and satellite-based observations, and typically include time intervals with clustered and scattered data.
For example, the sampling laws of surveys such as the {\it Hipparcos}\footnote{\url{http://hipparcos.esa.int}} \citep{ESA,Perryman} and the forthcoming {\it Gaia}\footnote{\url{http://gaia.esa.int}} \citep{PerrymanGaia} missions are characterized by gaps and clumps of measurements on time-scales much greater and smaller than the average sampling interval, respectively.\footnote{In the case of the {\it Hipparcos} data, sources were typically observed in sequences of 4 to 6 transits separated by 20 and 108~min and repeated every 3 to 5 weeks \citep{Eyer1994}.} }

A significant number of studies were devoted to the estimation of power spectra and modelling of irregularly sampled time series \citep[e.g.,][]{Carbonell,Koen,Vio}. 
\textcolor{black}{Various theoretical approaches to the problem of estimating the true values of irregularly sampled signals are described in the literature \citep[e.g.,][]{Rybicki, Scargle1989, Scargle1990}.}
\textcolor{black}{This paper evaluates pragmatically the effectiveness of some of the simplest solutions which can be implemented in a pipeline to process extremely large data volumes.
Big data constitute one of the current challenges in astronomy, with surveys like {\it Gaia}, the Panoramic Survey Telescope \& Rapid Response System\footnote{\texttt{http://pan-starrs.ifa.hawaii.edu/public}} \citep[Pan-STARRS,][]{Kaiser} and the Large Synoptic Survey Telescope\footnote{\texttt{http://www.lsst.org/lsst}} \citep[LSST,][]{Ivezic}, among others,
which require efficient algorithms to produce results that are as accurate as possible. 
Considering that per-cent level improvements can impact on decisions of a very large number of sources, tuning the balance between efficiency and accuracy is an important task of data processing.}

\textcolor{black}{
The objectives of the present study are statistical parameters (like moments and percentiles) of time series of deterministic signals.
If linear interpolation in time or phase is able to describe major features of signals, simple recipes can improve the accuracy of statistical parameters and that of subsequent analyses such as automated classification. The effect of the latter is tested on data related to periodic variable stars from the {\it Hipparcos} mission, which is one of the closest proxies for the sampling of {\it Gaia} (for which the improvement of a few per cent in classification accuracy can increase by millions the number of correctly classified variable sources).}

The \textcolor{black}{parameters considered for better accuracy include} descriptive statistics such as the mean, variance, higher central moments and robust equivalents: they can summarize essential features of signals in a few numbers employing straightforward computations, which makes them excellent precursors of more detailed  
analyses
like modelling and classification.
Herein, a simple \textcolor{black}{recipe to mitigate the} effects of irregular sampling on the characterization of a signal in terms of statistical parameters is presented. 
Separate works describe corrections of biases induced by small sample sizes and Gaussian uncertainties in the calculation of weighted moments and cumulants \citep{RimoldiniUnbiased,RimoldiniIntrinsic}.

Time series sampled irregularly or with varying candences may lead to very different estimates of statistical parameters for the same signal. 
For example, if a sinusoid is sampled mostly at maximum or minimum, the mean is offset by about the amplitude of the signal, the variance might be much smaller than expected (since most points sample the same region of the signal), and the distribution appears very skewed (the few non-clumped measurements form an asymmetric tail in the distribution of measurements). If the same clump of data was in proximity of the average level, instead, the mean and skewness values could  be close to the correct value (by serendipity), but the variance would  be much smaller than the true one.

Sampling-induced biases do not arise from sparsely sampled data only and they might manifest independently of the number of measurements. 
While time series with more measurements are generally associated with better 
coverage of signal features, the importance implicitly assigned to different parts of the signal by unweighted estimators is related to the relative frequency of measurements. 

In principle, the most accurate and precise statistical parameters could be inferred from the model of a signal. 
Alas, models are often complicated, in the attempt to describe the signal features under many circumstances, they might require lengthy processing and 
their accuracy cannot be guaranteed in all cases.
For example, Fourier series are  
well fitted to model periodic signals, but the description of sharp features, like those present in EA-type eclipsing binaries, requires a high number of harmonics, which can overfit smoother parts of the signal and cause unrealistic excursions in large intervals with no data \citep{Dubath}.

Linear interpolation is one of the simplest methods to approximate a model, assuming the signal features are sufficiently sampled.
This work estimates statistical parameters by averaging the linear interpolation of functions of time series measurements (depending on the specific estimator).
If the signal can be recognized in time domain, interpolation can be performed in time without requiring further information. 
If a sparsely sampled signal is primarily mono-periodic, it can be interpolated in phase by folding the time series with the corresponding period.
While the interpolated function might include profiles with spiky artefacts, its average is more robust and can be expressed 
as a weighted mean 
which assigns more relevance to scattered than clumped measurements.
Statistical parameters weighted by such a scheme were tested on data from the {\it Hipparcos} periodic catalogue and led to a significant general improvement in
the accuracy and precision of estimator values and automated classification results,
with respect to those obtained with the unweighted or error-weighted counterparts.

This paper is organized as follows. 
After the definition of the notation and terminology in Sec.~\ref{sec:notation}, the description of weighted estimators as averages of linear interpolations is presented in Sec.~\ref{sec:method}, with weights defined in time and phase, including adaptations for low signal-to-noise regimes and small sample sizes. 
The new weighting scheme is applied to time series from the {\it Hipparcos} periodic catalogue in Sec.~\ref{sec:hip}, which includes a comparison of the values of statistical parameters computed with different weighting schemes and their effect on automated classification. 
The conclusions are drawn in Sec.~\ref{sec:concl}, followed by a series of appendices with more details on the interpolation of (mono-)periodic signals (Appendix~\ref{app:periodicInterpolation}), the illustration of modelled light curves to verify the accuracy of estimators (Appendix~\ref{app:data}),  the definitions of the statistical parameters employed (Appendix~\ref{app:def}) and additional scatter plots of  estimators (Appendices~\ref{app:scatter} and \ref{app:scatterUNWEIGHTED}).

\section{Notation and terminology}
\label{sec:notation}
For a set of $n$ measurements $\mathbf{x}=(x_1,x_2,...,x_n)$, the following quantities are defined:
\begin{itemize}
\item[(i)]
The population central moments of order $r$ around the mean $\mu$ are denoted by $\mu_r$ and respective cumulants $\kappa_2=\mu_2$, $\kappa_3=\mu_3$ and $\kappa_4=\mu_4-3\mu_2^2$  \citep[e.g.,][]{Kendall}.
\item[(ii)]
Sample \textcolor{black}{weighted} central moments \textcolor{black}{are defined in terms of weights $w_i$ as} $m_r=\sum_{i=1}^n w_i (x_i-\bar{x})^r/W$, with respective cumulants $k_r$, where $\bar{x}=\sum_{i=1}^n w_i x_i/W$ and $W=\sum_{i=1}^n w_i$.
\item[(iii)]
Standardized skewness and kurtosis are $g_1=k_3/k_2^{3/2}$ and $g_2=k_4/k_2^2$, 
with population values $\gamma_1=\kappa_3/\kappa_2^{3/2}$ and $\gamma_2=\kappa_4/\kappa_2^2$, respectively. 
\item[(iv)]
Noise-unbiased estimates (sometimes called `denoised' for brevity) of central moments and cumulants \citep{RimoldiniIntrinsic} are denoted by an asterisk superscript.
\item[(v)]
No systematics or instrumental errors are considered herein and uncertainties are often referred to as errors.
\item[(vi)]
The accuracy of an estimator is related to its distance from the true value and thus combines the concepts of  bias and precision, while the classification accuracy rate is the ratio between the number of true positives and the total number of \textcolor{black}{stars} (of a given type).
\item[(vii)]
The precision of an estimator is quantified by its dispersion, while the classification precision rate is the ratio between the number of true positives of a given type and the total number of \textcolor{black}{stars} classified as such a type.
\end{itemize}

\section{Method}
\label{sec:method}
The proposed weighting scheme is derived from linear interpolation and thus all of its pros and cons are inherited by definition.
\textcolor{black}{In particular, applications are limited to deterministic signals and the characterization of stochastic processes is excluded.
Considering the great variety of signal shapes and sampling laws, simulations are suggested to better assess the accuracy gains and losses of the method based on a representative subset of the data and estimators under consideration, as illustrated in Sec.~\ref{sec:hip}.}
Strictly speaking, statistical estimators like the mean and moments are not defined for deterministic signals. However, such estimators can still give acceptable approximations when the time scale of features (or period) is much smaller than the time series duration. In the case of mono-periodic signals, the proposed weighting scheme leads to estimator values of the signal as if the latter was measured on a complete cycle.

\begin{figure}
\center
\includegraphics[width=\columnwidth]{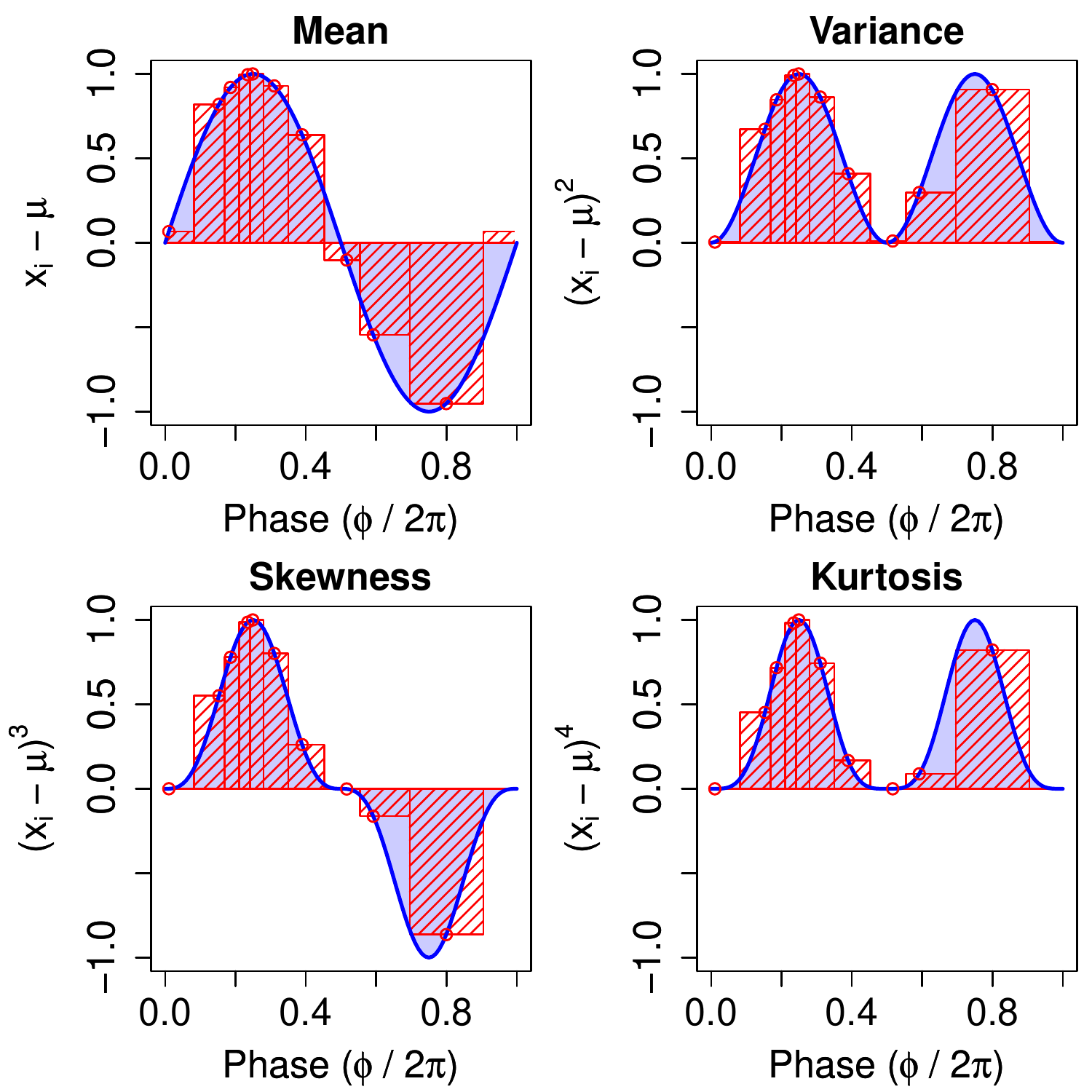}
\caption{A sinusoidal signal (blue curve) with mean $\mu$ is sampled unevenly by 10 measurements $x_i$ (red circles) and shown in the panel on the top-left hand side. The other panels illustrate different powers of deviations from the mean of the true signal (blue curve) and of the measurements (red circles). In particular, $(x_i-\mu)^r$ identifies terms associated with the central moments of order $r$: these estimator values are related to the areas enclosed by the blue curves and the zero level (shaded in blue). The weights defined by Eqs~(\ref{eq:pwi})--(\ref{eq:pwn}) define the variable widths of the shaded bars, which reduce the contribution of clustered measurements and increase the one of scattered data. In this case, although most measurements sample the first half of the signal, the increased weights assigned to the remaining data provides a better estimate of the area enclosed by the second half of the signal than using unweighted schemes.}
\label{fig:illustration}
\end{figure}

\textcolor{black}{Herein, the targeted} (population) value of a \textcolor{black}{generic} statistical parameter $\bar{\theta}$ of a continuous \textcolor{black}{deterministic} signal $x(t)$ in time~$t$ is computed by integrating \textcolor{black}{the estimator} function of such a signal \textcolor{black}{over the time series duration $T$}:
\begin{equation}
\textcolor{black}{\bar{\theta}=\frac{1}{T}\int_0^T \theta(x|t)\,{\mathrm d}t,  }
\label{eq:first}
\end{equation}
\textcolor{black}{where $\theta(x|t)$ is a function of the signal $x$ at a given time $t$. For example, in the case of a periodic source, time can be replaced by phase and $\bar{\theta}$ is represented by the mean areas shaded in blue in Fig.~\ref{fig:illustration}.} 
In \textcolor{black}{the} case of regular sampling in fine intervals,  
\textcolor{black}{replacing the integration with the} sum of discrete elements 
tends to the population result  for infinitely small  intervals. 
This limit is not necessarily satisfied when a continuous signal is sampled irregularly. 
The weighting scheme described in Sec.~\ref{sec:weights} recovers the property that the sum of discrete (weighted) terms approaches the result of the continuous function, in the limit of very dense (\textcolor{black}{although} not necessarily uniform) sampling.

Linear interpolation is a 
simple method to approximate a function with a broken line connecting the data points, 
provided the function is sufficiently sampled in time or phase.
In this section, such a function is represented by the expression of the additive terms of a statistical estimator. 
For example, a central moment of order $r$ is defined by the average of terms of the form $(x_i-\bar{x})^r$.
Results from linearly interpolating such a function are equivalent to those obtained from an effectively infinite regular sampling of the interpolated function.
In Sec.~\ref{sec:weights}, it is shown that the average of a linear interpolation \textcolor{black}{in time or phase} can be expressed as a weighted mean. 
\textcolor{black}{For example,} central moments of a sinusoidal signal are illustrated  in Fig.~\ref{fig:illustration} and they are related to the areas enclosed by $(\sin\phi)^r$, which are approximated by terms (\textcolor{black}{corresponding to the heights of bars}) with weights (\textcolor{black}{related to the widths of bars}) adapting to the sampling density in phase.

\textcolor{black}{Weighting can decrease the effective sample size,} 
since more importance is given to some measurements 
at the expense of other ones (e.g., a large sample with only a few  important elements 
will be similar to one with only these few elements) and exploiting correlations in the data with weights 
(e.g., assigning small weights to measurements separated by small intervals because their values are expected to be similar) might introduce biases if applied to expressions assuming independent data.
However, small biases could be justified by 
significant improvements in precision and a mixed weighting scheme which balances precision 
and accuracy depending on the signal-to-noise ($S/N$) level and sample size is described in Sec.~\ref{sec:mixed}.

\subsection{Weighting schemes}
\label{sec:weights}
If the targeted features of signals are sufficiently sampled in time, as for the data from the {\it Kepler} \citep{kepler} and {\it CoRoT} \citep{corot} missions, linear interpolation is expected to provide reliable approximations of 
\textcolor{black}{most deterministic signals}
and weights might be expressed in time domain \textcolor{black}{(see Sec.~\ref{sec:wtime})}. 
If typical time intervals between measurements are larger than the resolution needed to sample the signal (as it is often the case in the {\it Hipparcos} and {\it Gaia} surveys) but the latter is mono-periodic or dominated by a single period,  sampling  can be improved by folding the light curve with the value of the fundamental period and computing interpolations (or weights) in phase \textcolor{black}{(see Sec.~\ref{sec:wphase})}. 

The interpolation-based weighting scheme \textcolor{black}{naturally removes importance from clumped data (oversampling the same region of the signal) in favour of more scattered measurements (probing extended parts of the signal)}.
If gaps in time or phase occasionally cover a relevant fraction of a signal, measurements at the gap boundaries 
might not provide sufficient information on the features of the signal and the importance of such data could be limited, for example, to some maximum weight value. 
This strategy can extend the applicability of the weighting scheme \textcolor{black}{described in Sec.~\ref{sec:wtime}} to under-sampled \textcolor{black}{deterministic} signals, periodic or non-periodic, to mitigate statistical biases due to data clumps on scales smaller than the signal features \textcolor{black}{aimed at. For example, rare stellar bursts of short duration are expected to be associated with a skewed distribution of magnitudes. If sampling happens to be denser during one of the bursts, the value of the skewness can differ strongly, while appropriately weighting dense measurements could alleviate the problem. }

\subsubsection{Weights in time}
\label{sec:wtime}
If an estimator $\bar{\theta}$ is defined by the average of a function in time \textcolor{black}{as in Eq.~(\ref{eq:first})}, it can be approximated by the mean of the linear interpolation
of terms $\theta_i$ at times $t_i$, where $i\in(1,n)$ for a time series of $n$ measurements sorted in time, 
and it can be expressed as a weighted average as follows:
\begin{align}
\bar{\theta}
\approx&\, \frac{1}{t_n - t_1}\sum_{i=1}^{n-1} \int_{t_i}^{t_{i+1}} \left[\theta_i+\frac{\theta_{i+1}-\theta_i}{t_{i+1}-t_i}\left(t-t_i\right) \right] {\mathrm d}t \label{eq:interpolation}\\
=&\, \frac{1}{t_n - t_1}\sum_{i=1}^{n-1} \frac{\theta_{i}+\theta_{i+1}}{2} \left(t_{i+1}-t_i\right) \\
=&\, \frac{1}{2(t_n - t_1)}\left[\sum_{i=1}^{n-1}  \theta_{i} \left(t_{i+1}-t_{i}\right) +\sum_{i=2}^{n} \theta_i\left(t_{i}-t_{i-1}\right) \right] \\
=&\, \frac{1}{2(t_n - t_1)}\left[\sum_{i=2}^{n-1}  \theta_{i} \left(t_{i+1}-t_{i-1}\right) + \right.\nonumber \\
& \left.~~~~~~~~~~~~~~~~~~ +\theta_1\left(t_{2}-t_{1}\right) +\theta_n \left(t_{n}-t_{n-1}\right)  \rule{0cm}{0.5cm}\right] \\
=&\,\frac{1}{W} \sum_{i=1}^{n} w_i \, \theta_i,
\label{eq:lambdaTheta}
\end{align}
where
\begin{align}
&w_i=t_{i+1}-t_{i-1}~~~~~\forall i \in (2,n-1) \label{eq:twi}\\
&w_1 = t_{2}-t_{1}\\
&w_n =t_{n}-t_{n-1}, \label{eq:twn}
\end{align}
and
\begin{align}
W=\sum_{i=1}^n w_i=2(t_n-t_1).
\end{align}
When  differences between successive times $t_i$ are too large for sensible interpolation, they might be limited to some maximum interval $\Delta t_{\max}$ as follows:
\begin{align}
w_i=&\min\left\{t_{i+1}-t_i, \, \Delta t_{\max} \right\}+ \nonumber \\
&+\min\left\{t_{i}-t_{i-1}, \, \Delta t_{\max} \right\}~~~~~\forall  i \in (2,n-1) \label{eq:twi_delta}\\
w_1=&\min\left\{t_{2}-t_1, \, \Delta t_{\max} \right\}\\
w_n=&\min\left\{t_{n}-t_{n-1}, \, \Delta t_{\max} \right\} \label{eq:twn_delta}
\end{align}
and, of course, then $W\leq 2(t_n-t_1)$.

\subsubsection{Weights in phase}
\label{sec:wphase}
If sampling in time 
is sparse
 but the data exhibit periodicity, the time series can be folded with the dominant period 
to increase the average sampling rate in phase  
(by a factor \textcolor{black}{of the order of} the duration of the time series divided by the period).  
  
Since interpolation of phase-sorted data is carried over from the last to the first point in phase, weights in phase $\phi$  corresponding to Eqs~(\ref{eq:twi_delta})--(\ref{eq:twn_delta}) become (see Appendix~\ref{app:periodicInterpolation}):
\begin{align}
w_i=&\min\left\{\phi_{i+1}-\phi_i, \, \Delta\phi_{\max} \right\}+ \nonumber \\
&+\min\left\{\phi_{i}-\phi_{i-1}, \, \Delta\phi_{\max} \right\}~~~~~\forall  i \in (2,n-1) \label{eq:pwi_delta}\\
w_1=&\min\left\{\phi_{2}-\phi_1, \, \Delta\phi_{\max} \right\}+ \nonumber \\
&+\min\left\{\phi_{1}-\phi_{n}+2\pi, \, \Delta\phi_{\max} \right\} \\
w_n=&\min\left\{\phi_{n}-\phi_{n-1}, \, \Delta\phi_{\max} \right\}+ \nonumber \\
&+\min\left\{\phi_{1}-\phi_{n}+2\pi, \, \Delta\phi_{\max} \right\} \label{eq:pwn_delta}
\end{align}
with $W\leq 4\pi$. 

\subsubsection{\textcolor{black}{Examples of mixed weighting schemes}}
\label{sec:mixed}
In order to avoid interpolating large noise fluctuations, 
weights could be set to reduce to inverse-squared uncertainties at low $S/N$, which proved more precise in the simulations described by \citet{RimoldiniIntrinsic}. The transition between high and low $S/N$ regimes could be pursued with \textcolor{black}{weights of the form
\begin{align}
&w_i= h(S/N|a,b)\,\frac{w'_i}{\sum_{j=1}^n w'_j}+ \left[1-h(S/N|a,b)\right]\frac{\epsilon_i^{-2}}{\sum_{j=1}^n \epsilon_j^{-2}} \label{eq:mixed},
\end{align}
where $\epsilon_i$ and $w'_i$ denote the uncertainty and interpolation-based weight (introduced in the preceding paragraphs) associated with the $i$-th measurement, respectively, and
\begin{align}
h(S/N|a,b)=\frac{1}{1+e^{-(S/N-a)/b}}~~~\mbox{for}~~a,b>0.
\label{eq:h}
\end{align}
The family of functions defined in Eq.~(\ref{eq:h}) is only one example of many possible alternatives with the same limit behaviours. The dependence on the $S/N$ ratio and a set of tuning parameters (such as $a,b$) make it possible to control a mixed weighting scheme and in particular regulate the transition of weights from $w'_i$ to $\epsilon_i^{-2}$ in the limits of high and low $S/N$ ratios, respectively.
In the same spirit, a further} function of the form $h(n|a',b')$ could  be used to reduce the relevance of interpolation-based weights for small sample sizes $n$
\textcolor{black}{(as alternative to weights limited by maximum interpolation intervals):}
\begin{align}
&w'_i= h(n|a',b')\,\frac{w''_i}{\sum_{j=1}^n w''_j}+ \left[1-h(n|a',b')\right]/n ,
\end{align}
where $w''_i$ is defined by Eqs~(\ref{eq:twi})--(\ref{eq:twn}) or (\ref{eq:pwi})--(\ref{eq:pwn}).
Tuning parameters $a,a',b,b'$ offer the possibility to reach a compromise solution between precision and accuracy at high and low values of $n$ and $S/N$ ratios, according to the specific set of estimators,  signals, sampling, errors,  sample sizes and their distributions in the data. 
\textcolor{black}{The determination of such parameters might involve the maximization of the overall accuracy of the set of estimators under consideration, employing data simulated in a context similar to the  one intended for analysis.
The effect of such tuning parameters on simple simulated signals is illustrated in \citet{RimoldiniUnbiased,RimoldiniIntrinsic} as a function of sample size and $S/N$ ratio, while an application to real data is presented in the next section.}

\section{{\em Hipparcos} periodic variable stars}
\label{sec:hip}
The effect of  interpolation-based weights on statistical parameters was explored for a realistic distribution of signals, represented by time series from the {\it Hipparcos} catalogue of periodic variable stars.
The {\it Hipparcos} mission \citep{ESA} performed astrometric and photometric measurements of the brightest sources in the sky.
The full catalogue \citep{Perryman} contains 118\,204 sources with photometry. Among the 11\,597 stars identified as variables, a reliable period could be computed (or was consistent with the one from literature) for  2\,712 objects, which were published in the periodic catalogue \citep[Vol.~11][]{ESA,EyerThesis}. 
This set of sources, which contained light variations  dominated mostly by single periods, was chosen to illustrate the application of one of the weighting schemes described in Sec.~\ref{sec:method}. 

\subsection{Statistical parameters}
\label{sec:pars}
In order to assess the accuracy of estimators with respect to the (unknown) real signal, the latter was assumed to be represented by a model of the  time series and \textcolor{black}{the magnitude measurements} were simulated employing \textcolor{black}{the true} uncertainties as Gaussian random variables (around the model) \textcolor{black}{at the phases given by the real data}.
Only good quality measurements were accepted, flagged by the field HT4 as zero or one \citep[see Vol.~1 of][]{ESA}, which reduced the number of sources considered to 2683.\footnote{The {\it Hipparcos} identifiers of sources with no good quality measurements (i.e., with field HT4 $> 1$ only) were: 1196, 10027, 17878, 20570, 24019, 25673, 39084, 42715, 42726, 46502, 52538, 53937, 58112, 60904, 61997, 63125, 69582, 72583, 88905, 90026, 93595, 93724, 96007, 99675, 102246, 102409, 112317, 112470 and 118188.}

\subsubsection{Light-curve models}
Time series were folded with the period provided by the {\it Hipparcos} catalogue\footnote{The mean period \citep[field P11 of][]{ESA} derived from the {\it Hipparcos} data was employed. If this was not available, the period from the literature (listed in field P18) was considered.} and modelled by a cubic smoothing spline \citep[the \texttt{smooth.spline} function from the \texttt{stats} package in R,][]{R}. The condition of periodicity at the boundaries was approximated by replicating a whole cycle of folded data before and after the cycle considered as reference for the `true' statistical parameters and for the generation of simulated data. The smoothing parameter was estimated from the data with a generalized cross-validation method \citep[e.g.,][]{GCV} and the degrees of freedom were adjusted in special cases, to mitigate over-fitting highly clumped data with large gaps as well as under-fitting the profile of eclipsing binaries.\footnote{The number of degrees of freedom \texttt{df} depended on the unweighted standardized sample skewness $\mathcal{S}$ of the data, which identified light variations typical of eclipsing binaries, and the ratio $\mathcal{R}$ between the median and the third largest gaps in phase, to better deal with clumped data with large gaps. Denoting by $n$ the number of measurements in a time series, \texttt{df} equalled 15, 24 and 36 for $\mathcal{R}$ less than $6/n$, $10/n$ and $20/n$, respectively, provided $\mathcal{S}<1.5$. For greater values of $\mathcal{S}$, \texttt{df} was set to the smallest integer not less than $3n/5$. The quoted values of \texttt{df} take into account the replicated data at each of the extremes of the folded light curve (to induce quasi-periodic boundary conditions on the model of the central cycle).}
The combination of large gaps and sharp features did not lead to accurate models  and other methods 
might help avoid modelling artefacts in gapped data, although these are out of the scope of this application.
Also, cases in which sparse sampling missed important features could not  be improved. Nevertheless,  the smoothed best-fitting curves seemed to capture the relevant shapes of  true signals in most cases.

The resulting models are presented together with the original and simulated data in Appendix~\ref{app:data}. The difficulty to achieve accurate models for all sources and the need of more complex modelling techniques were confirmed. While models were sometimes not ideal, most of them were of sufficient quality to supply a realistic distribution of the relevant  features of the {\it Hipparcos} periodic variable stars. 
Differences from data of other surveys were expected to be  greater than those due to modelling inaccuracies.
Less than one per cent of all sources (23 time series)\footnote{The {\it Hipparcos} identifiers of the 23 sources removed from consideration because of poor modelling were: 1901, 4279, 21600, 23416, 23453, 25591, 32397, 40853, 42853, 45094, 48054, 58854, 61281, 68064, 73533, 76152, 89579, 90313, 95611, 96739, 104483, 108317 and 112928.} with significant modelling artefacts in gaps with no data were removed (many of them were EA-type eclipsing binaries  with large data gaps), thus 2660 objects were included in the assessment of statistical parameters. The number of measurements per time series ranged from 18 to 331, with $S/N$ ratios from 0.2 to  116, according to the definition in Eq.~(\ref{eq:snr}).

\subsubsection{Weighting scheme}
The {\it Hipparcos} light curves were folded with the catalogue period and interpolation-based weights were computed in phase.
For brevity, such weights are referred to as `phase weights', while inverse-squared uncertainties are called `error weights'.

The estimators employed herein included weighted moments and cumulants corrected for biases from Gaussian uncertainties \citep[called `noise-unbiased' or `denoised' estimators; see][]{RimoldiniIntrinsic} and some robust weighted measures, as defined in Appendix~\ref{app:def}.
Simulations in \citet{RimoldiniUnbiased,RimoldiniIntrinsic} suggested that noise-unbiased phase-weighted sample moments can be more accurate for $S/N>2$ and sample sizes $n>20$ with respect to other schemes, while error weighting appeared the most appropriate option for noisy signals. 
Thus, the weighting scheme chosen for this application combined error and phase information as described by Eqs~(\ref{eq:mixed})  and (\ref{eq:pwi})--(\ref{eq:pwn}). 
The balance between phase weights (at high $S/N$) and error weights (at low $S/N$) was controlled by the parameters $a=3$ and $b=1.2$, which provided a satisfactory overall accuracy for the set of estimators and light-curve shapes of the {\it Hipparcos} periodic variable stars.
The signal-to-noise ratio $S/N$ was estimated from the true (model) signal variance $\mu_2$ and the average of squared measurement uncertainties $\epsilon_i$ as follows:
\begin{equation}
S/N=\left[\frac{\mu_2}{n^{-1}\sum_{i=1}^{n} \epsilon_i^2} \right]^{1/2}.
\label{eq:snr}
\end{equation}

\subsubsection{Results}
\begin{figure*}
\begin{minipage}{\columnwidth}
\center
~~~~~~~~~~~~~(a)\\
\vspace{-0.25cm}
\includegraphics[width=\columnwidth]{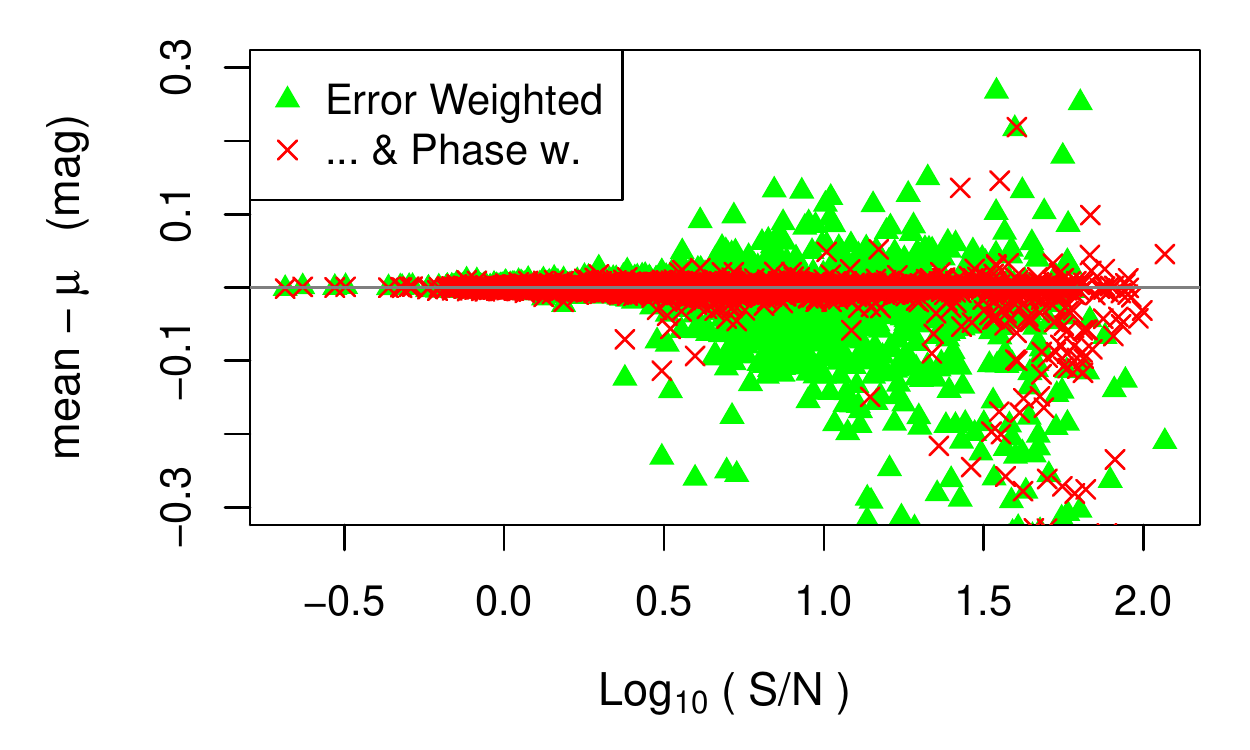}\\
~~~~~~~~~~~~~(c)\\
\vspace{-0.25cm}
\includegraphics[width=\columnwidth]{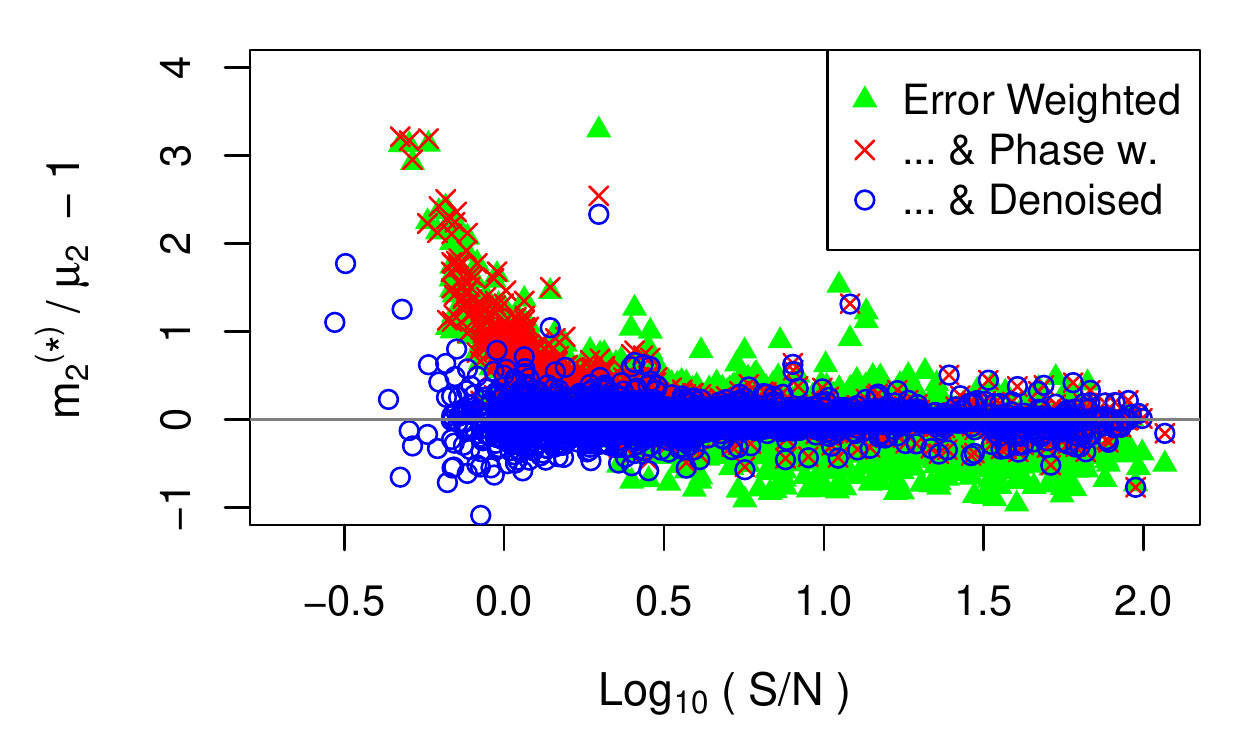}\\
~~~~~~~~~~~~~(e)\\
\vspace{-0.25cm}
\includegraphics[width=\columnwidth]{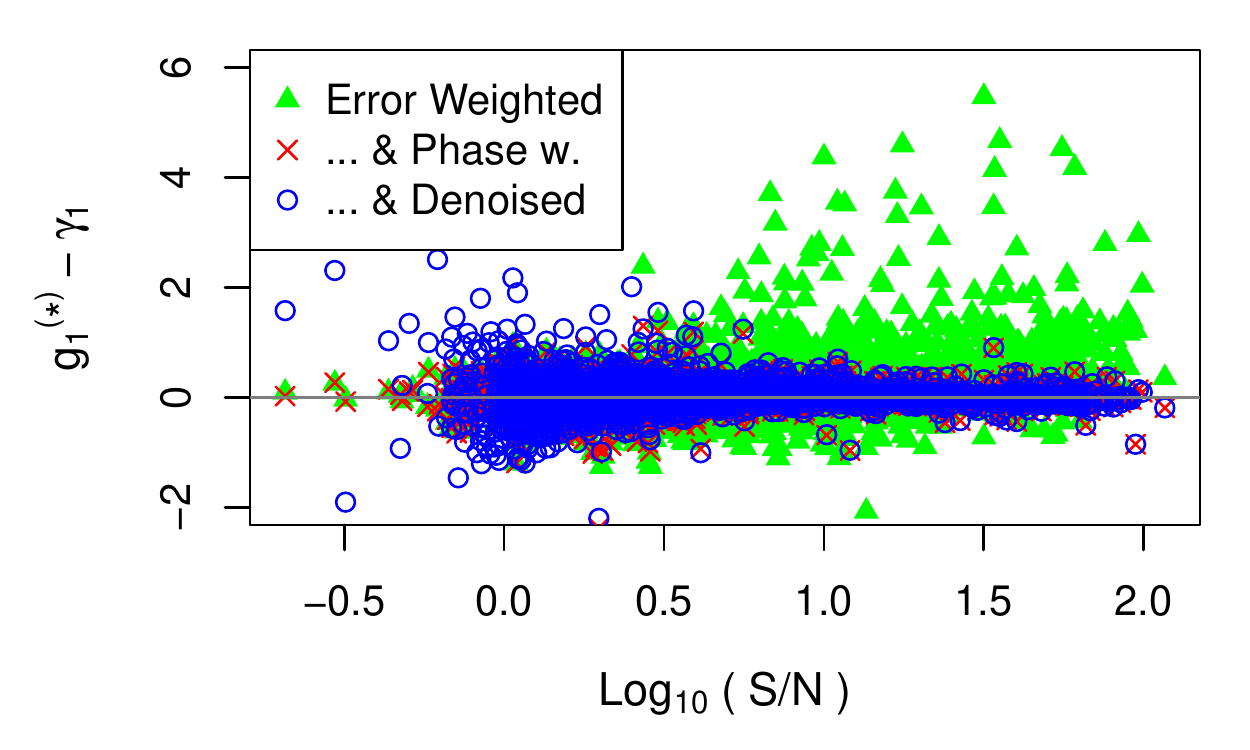}
\end{minipage}
\begin{minipage}{\columnwidth}
\center
~~~~~~~~~~~~~(b)\\
\vspace{-0.25cm}
\includegraphics[width=\columnwidth]{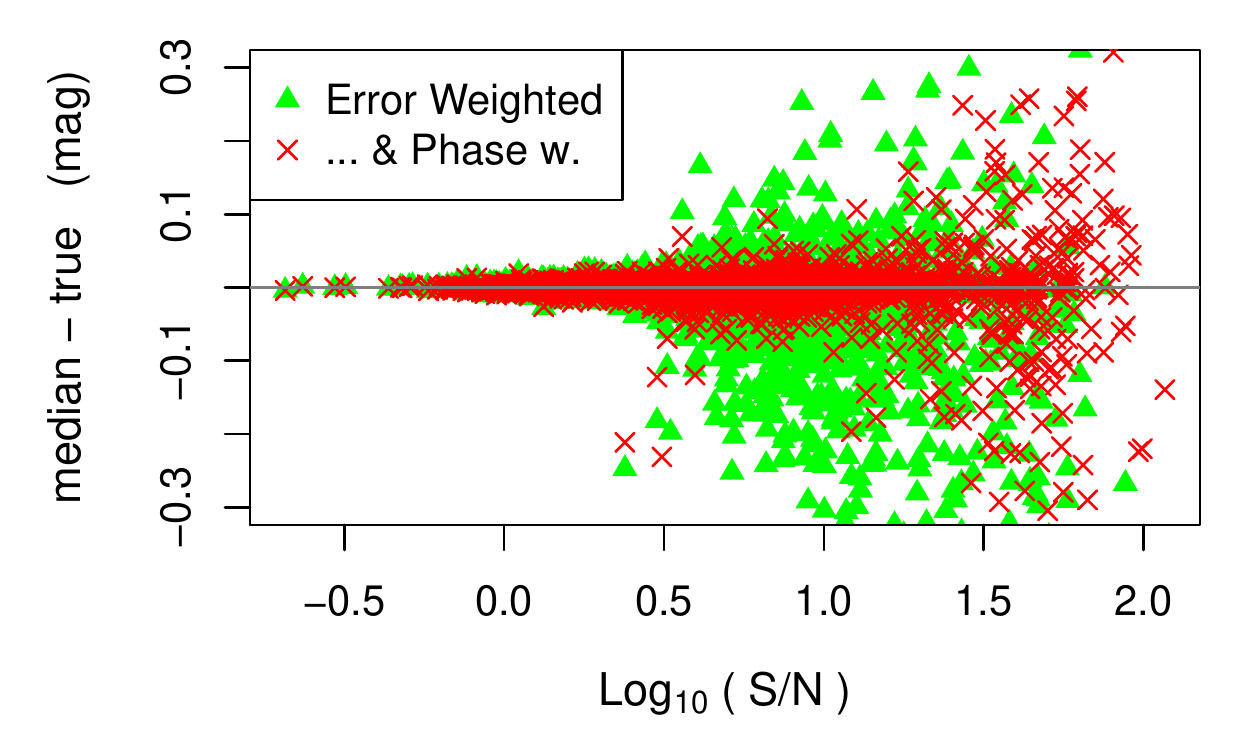}\\
~~~~~~~~~~~~~(d)\\
\vspace{-0.25cm}
\includegraphics[width=\columnwidth]{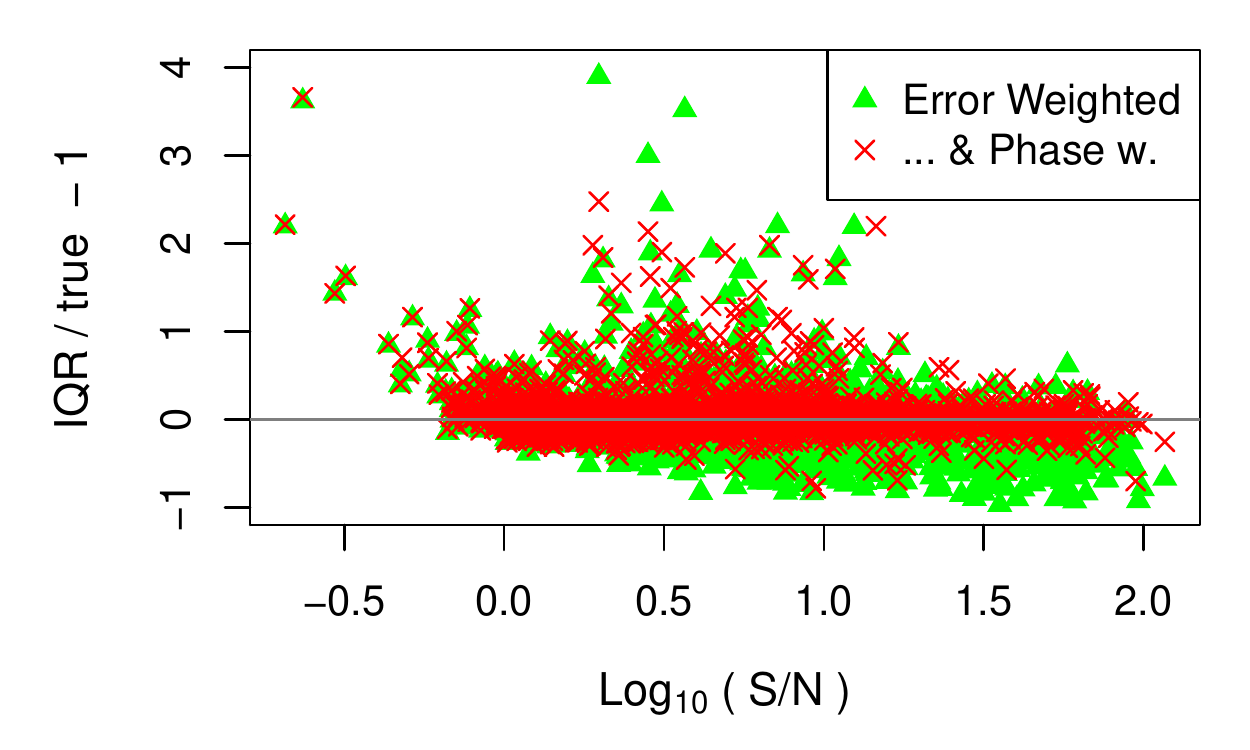}\\
~~~~~~~~~~~~~(f)\\
\vspace{-0.25cm}
\includegraphics[width=\columnwidth]{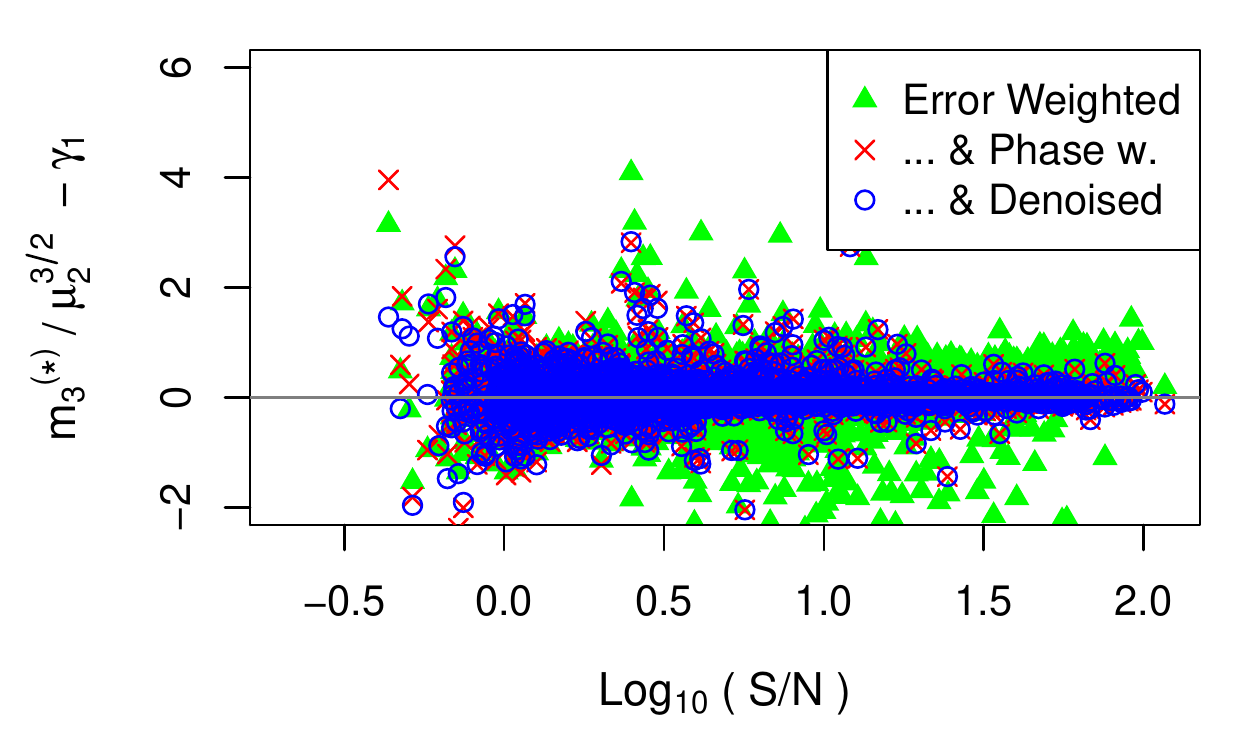}
\end{minipage}
\caption{Deviations of the mean, median, variance, interquartile range (IQR) and skewness from their true values, as a function of $S/N$ ratio. Green triangles denote error-weighted estimators, red crosses indicate error-phase weighted estimators and blue circles represent noise-unbiased error-phase weighted estimators. In the case of the median and IQR, `true' refers to the true median and IQR values, respectively.}
\label{fig:noise1}
\end{figure*}

\begin{figure*}
\begin{minipage}{\columnwidth}
\center
~~~~~~~~~~~~~(a)\\
\vspace{-0.25cm}
\includegraphics[width=\columnwidth]{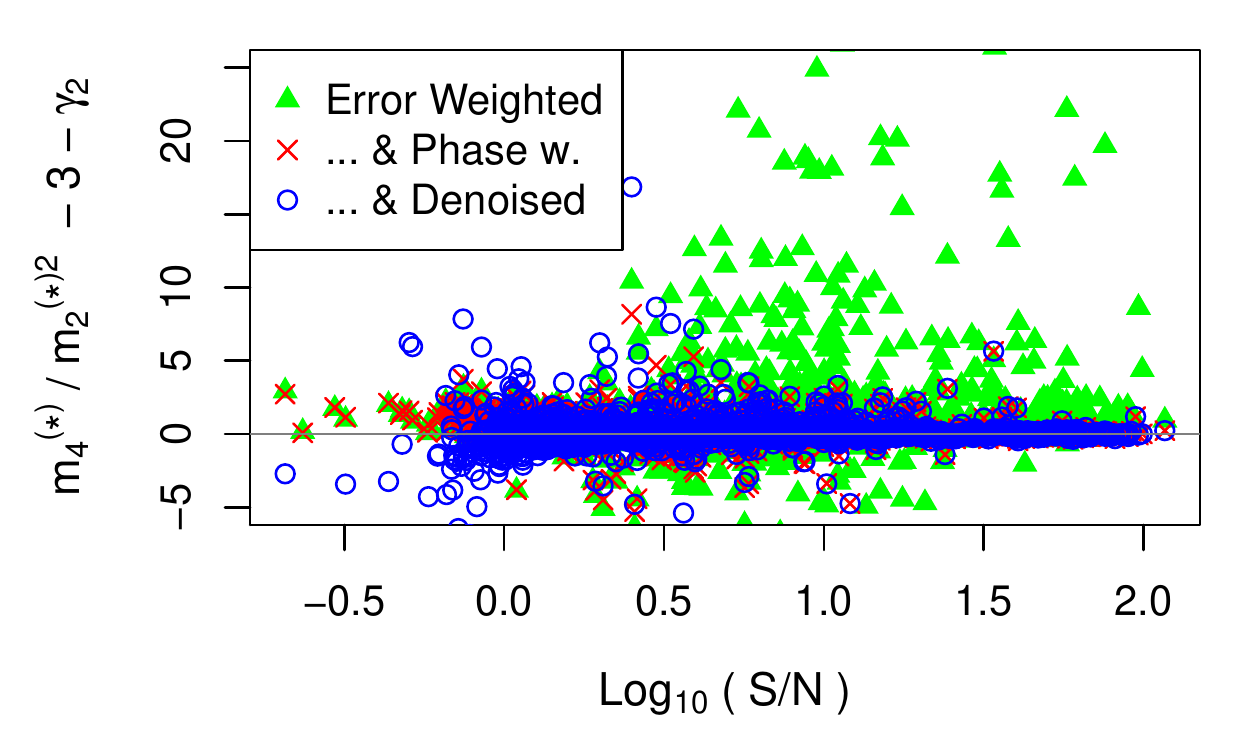}\\
~~~~~~~~~~~~~(c)\\
\vspace{-0.25cm}
\includegraphics[width=\columnwidth]{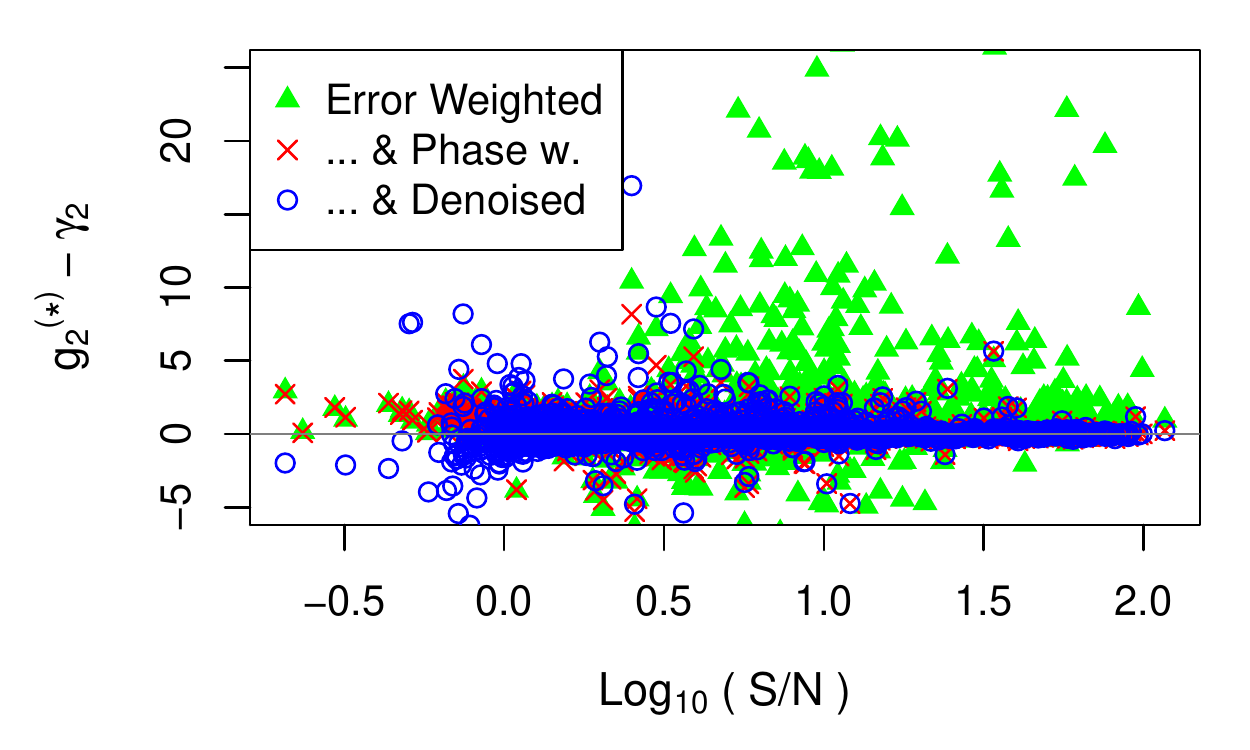}
\end{minipage}
\begin{minipage}{\columnwidth}
\center
~~~~~~~~~~~~~(b)\\
\vspace{-0.25cm}
\includegraphics[width=\columnwidth]{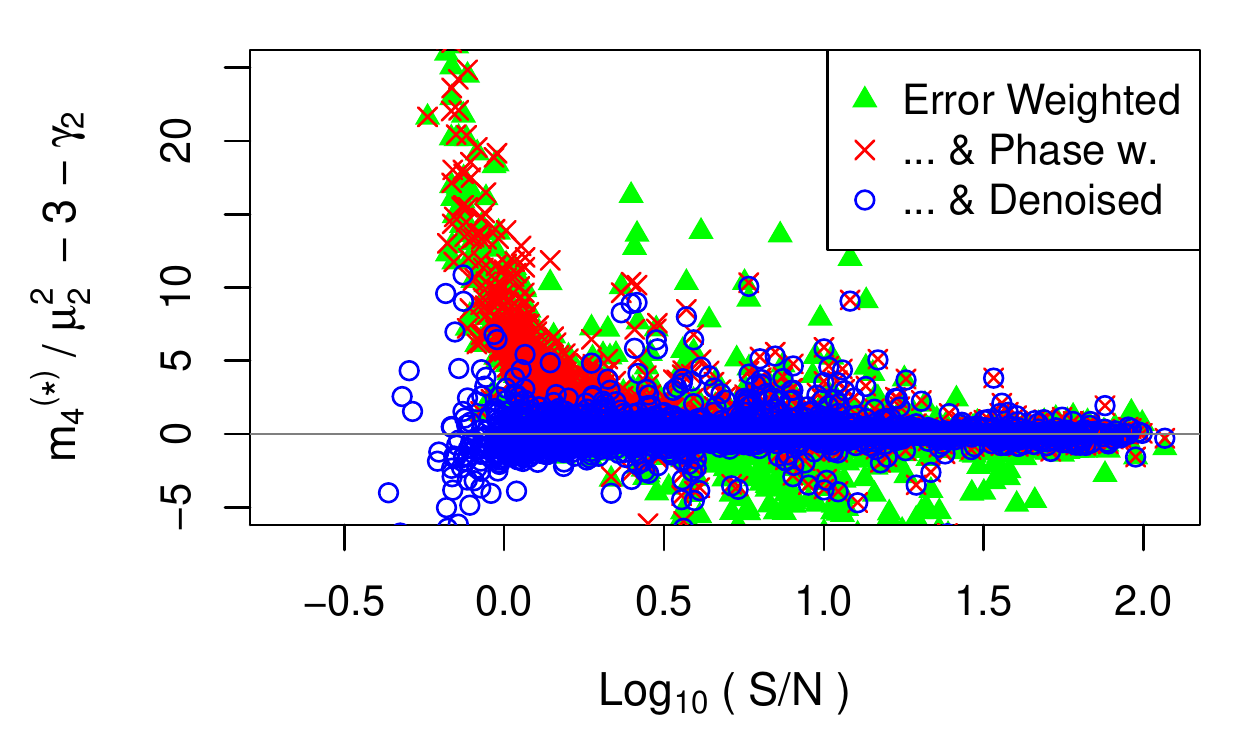}\\
~~~~~~~~~~~~~(d)\\
\vspace{-0.25cm}
\includegraphics[width=\columnwidth]{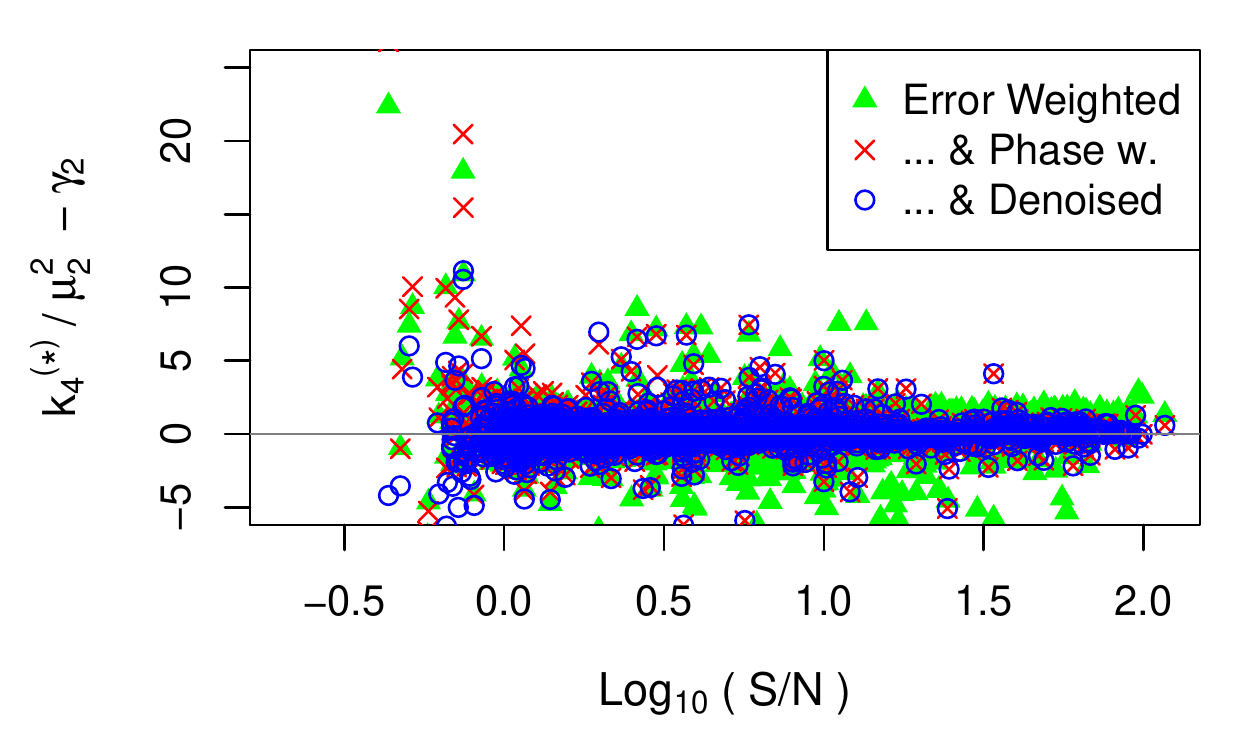}
\end{minipage}
\caption{Deviations of the kurtosis moment and cumulant from their true values, as a function of $S/N$ ratio. Green triangles denote error-weighted estimators, red crosses indicate error-phase weighted estimators and blue circles represent noise-unbiased error-phase weighted estimators.}
\label{fig:noise2}
\end{figure*}

The deviations of error-weighted sample statistical parameters from population values (computed from a fine regular sampling of the models) are compared to the noise-unbiased error-phase weighted counterparts as a function of $S/N$ ratio in Figs~\ref{fig:noise1}--\ref{fig:noise2}.
Error-phase weighted (and noise-unbiased, when applicable) estimators were generally more accurate than error-weighted estimators.
Since larger light variations were correlated with higher $S/N$ levels, the effect of the correlation between errors and magnitudes \citep[e.g., see][]{VanLeeuwen} was visible when weighting by errors at high $S/N$: the mean was biased towards brighter measurements, the variance tended to be smaller than the real value and the scatter of higher moments around the true values was greater. 
Such effects were significantly alleviated by error-phase weights.
Also, the accuracy of the noise-unbiased variance $m_2^*$ and kurtosis moment $m_4^*$ were much improved at $S/N<3$. In the particular case of the standardized kurtosis, the value of $m_4/m_2^2-3-\gamma_2$ equalled the one of the cumulant $k_4/m_2^2-\gamma_2$ by definition, for any weighting scheme (unlike the noise-unbiased counterparts).\footnote{The deviations from model values of the standardized noise-unbiased kurtosis moment and cumulant differ because $k^*_4$ involves a term in $(m^2_2)^*$, which does not simplify after the normalization by $(m^*_2)^2$.}

Estimators which provide similar information (such as the mean and median, variance and interquartile range, skewness and kurtosis standardized by the estimated or true variance) are compared in scatter plots in Appendix~\ref{app:scatter}. Although contributions from different $S/N$ ratios are not distinguished, error-phase weighted (and noise-unbiased, when applicable) estimators are always associated with more strongly peaked distributions around the true values than the error-weighted analogues.

\begin{figure*}
\begin{minipage}{\columnwidth}
\center
~~~~~~~~~(a)\\
\vspace{-0.2cm}
\includegraphics[width=\columnwidth]{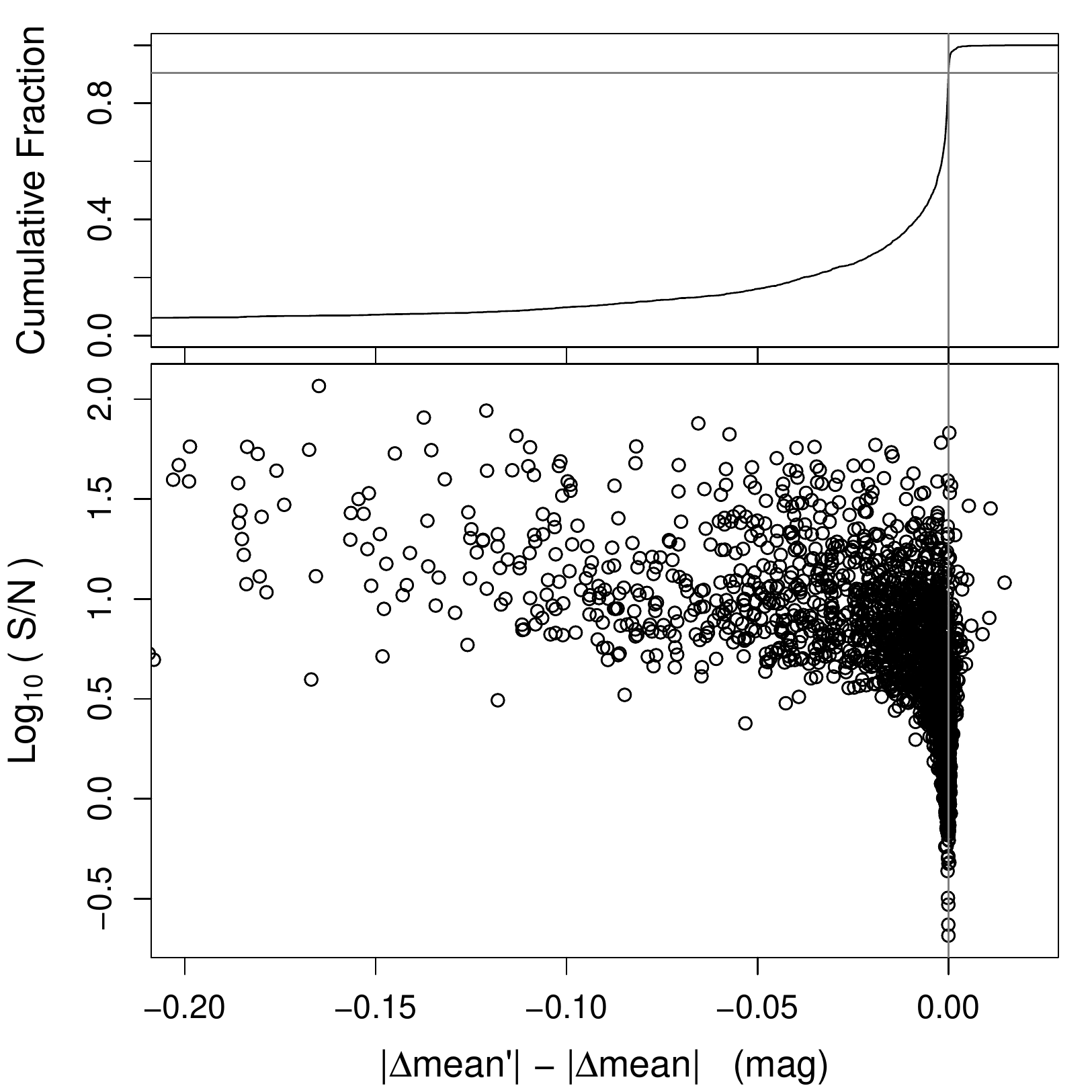}\\
~~~~~~~~~(c)\\
\vspace{-0.2cm}
\includegraphics[width=\columnwidth]{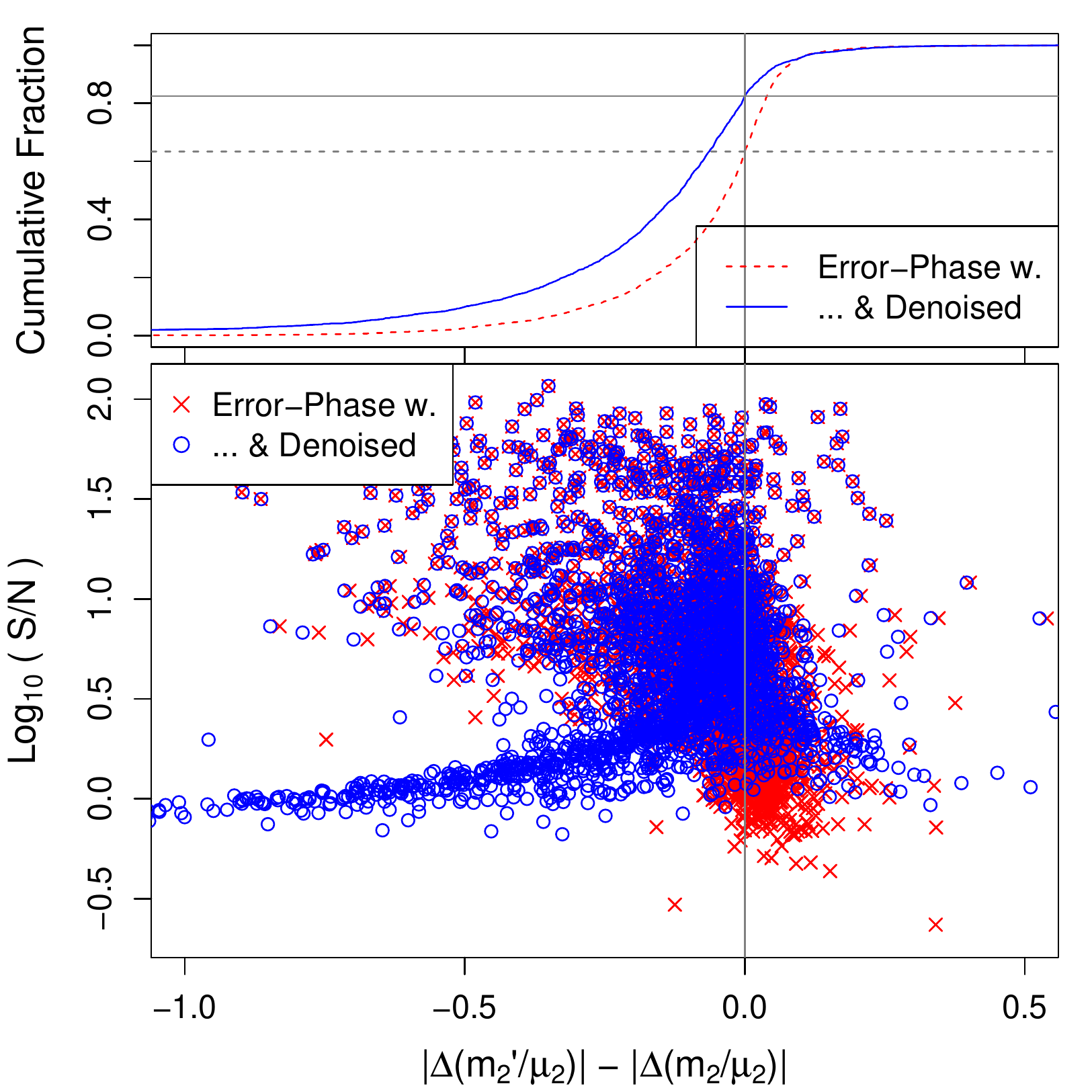}
\end{minipage}
\begin{minipage}{\columnwidth}
\center
~~~~~~~~~(b)\\
\vspace{-0.2cm}
\includegraphics[width=\columnwidth]{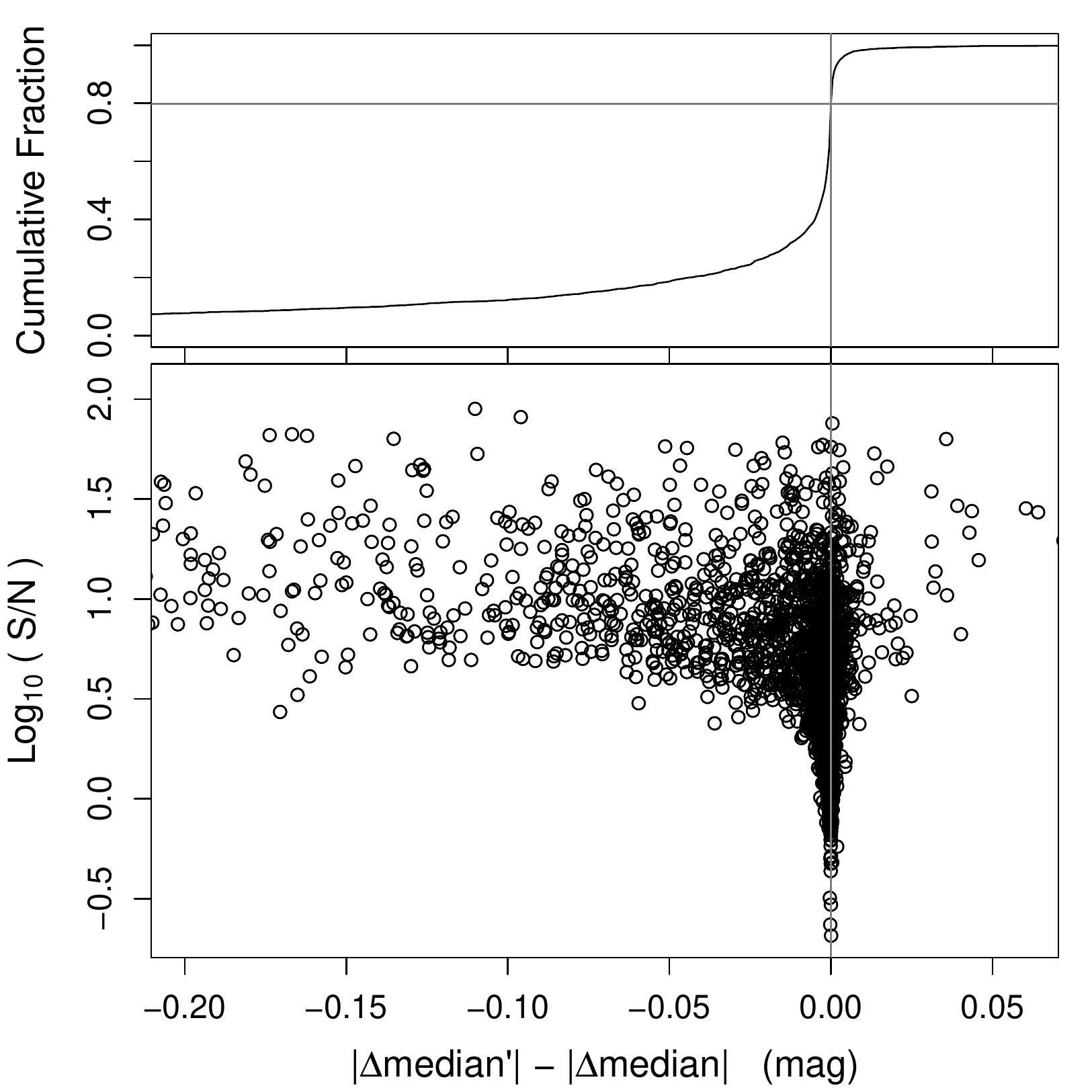}\\
~~~~~~~~~(d)\\
\vspace{-0.2cm}
\includegraphics[width=\columnwidth]{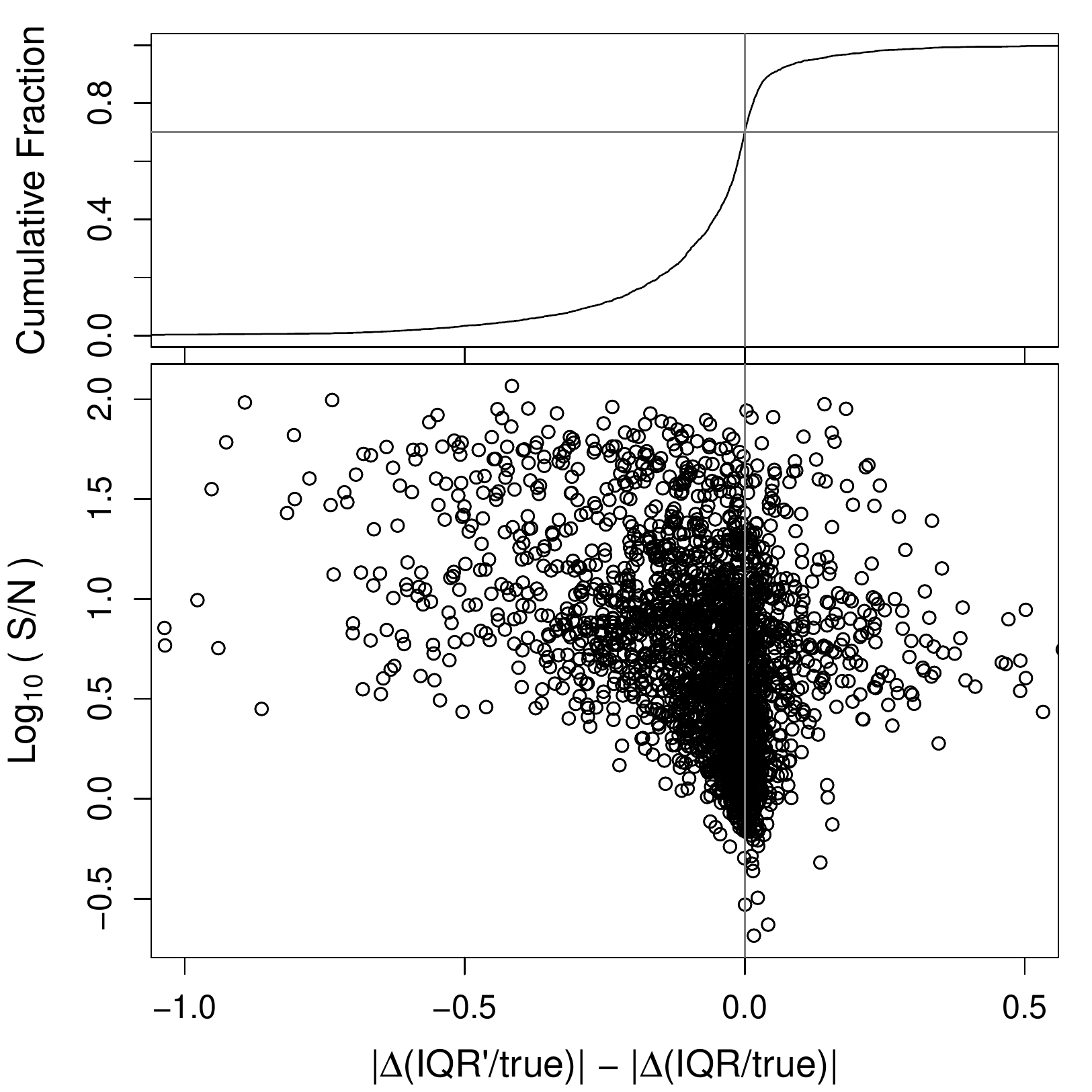}
\end{minipage}
\caption{The absolute deviations from population values of error-phase weighted estimators (red crosses and dashed lines, or black circles and black solid lines) with additional denoising, when applicable (blue circles and blue solid lines), are denoted by a prime symbol and their difference from the error-weighted counterparts is presented as a function of $S/N$ ratio.  Negative values indicate improvements in accuracy with respect to error-weighted estimators. The fraction and magnitude of improved cases can be inferred from the cumulative distributions in the top panels. 
The following estimators are included: (a) mean, (b) median, (c) variance and (d) interquartile range.}
\label{fig:quantile1}
\end{figure*}

\begin{figure*}
\begin{minipage}{\columnwidth}
\center
~~~~~~~~~(a)\\
\vspace{-0.2cm}
\includegraphics[width=\columnwidth]{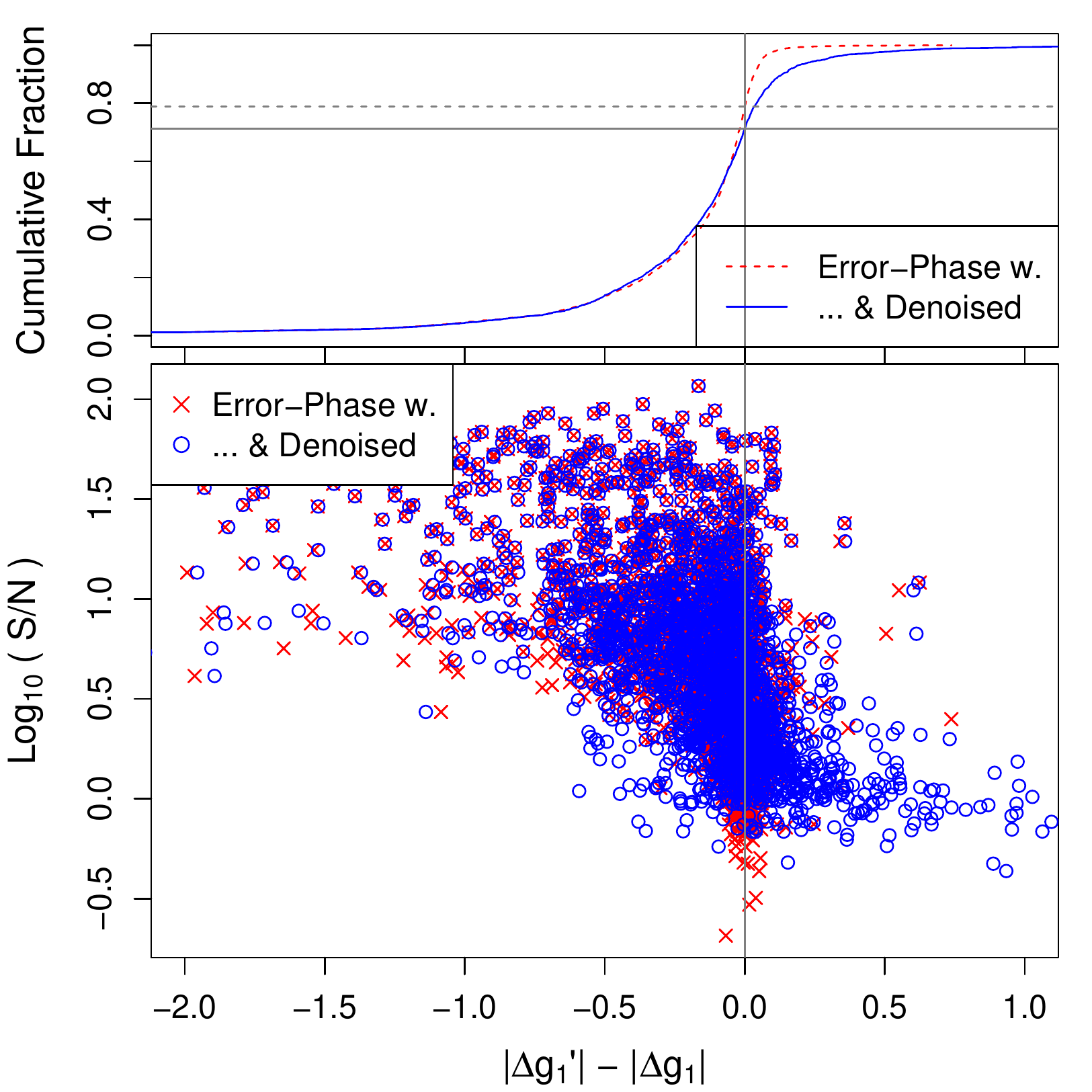}\\
~~~~~~~~~(c)\\
\vspace{-0.2cm}
\includegraphics[width=\columnwidth]{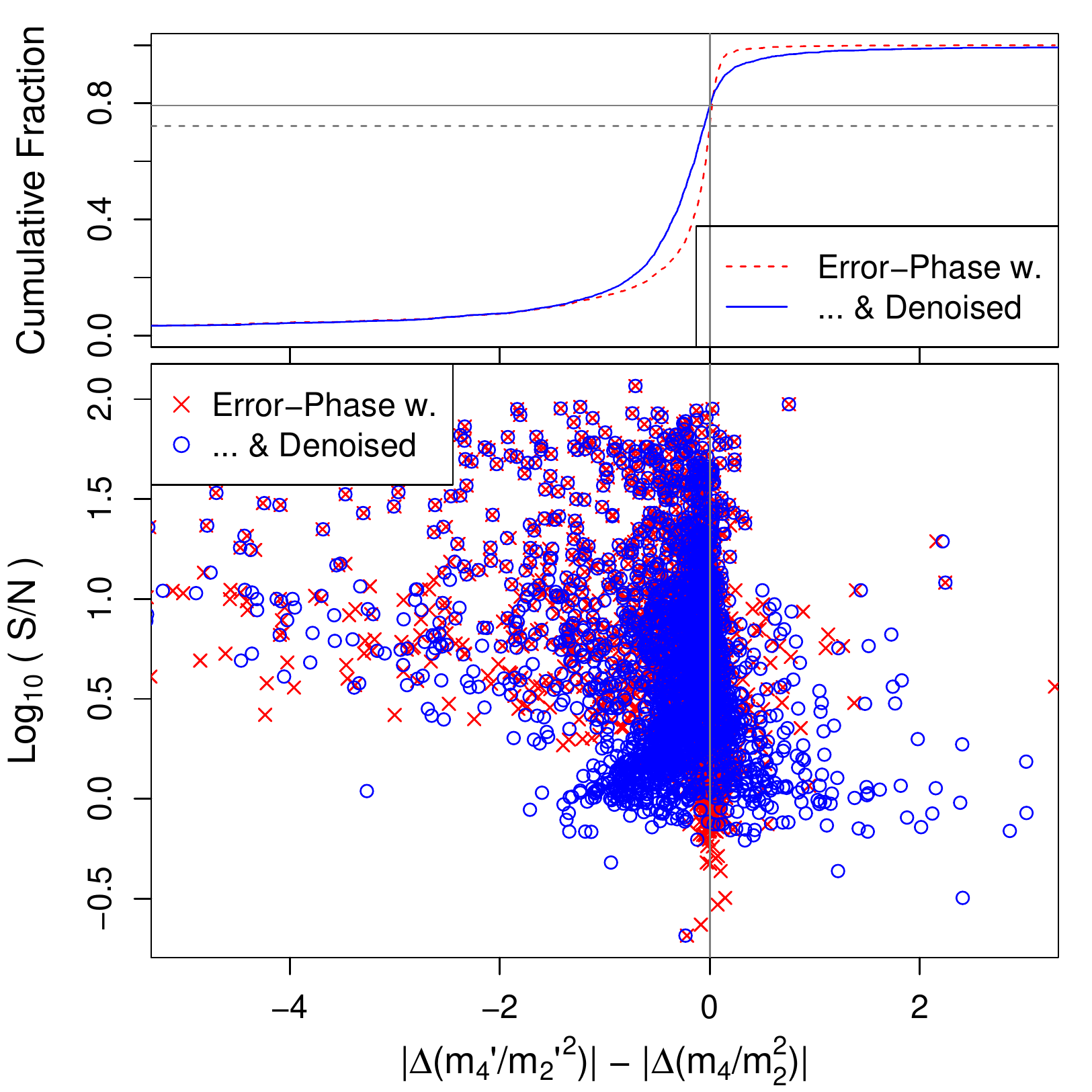}
\end{minipage}
\begin{minipage}{\columnwidth}
\center
~~~~~~~~~(b)\\
\vspace{-0.2cm}
\includegraphics[width=\columnwidth]{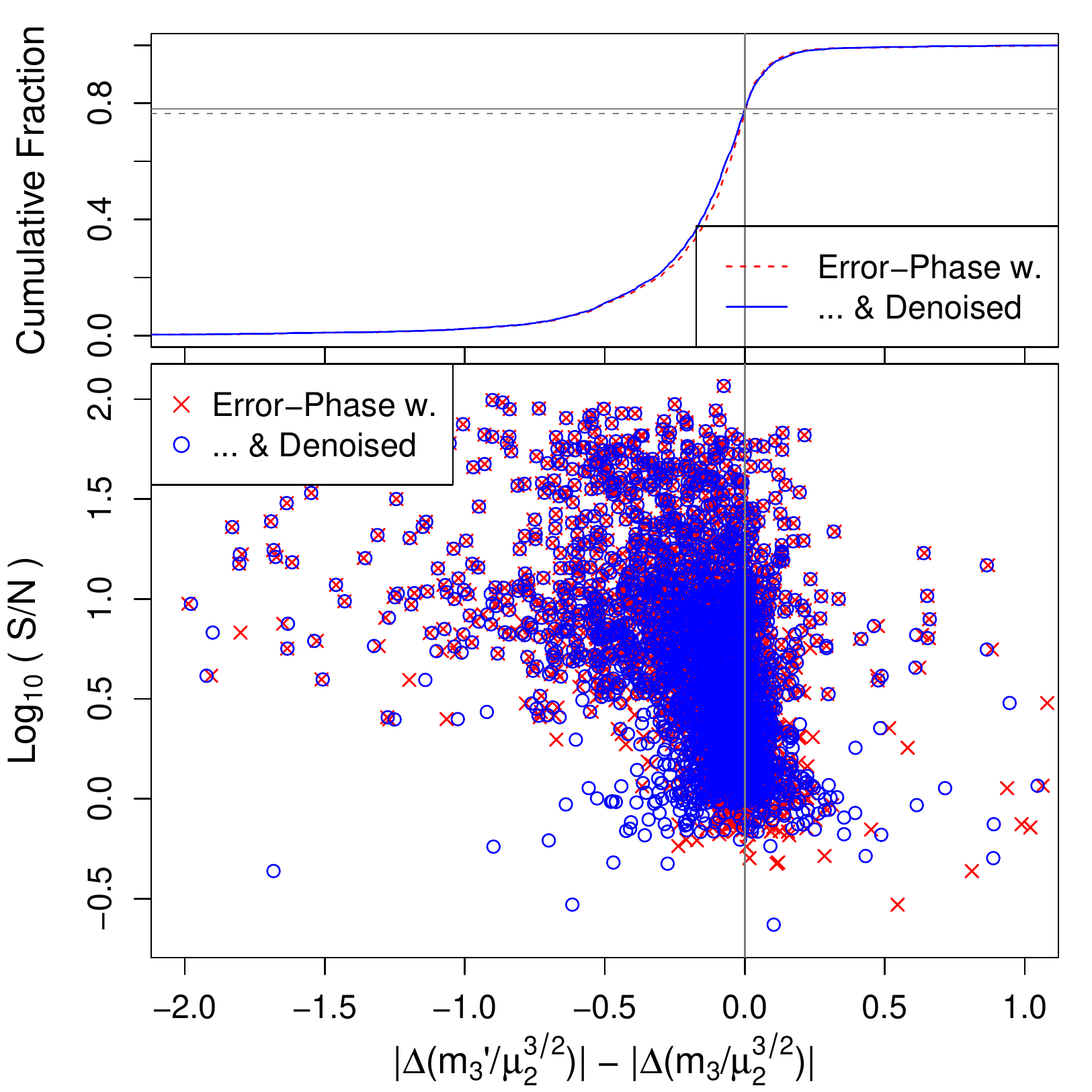}\\
~~~~~~~~~(d)\\
\vspace{-0.2cm}
\includegraphics[width=\columnwidth]{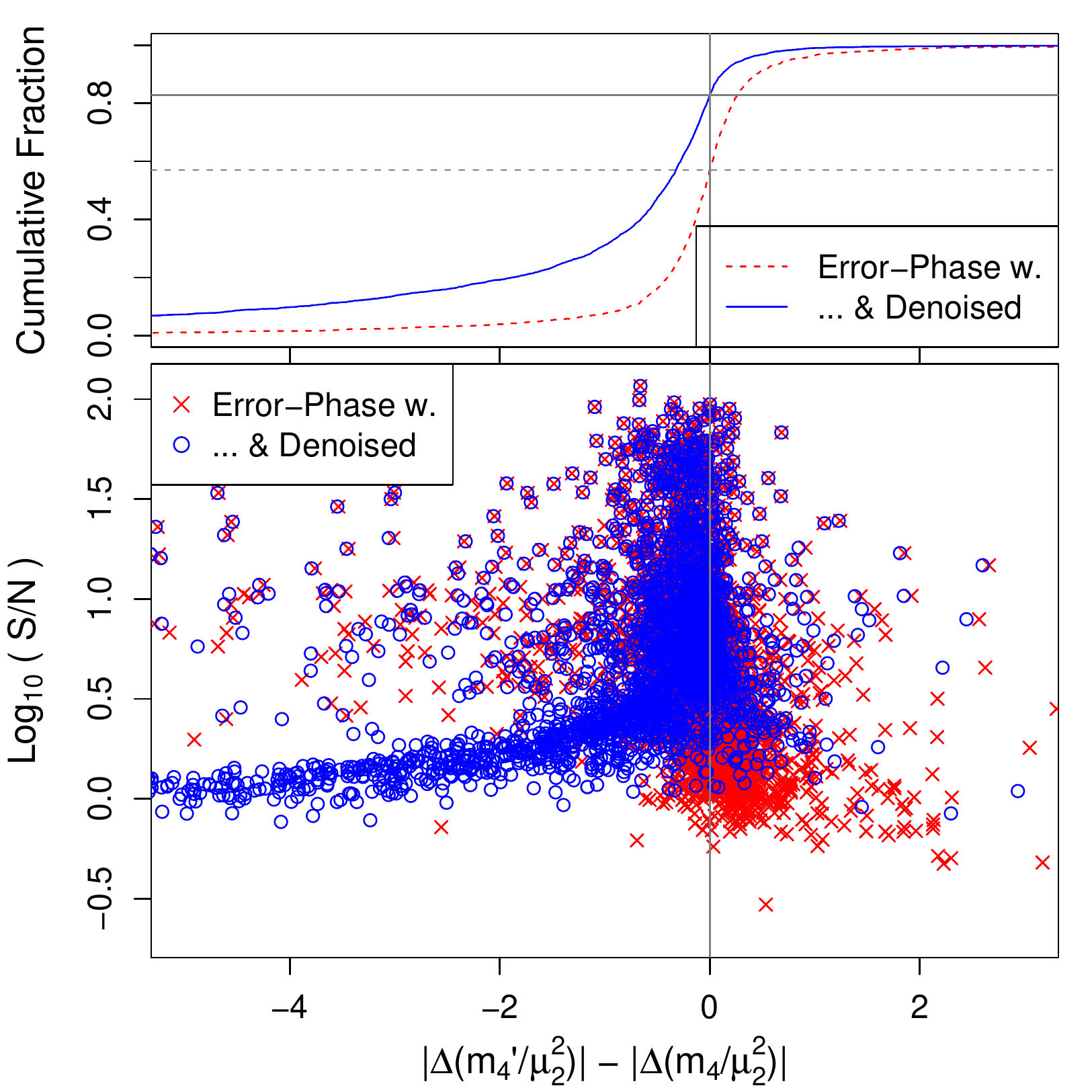}
\end{minipage}
\caption{The absolute deviations from population values of error-phase weighted estimators (red crosses and dashed lines) with additional denoising (blue circles and solid lines) are denoted by a prime symbol and their difference from the error-weighted counterparts is presented as a function of $S/N$ ratio.  Negative values indicate improvements in accuracy with respect to error-weighted estimators.  The fraction and magnitude of improved cases can be inferred from the cumulative distributions in the top panels.
The skewness (a,~b) and kurtosis (c,~d) moments are standardized by the estimated and true variances.}
\label{fig:quantile2}
\end{figure*}

\begin{figure*}
\begin{minipage}{\columnwidth}
\center
~~~~~~~~~(a)\\
\vspace{-0.2cm}
\includegraphics[width=\columnwidth]{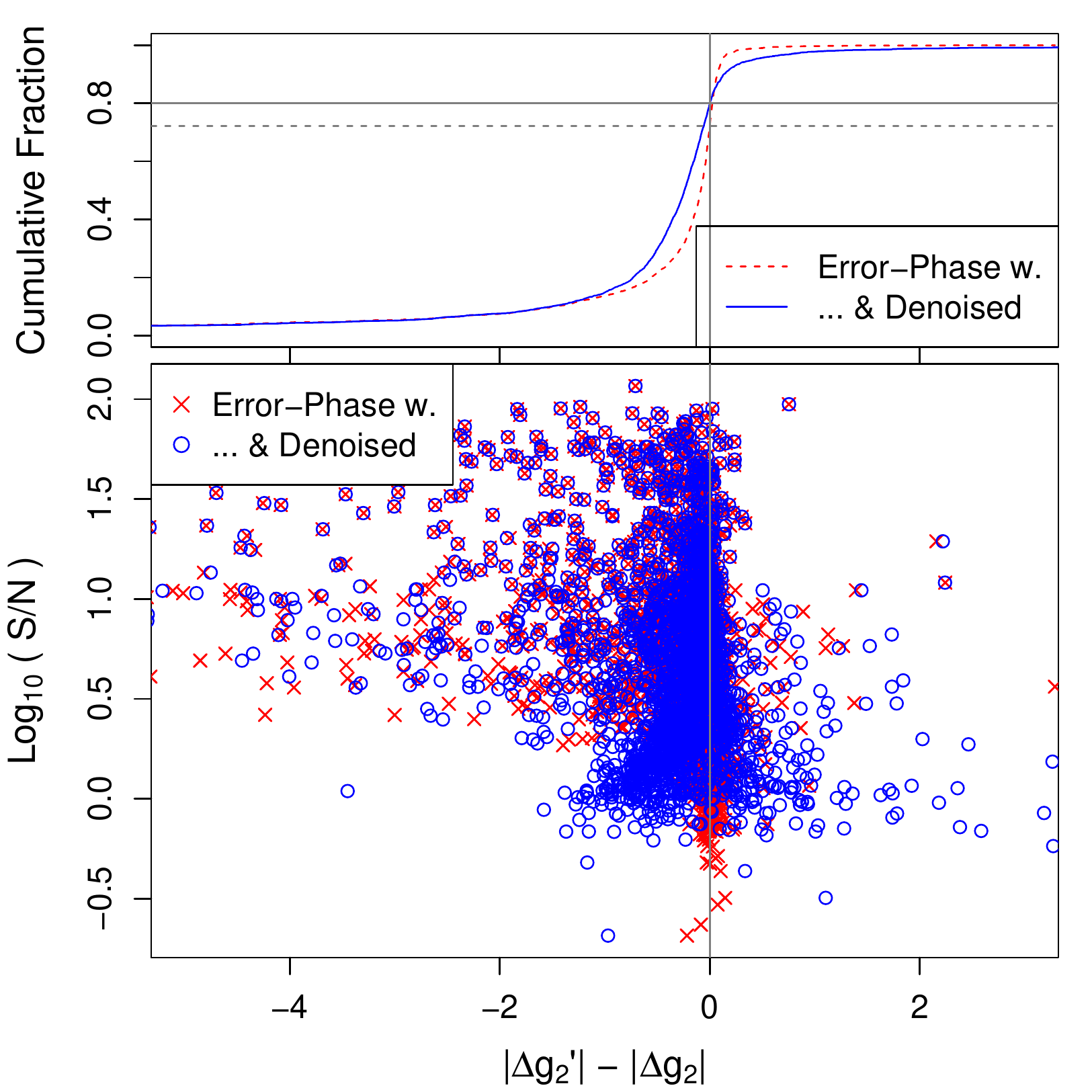}
\end{minipage}
\begin{minipage}{\columnwidth}
\center
~~~~~~~~~(b)\\
\vspace{-0.2cm}
\includegraphics[width=\columnwidth]{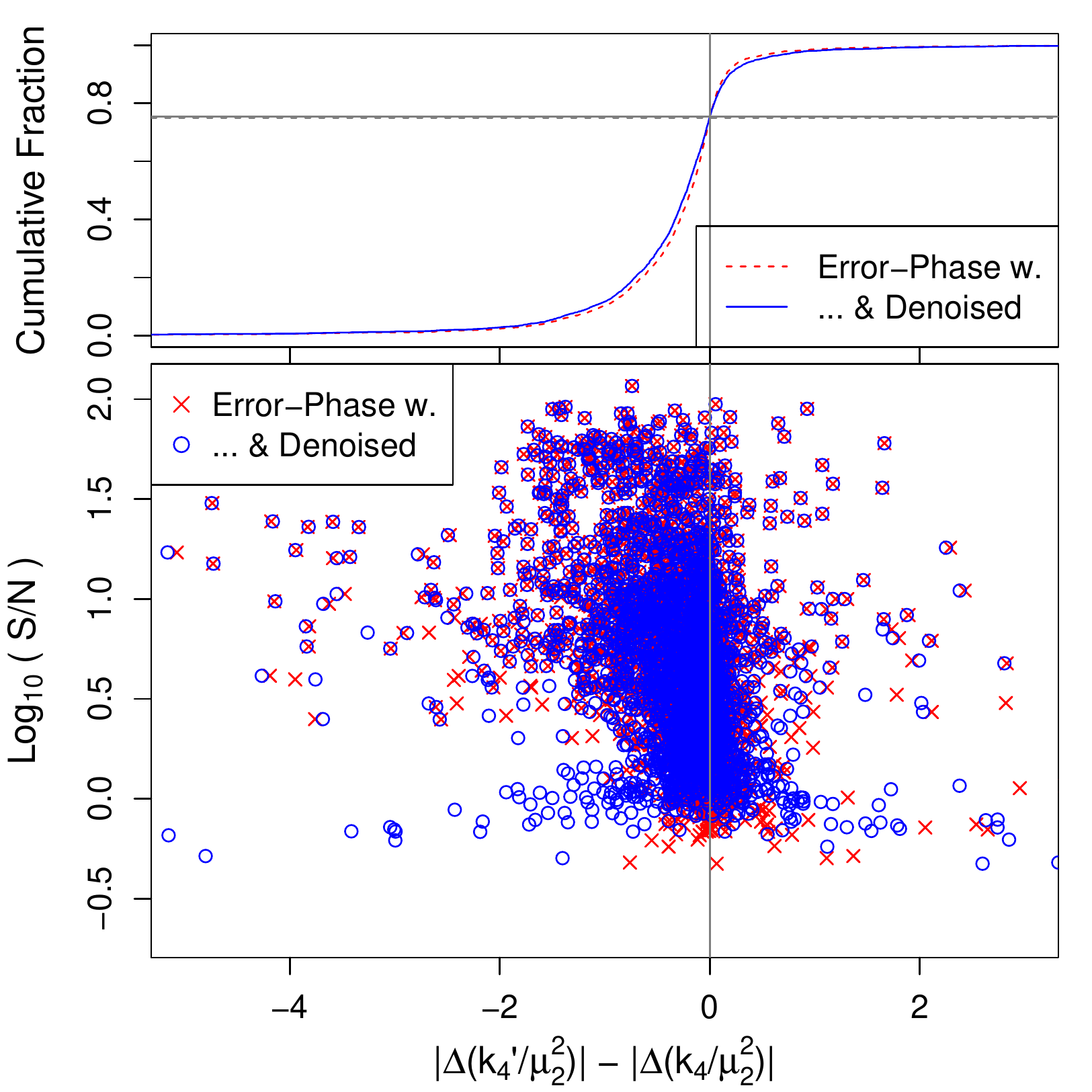}
\end{minipage}
\caption{The absolute deviations from population values of error-phase weighted estimators (red crosses and dashed lines) with additional denoising (blue circles and solid lines) are denoted by a prime symbol and their difference from the error-weighted counterparts is presented as a function of $S/N$ ratio.  Negative values indicate improvements in accuracy with respect to error-weighted estimators. The fraction and magnitude of improved cases can be inferred from the cumulative distributions in the top panels.
The kurtosis cumulants are standardized by the estimated and true variances in panels (a) and (b), respectively.}
\label{fig:quantile3}
\end{figure*}

In order to quantify the effect of error-phase weighting and additional denoising with respect to simple error weighting, the top panels in Figs~\ref{fig:quantile1}--\ref{fig:quantile3} indicate the fraction of sources associated with estimators improved  by phase weights and noise correction. 
The accuracy of a generic estimator $\mathcal E$ is assessed by its distance $|\Delta\mathcal E|=|\mathcal E-\eta |$ from the true value $\eta$. 
The abscissas in Figs~\ref{fig:quantile1}--\ref{fig:quantile3} represent the difference between the distance of error-phase weighted (and noise-unbiased, when applicable) estimators and the one of error-weighted estimators from the correct values: negative differences 
indicate smaller distances to the true values and thus
improved accuracy
with respect to error-weighted estimators.
The accuracy of error-phase weighted and noise-unbiased estimators improved in 70-to-90 per cent of the cases and deteriorations were typically smaller in  frequency and magnitude than improvements.  In the singular case of the standardized skewness $g_1$, denoising worsened almost 8 per cent of the estimates with respect to error-phase weighting. As apparent in Fig.~\ref{fig:quantile2}a, the accuracy of the standardized noise-unbiased  skewness worsened at low $S/N$ ratios.
This was expected because the noise-unbiased variance might be easily underestimated when uncertainties are of the same order as the signal, and the normalization of the skewness by a much smaller number than the correct one could rapidly degrade the precision of the estimator. 
A similar (less pronounced) tendency at low $S/N$ ratios was also observed in the case of the standardized kurtosis.

As noted in \cite{RimoldiniIntrinsic}, 
the noise-unbiased variance can become too small at low $S/N$ levels, overestimating the standardized noise-unbiased  skewness and kurtosis, or leading to an undetermined skewness value in case of non-positive variance.
On the other hand, when the variance is not corrected by noise biases, it is generally overestimated, and if the true values of skewness or kurtosis are sufficiently close to zero, the standardized noise-{\em biased}  skewness and kurtosis become more accurate (by serendipity).

\subsection{Automated classification}
The effect of statistical parameters weighted by different  schemes on automated classification was  assessed by comparing the classification accuracy and precision as a function of  variability type employing unweighted, error-weighted, phase-error weighted and noise-unbiased phase-error weighted estimators.

\subsubsection{Attributes and variability types}
Automated classification of stellar variability types was pursued with a set of attributes which characterized features of different classes. Some studies employed only information from light-curve modelling \citep{DebosscherA,DebosscherB,BlommeA,BlommeB}, with additional statistical parameters \citep{Richards} or colour information \citep{Sarro}, while attributes from modelling, statistical and astrophysical quantities were used in \citet{Dubath,RimoldiniUnsolved}.

The list of attributes employed for classification herein was restricted to descriptive statistics (mean, median, variance, interquartile range, normalized and non-normalized skewness and kurtosis), in order to test the 
effect of different weighting schemes on classification.
Only variability types which could be identified by the distribution of measurements were included in the training set, such as
eclipsing binaries, 
RR Lyrae, Mira, $\delta$ Cephei and Alpha$^2$ Canum Venaticorum stars, accounting for over 60~per cent of all sources in the {\it Hipparcos} periodic catalogue.

In these classification experiments, statistical parameters did not depend on modelling and were computed on the original (not simulated) data, which thus included the  sources  previously excluded because of modelling issues. The same quality flags, periods and weighting schemes were applied as described in 
Sec.~\ref{sec:pars}.
Since the true signal variance was unknown, the $S/N$ ratio was estimated by substituting the unknown $\mu_2$ in Eq.~(\ref{eq:snr}) with the noise-unbiased phase-weighted sample variance $m^*_2$ and by weighting the average of squared errors in the denominator with the same weights for consistency:
\begin{equation}
S/N=\left[\frac{\sum_{i=1}^{n}w_i (x_i-\bar{x})^2+\sum_{i=1}^{n}w_i^2  \epsilon_i^2/W}{\sum_{i=1}^{n} w_i \epsilon_i^2} -1\right]^{1/2}.
\end{equation}
The $S/N$ ratios of the  sources selected for classification 
(as described in Sec.~\ref{sec:dataselection}) 
spanned a range from 0.4 to 140. 

\subsubsection{Data selection}
\label{sec:dataselection}
Sources from the {\it Hipparcos} periodic catalogue were cross-matched with classifications from literature as available in
the Variable Star Index \citep*{Watson}\footnote{See \href{http://www.aavso.org/vsx/index.php?view=about.top}{http://www.aavso.org/vsx/index.php?view=about.top} for more details on how literature information has been selected, maintained and revised. A comprehensive list of variability types, labels and their descriptions is available at  \href{http://www.aavso.org/vsx/help/VariableStarTypeDesignationsInVSX.pdf}{http://www.} \href{http://www.aavso.org/vsx/help/VariableStarTypeDesignationsInVSX.pdf}{aavso.org/vsx/help/VariableStarTypeDesignationsInVSX.pdf}}
of the American Association of Variable Star Observers (AAVSO) with the nearest object within 1~arcsec.\footnote{Only one source, Hip~31400, was not considered because  associated with different classes at very similar angles from the direction of the {\it Hipparcos} source.}

Uncertain classifications (with class labels followed by the mark `:') were excluded, unless the uncertainty referred to properties of eclipsing binaries other than light-curve shapes, such as the physical characteristics, the luminosity class of the components, or the degree of filling of the inner Roche lobes. Objects associated with combinations of different classes (with labels joined by the symbol `+') were not addressed herein. 
The subset of variability types chosen to test classification with statistical parameters are listed in Table~\ref{tab:ts} and include 1605 sources associated with class labels EA, EB, EW, ACV, M,  RRAB, RRC, DCEP and DCEPS, defined in Table~\ref{tab:ts} together with the corresponding sample sizes.
Sources which poorly represented their class were not removed from the training set in order to avoid the introduction of selection biases (e.g., by favouring objects with higher $S/N$ ratios) and because the focus of this application was related to the {\em relative} classification accuracy employing attributes with different weighting schemes. 

\begin{table}
 \centering
 \caption{The training set employed to test the effect of descriptive statistics weighted by different schemes on classification includes 9 variability types and 1605 sources from the {\it Hipparcos} periodic catalogue.}
 \label{tab:ts}
 \begin{tabular}{llr}
  \hline
Variability Type&Label&Number\\
\hline
Eclipsing Binary: Algol type&EA&409\\
~~~~~~~~Beta Lyrae type&EB&208\\
~~~~~~~~W Ursae Majoris type&EW&170\\
Alpha$^2$ Canum Venaticorum&ACV&183\\
Mira Ceti&M&218\\
RR Lyrae: Asymmetric light curve&RRAB&139\\
~~~~~~~~Nearly symmetric light curve&RRC&28\\
Delta Cephei &DCEP&220\\
~~~~~~~~First overtone pulsators&DCEPS&30\\
\\
& Total:&1605\\
\hline
\end{tabular} 
\end{table}

\subsubsection{Training-set attributes}
\begin{figure*}
\begin{minipage}{\columnwidth}
\center
~~~~~~~~~(a)\\
\vspace{-0.2cm}
\includegraphics[width=\columnwidth]{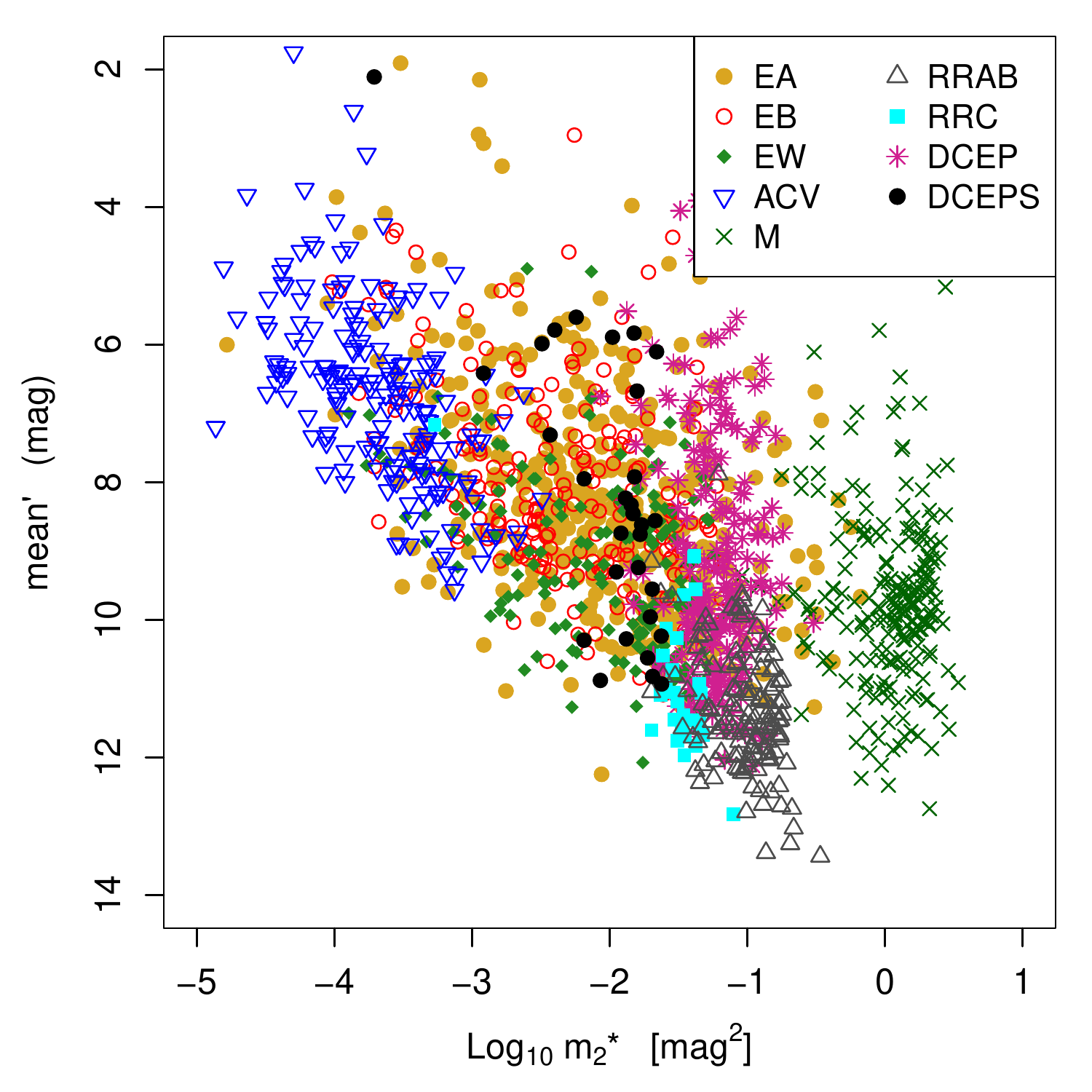}\\
~~~~~~~~~(c)\\
\vspace{-0.2cm}
\includegraphics[width=\columnwidth]{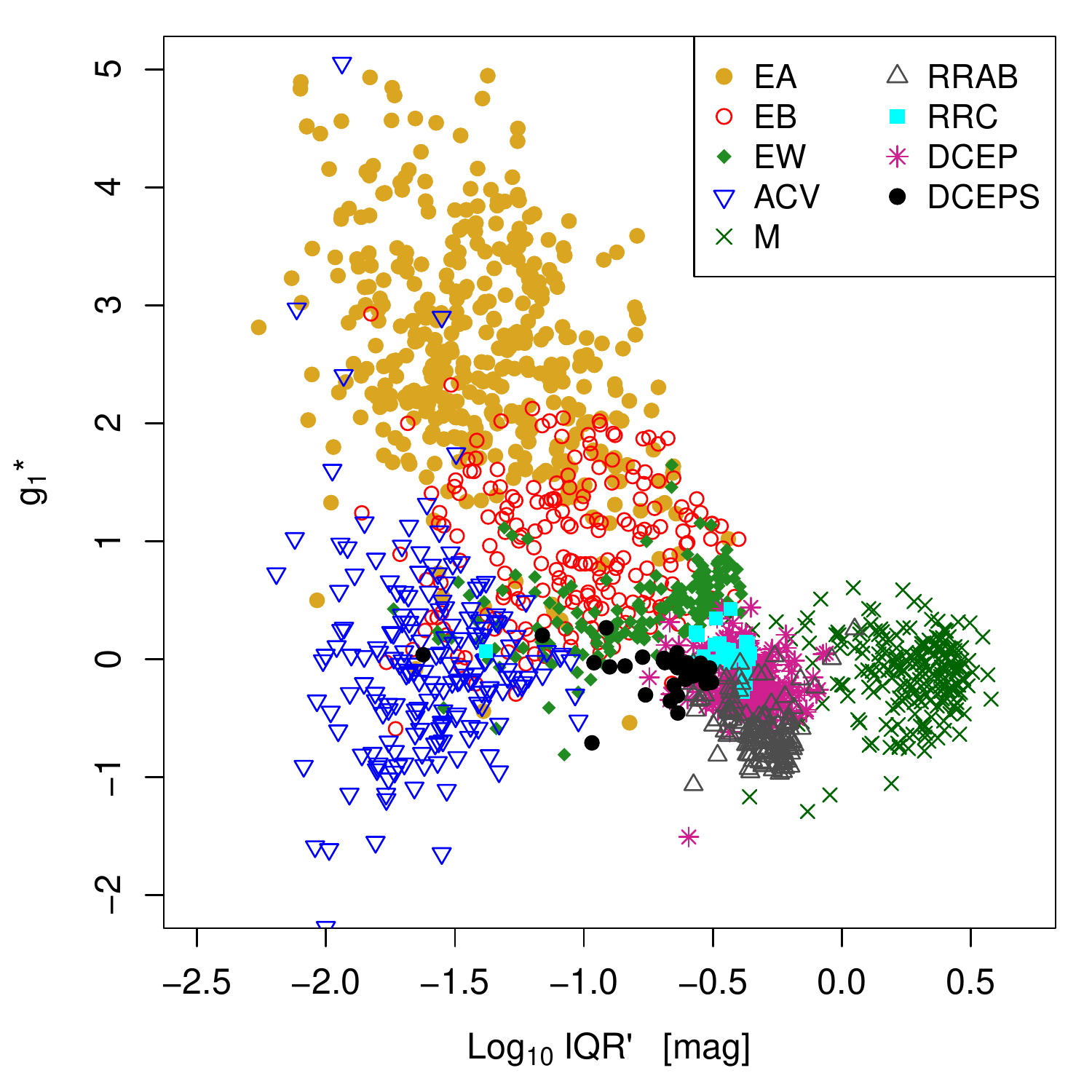}
\end{minipage}
\begin{minipage}{\columnwidth}
\center
~~~~~~~~~(b)\\
\vspace{-0.2cm}
\includegraphics[width=\columnwidth]{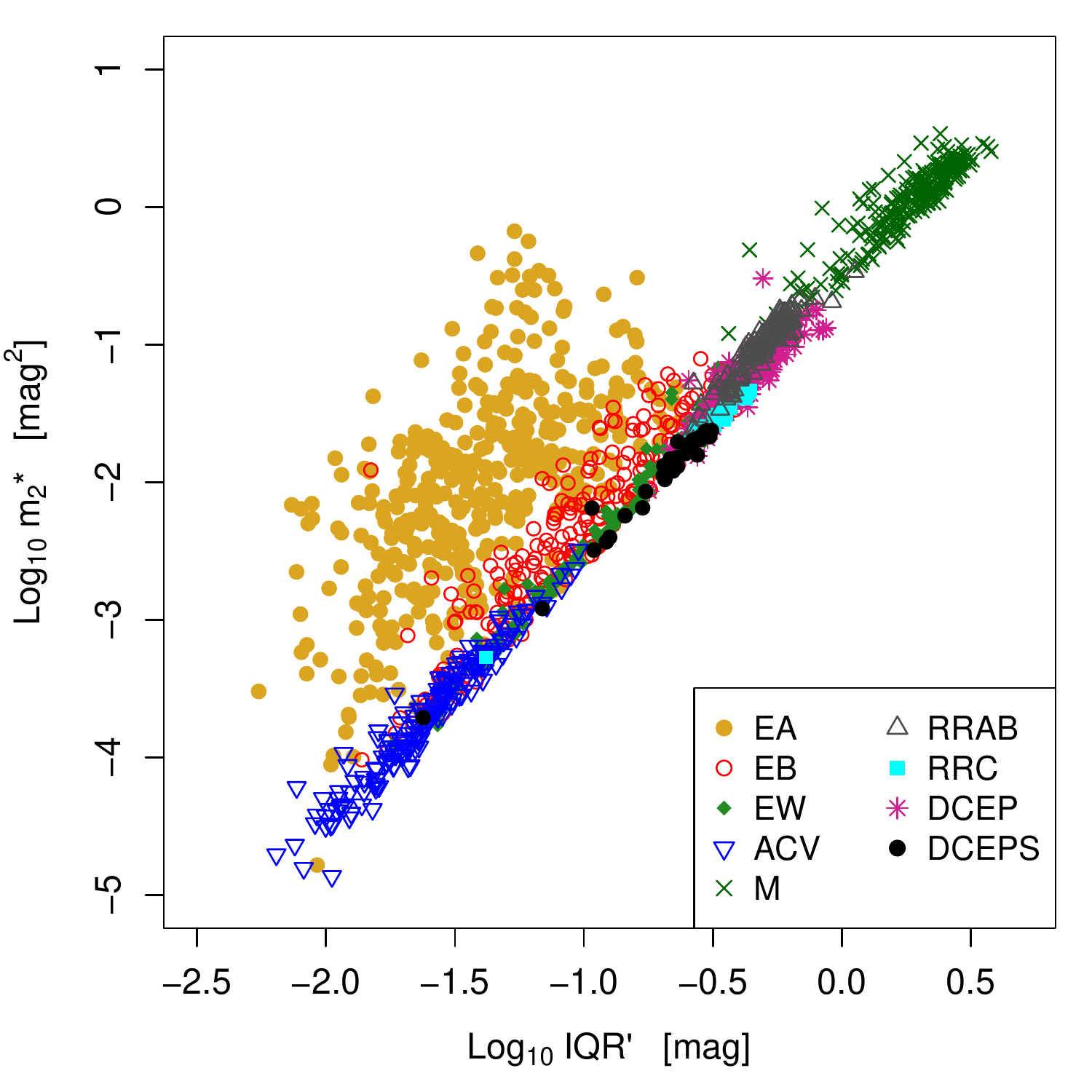}\\
~~~~~~~~~(d)\\
\vspace{-0.2cm}
\includegraphics[width=\columnwidth]{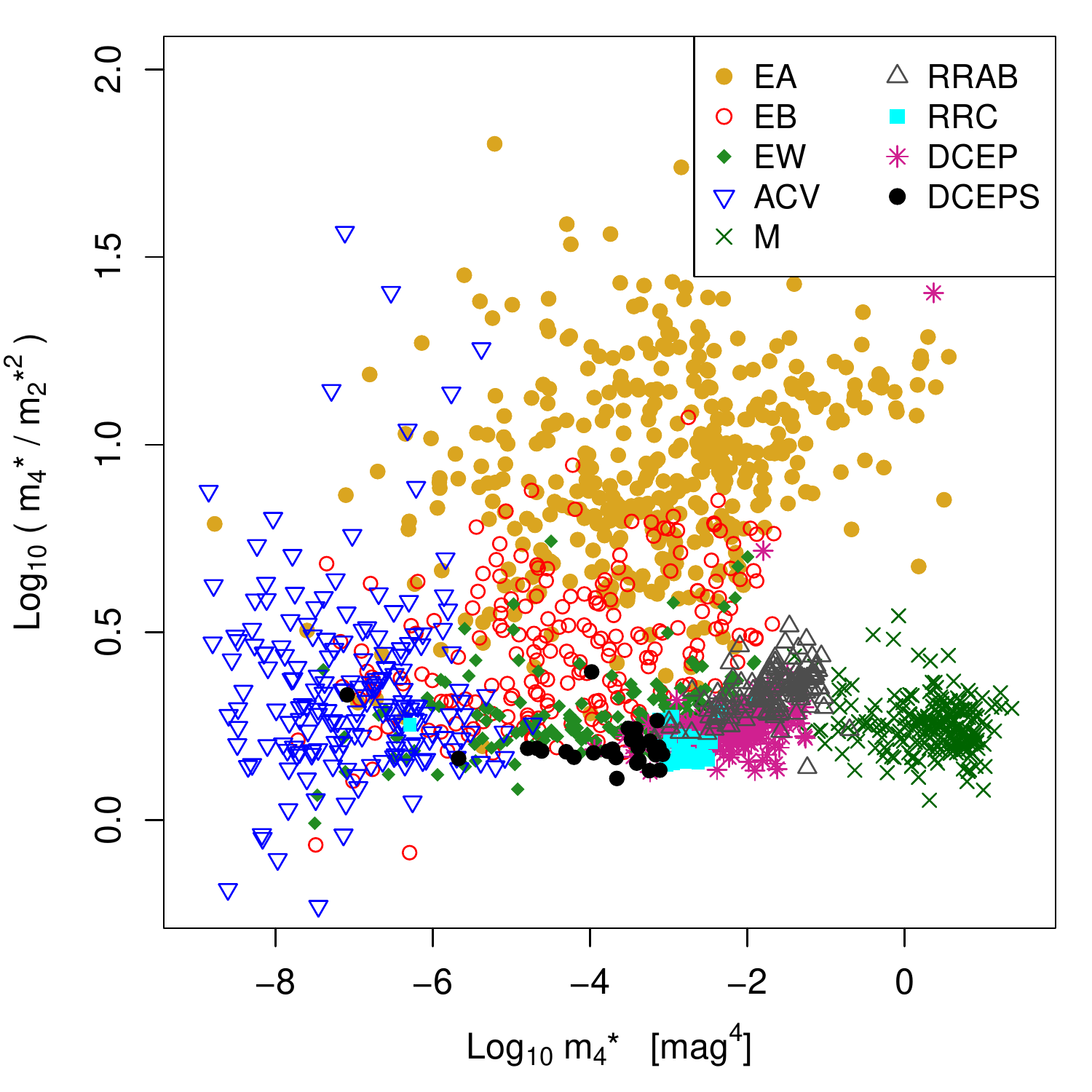}
\end{minipage}
\caption{A selection of estimators employed for classification is illustrated as a function of variability type.
All of the estimators are phase-error weighted (primed) and some are also noise-unbiased (starred), as defined in Appendix~\ref{app:def}. 
Class labels are described in Table~\ref{tab:ts} and denoted by symbols as shown in the legend of each panel.}
\label{fig:scatter}
\end{figure*}

The distributions of a selection of error-phase weighted (and noise-unbiased, when applicable) statistical parameters as a function of variability type are presented in Fig.~\ref{fig:scatter}. 
The mean or median magnitudes of sources at broadly different distances from the observer do not provide information on the intrinsic source properties.
However, the correlation between magnitude and noise, coupled with the interquartile range or variance (as shown in Fig.~\ref{fig:scatter}a), is related to the $S/N$ level and thus to an observational selection.
Figure~\ref{fig:scatter}b illustrates clearly the difference between robust and non-robust estimators for the same quantity, such as the interquartile range versus the variance, which proves  effective at separating eclipsing binaries (mostly EA and some fraction of EB types).
The interquartile range and standardized skewness are presented in Fig.~\ref{fig:scatter}c: the skewness separates the eclipsing binary subtypes from most other classes, which are better distinguished by the interquartile range. 
A similar scatter plot is shown in  Fig.~\ref{fig:scatter}d in terms of normalized and non-normalized kurtosis moments, with the difference that some classes occupy different relative loci (such as the EA types and the RRAB with respect to DCEP variables).

The estimators and variability types illustrated in Fig.~\ref{fig:scatter} are also presented  for the unweighted case in Appendix~\ref{app:scatterUNWEIGHTED}.
The distributions of most unweighted estimators in Fig.~\ref{fig:scatterUNWEIGHTED} tend to be more scattered than the ones shown in Fig.~\ref{fig:scatter}. 
In the case of ACV stars, instead, the unweighted skewness and kurtosis are less scattered, because these sources 
have very low $S/N$ ratios (typically $S/N<2.5$) and the correction of noise biases at such $S/N$ levels can decrease significantly the precision of higher moments \citep{RimoldiniIntrinsic}.
Another noticeable difference between Figs~\ref{fig:scatter} and \ref{fig:scatterUNWEIGHTED} is related to the larger skewness $g_1$ of Mira stars in the unweighted case, which can be understood by the systematic difference in the uncertainties associated with faint and bright measurements \citep[e.g.,][]{VanLeeuwen}. 
Such a difference is enhanced by the large amplitudes of light variations of Mira types, so that greater uncertainties on the faint side of the signal generate more faint than bright `outliers', biasing the unweighted skewness towards greater values.

\subsubsection{Classification method}
Automated classification tests were performed employing random forests  \citep{Breiman}.\footnote{This work employed the {\sc randomForest} package implemented in R \citep{R} by \citet{Rrf}.}
Random forest is an accurate tree-based classification method \citep[e.g., see][]{Richards}, quite robust to outliers and strongly correlated attributes.
The accuracy of the classifier was estimated from a subset of sources (about one-third of all objects)  randomly omitted from the learning process and thus called `out-of-bag' sources.
The importance of an attribute for classification was measured by the mean decrease in accuracy after permuting the values of that attribute in the out-of-bag sources  \citep[see][for more details]{RimoldiniUnsolved}. 

The random forest classifier was trained with estimators employing a single weighting scheme per run, i.e., unweighted, error-weighted, error-phase weighted  or noise-unbiased error-phase weighted.
For each weighting scheme, random forest was executed 1000 times with 500 trees and the classification accuracy and precision rates 
were aggregated after each run in order to assess their mean and dispersion.

\subsubsection{Results}

\begin{figure}
\center
\includegraphics[width=\columnwidth]{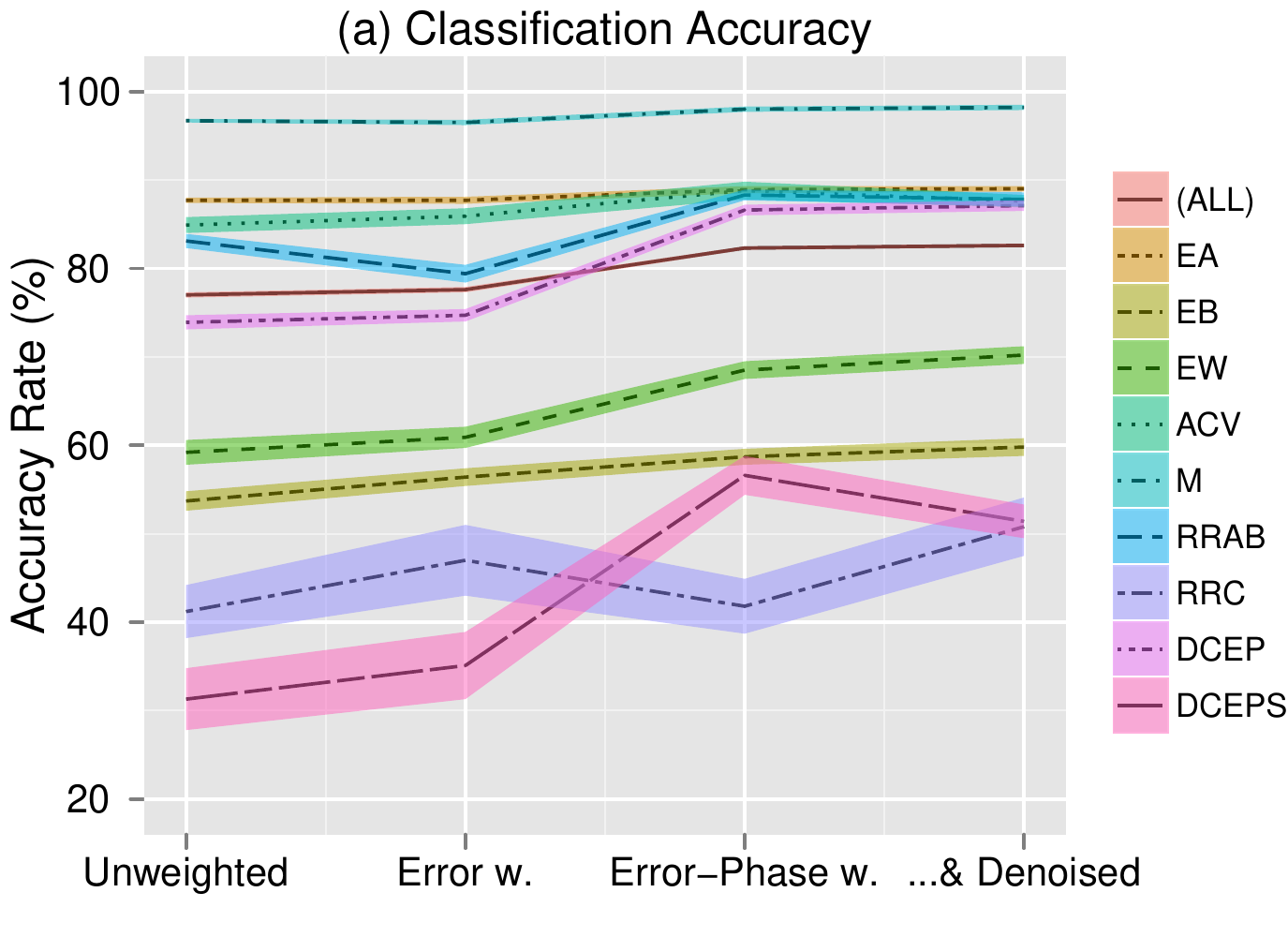}\\
\includegraphics[width=\columnwidth]{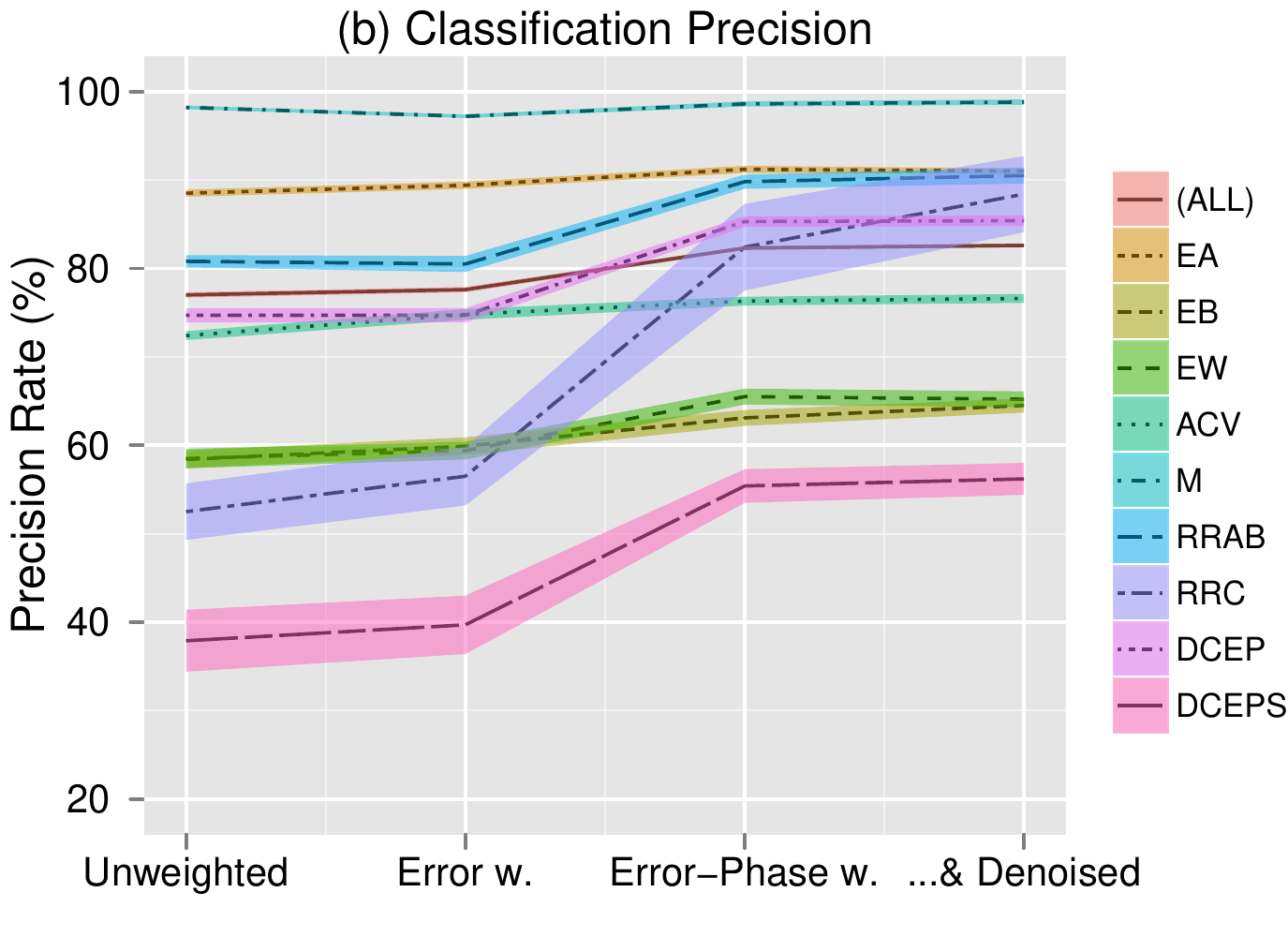}
\caption{Classification accuracy and precision rates are shown in panels (a) and (b), respectively, as a function of variability type and attribute weighting scheme. The values are obtained from the reclassification of training-set sources from the {\it Hipparcos} periodic catalogue and are listed in Tables~\ref{tab:res1} and \ref{tab:res2}. Class labels are defined in Table~\ref{tab:ts} and colour coded as shown in the legend of each panel. 
}
\label{fig:classResults}
\end{figure}

The mean classification accuracy and precision rates  are listed as a function of variability type and attribute weighting scheme in Tables~\ref{tab:res1} and~\ref{tab:res2}, and illustrated in Figs~\ref{fig:classResults}a and \ref{fig:classResults}b, respectively.
Training-set sources  
were reclassified by sole statistical parameters with an overall accuracy from 77.0 to 82.6 per cent on average, depending on the weighting scheme adopted. Unweighted estimators led to the smallest accuracy level, closely followed by error-weighted estimators at 77.6 per cent. The introduction of phase weights improved the accuracy to 82.3 per cent and noise-unbiased error-phase weighted estimators achieved the best overall classification accuracy of 82.6 per cent (with uncertainty at the level of 0.2 to 0.3 per cent).
The overall trends were generally reflected by single variability types 
with only a few  exceptions. 
In particular, error-weighted estimators of RRAB types led to significantly worse classification accuracy than unweighted estimators. 
This was explained by the relatively large amplitude of RRAB variables, which enhanced the systematic difference of uncertainties between bright and faint measurements in the light curve: weighting by errors decreased the importance of faint measurements, increasing the similarity (and thus confusion) with DCEP and M-type light curves.
In the case of DCEPS variables, denoising was marginally counterproductive for accuracy, although still significantly better with respect to the unweighted or error-weighted schemes.  
This was expected to be related to the increased confusion with the noise-unbiased estimators of EW types, which were more numerous and extended to lower $S/N$ ratios than the DCEPS stars: while the distributions of the standardized error-phase weighted kurtosis moments of DCEPS and EW types could just be  separated,  
the additional denoising decreased the values of standardized kurtosis moments more for the EW than the DCEPS stars,
leading to overlapping distributions dominated by the EW types.

\begin{table}
 \centering
 \caption{Classification accuracy rates (per cent values) from the reclassification of training-set sources from the {\it Hipparcos} periodic catalogue are listed as a function of variability type and attribute weighting scheme. Class labels are defined in Table~\ref{tab:ts}. Accuracies and corresponding uncertainties represent average values from 1000 runs of random forests employing 500 trees.}
 \label{tab:res1}
 \begin{tabular}{@{}l@{~~}c@{~~~}c@{~~~}c@{~~~}c@{}}
  \hline
 Var. & & Error & Error-Phase & Err.-Ph. w. \\
Type&Unweighted&Weighted&Weighted& \& Denoised\\
\hline
EA & 87.7 $\pm$ 0.3 & 87.7 $\pm$ 0.4 & 88.9 $\pm$ 0.4 & 89.0 $\pm$ 0.3\\
EB & 53.7 $\pm$ 1.1 & 56.4 $\pm$ 1.0 & 58.7 $\pm$ 0.9 & 59.8 $\pm$ 1.0\\
EW & 59.2 $\pm$ 1.4 & 60.9 $\pm$ 1.2 & 68.5 $\pm$ 1.0 & 70.2 $\pm$ 1.0\\
ACV & 84.9 $\pm$ 0.9 & 85.9 $\pm$ 0.9 & 88.8 $\pm$ 1.0 & 87.6 $\pm$ 0.7\\
M & 96.7 $\pm$ 0.2 & 96.5 $\pm$ 0.3 & 98.0 $\pm$ 0.3 & 98.2 $\pm$ 0.3\\
RRAB & 83.1 $\pm$ 0.8 & 79.4 $\pm$ 1.0 & 88.3 $\pm$ 0.6 & 87.8 $\pm$ 0.8\\
RRC & 41.2 $\pm$ 3.0 & 47.0 $\pm$ 4.0 & 41.8 $\pm$ 3.1 & 50.8 $\pm$ 3.3\\
DCEP & 73.9 $\pm$ 0.8 & 74.7 $\pm$ 0.7 & 86.6 $\pm$ 0.6 & 87.1 $\pm$ 0.6\\
DCEPS & 31.3 $\pm$ 3.5 & 35.1 $\pm$ 3.8 & 56.6 $\pm$ 2.2 & 51.4 $\pm$ 1.9\\
\\
ALL & 77.0 $\pm$ 0.3 & 77.6 $\pm$ 0.3 & 82.3 $\pm$ 0.2 & 82.6 $\pm$ 0.2\\
\hline
\end{tabular} 
\end{table}

Precision rates of classification per variability type 
showed successive improvements (consistent within uncertainties) from the unweighted to the error-weighted, error-phase weighted and noise-unbiased schemes. A single exception was related to Mira stars: their classification precision with error-weighted estimators was about one per cent worse than with unweighted attributes, as a consequence of the increased contamination by RRAB stars (as explained above).

\begin{table}
 \centering
 \caption{Classification precision rates (per cent values) from the reclassification of training-set sources from the {\it Hipparcos} periodic catalogue are listed as a function of variability type and attribute weighting scheme. Class labels are defined in Table~\ref{tab:ts}. Precisions and corresponding uncertainties represent average values from 1000 runs of random forests employing 500 trees.}
 \label{tab:res2}
 \begin{tabular}{@{}l@{~~}c@{~~~}c@{~~~}c@{~~~}c@{}}
  \hline
 Var. & & Error & Error-Phase & Err.-Ph. w. \\
Type&Unweighted&Weighted&Weighted& \& Denoised\\
\hline
EA & 88.5 $\pm$ 0.4 & 89.4 $\pm$ 0.4 & 91.2 $\pm$ 0.4 & 91.0 $\pm$ 0.3\\
EB & 58.4 $\pm$ 1.0 & 59.9 $\pm$ 1.0 & 63.1 $\pm$ 0.9 & 64.5 $\pm$ 0.8\\
EW & 58.5 $\pm$ 1.1 & 59.4 $\pm$ 1.0 & 65.5 $\pm$ 0.9 & 65.2 $\pm$ 0.9\\
ACV & 72.4 $\pm$ 0.5 & 74.8 $\pm$ 0.6 & 76.3 $\pm$ 0.5 & 76.6 $\pm$ 0.5\\
M & 98.2 $\pm$ 0.2 & 97.2 $\pm$ 0.2 & 98.6 $\pm$ 0.3 & 98.8 $\pm$ 0.3\\
RRAB & 80.8 $\pm$ 0.7 & 80.5 $\pm$ 0.9 & 89.8 $\pm$ 0.8 & 90.5 $\pm$ 0.9\\
RRC & 52.5 $\pm$ 3.2 & 56.5 $\pm$ 3.3 & 82.4 $\pm$ 4.9 & 88.4 $\pm$ 4.3\\
DCEP & 74.7 $\pm$ 0.8 & 74.7 $\pm$ 0.8 & 85.3 $\pm$ 0.6 & 85.4 $\pm$ 0.6\\
DCEPS & 37.9 $\pm$ 3.5 & 39.7 $\pm$ 3.3 & 55.4 $\pm$ 1.9 & 56.2 $\pm$ 1.8\\
\\
ALL& 77.0 $\pm$ 0.3 & 77.6 $\pm$ 0.3 & 82.3 $\pm$ 0.2 & 82.6 $\pm$ 0.2\\
\hline 
\end{tabular} 
\end{table}

\begin{figure}
\includegraphics[width=\columnwidth]{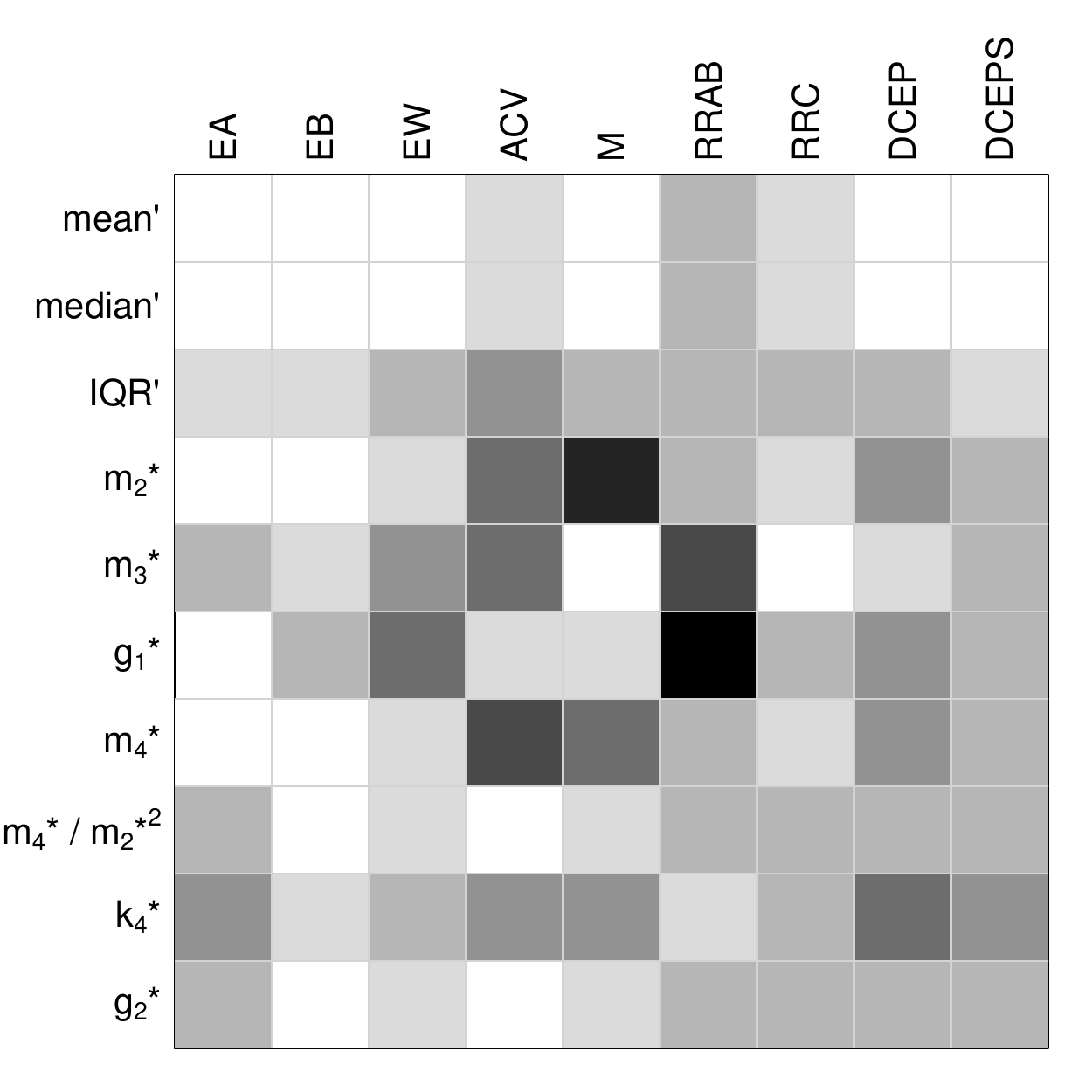}
\caption{The  importance of  attributes is illustrated as a function of variability type, with darker grey levels indicating more important attributes for the identification of a given class.
Attributes are phase-error weighted (primed) and some also noise-unbiased (starred), as defined in Appendix~\ref{app:def}. Class labels are described in Table~\ref{tab:ts}.}
\label{fig:importance}
\end{figure}

The importance of phase-weighted (and noise-unbiased, when applicable) attributes as a function of variability type from a run of random forest is depicted in Fig.~\ref{fig:importance}, with darker cells indicating more important attributes for a given class.
For example, the skewness confirmed to be particularly useful to distinguish eclipsing binaries 
and the asymmetric light curves of RRAB stars. Also, large and small amplitude variables, such as ACV and M types, could easily be  separated by the variance.

\begin{figure}
\includegraphics[width=\columnwidth]{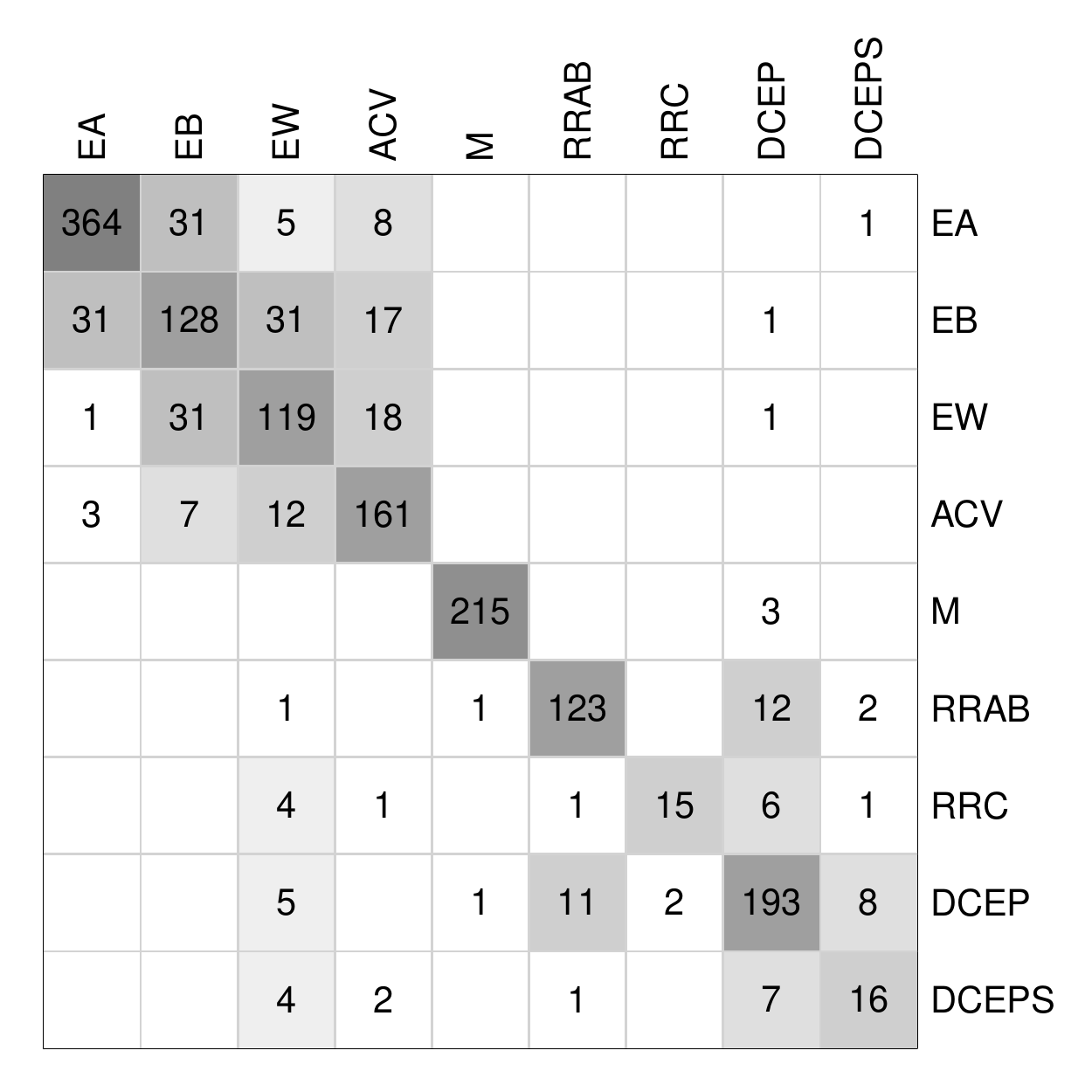}
\caption{The confusion matrix of training-set sources reclassified by random forests with 500 trees and employing error-phase weighted (and noise-unbiased, when applicable) estimators. The overall classification accuracy is 83 per cent and reaches 92 per cent after aggregating the subtypes of eclipsing binary, RR Lyrae and Delta Cephei into their superclasses. 
Rows and columns refer to literature and predicted types, respectively. Class labels are defined in Table~\ref{tab:ts}.}
\label{fig:CM}
\end{figure}

The confusion matrix from a run of random forest, 
employing error-phase weighted (and noise-unbiased, when applicable) estimators,
is shown in Fig.~\ref{fig:CM}. 
The overall classification accuracy was 83 per cent and reached 92 per cent after aggregating the subtypes of eclipsing binary, RR Lyrae and Delta Cephei into their superclasses. 
The improvement in classification accuracy, with respect to results employing unweighted estimators, was about 6 per cent in both cases of separated and aggregated variability subtypes.
The level of confusion between eclipsing binaries of EA, EB and EW types was expected from the natural overlap in their definitions. 
The misclassification of many ACV types as eclipsing binaries was attributed to the restricted set of attributes employed herein \citep[e.g., the attribute `QSO variations' in][proved to be useful in this context]{RimoldiniUnsolved}.
Similarly, the confusion between RR Lyrae and Delta Cephei variables would not have occurred if the period was included in the set of classification attributes. 
On the other hand, simple amplitude estimators such as variance and interquartile range were sufficient to separate Mira stars from most other classes with high accuracy  (as apparent in Fig.~\ref{fig:scatter}).

\section{Conclusions} 
\label{sec:concl}
A simple weighting scheme based on linear interpolation, applicable in phase or time, was proposed in order to improve the accuracy of descriptive statistics of unevenly sampled time series.
Weights represented by time intervals could be applied to well-sampled \textcolor{black}{deterministic} signals. 
For sparsely sampled but periodic time series (dominated by a single period), light curves could be folded with the fundamental periods and weights expressed in terms of phase intervals.

The {\it Hipparcos} catalogue of periodic variables represented a suitable test bed of unevenly sampled light curves with realistic distributions of variability types to investigate the accuracy of estimators weighted by different schemes.
Noise-unbiased estimators weighted by phase intervals (or inverse-squared uncertainties at low $S/N$ ratios) improved the accuracy for 70-to-90 per cent of the sources with respect to the values of  the error-weighted counterparts.  Deteriorations in accuracy were observed in 10-to-30 per cent of the cases, with magnitudes typically smaller than the ones corresponding to improvements.

Automated classification experiments confirmed that the best overall results were achieved employing the set of the most accurate attributes, i.e., the noise-unbiased  estimators weighted by phase intervals and inverse-squared uncertainties at high and low $S/N$ ratios, respectively. The overall improvement in classification accuracy with respect to the result employing unweighted estimators was about 6 per cent and most of it was related to the introduction of 
phase weights which adapted to the varying sampling density of the light curve.

\section*{Acknowledgments}
The author is grateful to M. S\"uveges, P. Dubath \textcolor{black}{and the anonymous referee} for useful comments on the original manuscript.
\textcolor{black}{This work made use of the Hipparcos catalogue obtained from the European Space Agency astrometric mission {\it Hipparcos}, the VizieR service, operated at the Centre de Donn\'ees astronomiques de Strasbourg (Strasbourg, France) and the International Variable Star indeX (VSX), operated by the American Association of Variable Star Observers (Cambridge, Massachusetts, USA).}


\appendix

\section{Periodic interpolation}
\label{app:periodicInterpolation}
Substituting time-sorted data with phase-sorted measurements in Eq.~(\ref{eq:interpolation}) and adding the interpolation term 
from the last to the first point in phase, it follows:
\begin{align}
\bar{\theta} 
\approx&\, \frac{1}{2\pi}\sum_{i=1}^{n} \frac{\theta_{i}+\theta_{i+1}}{2} \left(\phi_{i+1}-\phi_i\right) \\
=&\, \frac{1}{4\pi}\left[\sum_{i=1}^{n}  \theta_{i} \left(\phi_{i+1}-\phi_{i}\right) +\sum_{i=2}^{n+1} \theta_i\left(\phi_{i}-\phi_{i-1}\right) \right] \\
=&\,  \frac{1}{4\pi} \left[\sum_{i=2}^{n}  \theta_{i}\left(\phi_{i+1}-\phi_{i-1}\right)+  \right.\nonumber \\
& \left.~~~~~~~~ +\theta_1\left(\phi_{2}-\phi_{1}\right) +\theta_{n+1} \left(\phi_{n+1}-\phi_{n}\right) \rule{0cm}{0.5cm}\right] \\
=&\,\frac{1}{W} \sum_{i=1}^{n} w_i \, \theta_i,
\end{align}
where $\theta_{n+1}=\theta_1$ and $\phi_{n+1}=\phi_1+2\pi$, so that
\begin{align}
&w_i=\phi_{i+1}-\phi_{i-1}~~~~~~~~\forall  i \in (2,n-1) \label{eq:pwi}\\
&w_1 = \phi_2 - \phi_n +2\pi\\
&w_n =\phi_1- \phi_{n-1}+2\pi, \label{eq:pwn}
\end{align}
and
\begin{equation}
W=\sum_{i=1}^n w_i=4\pi.
\end{equation}
When  differences between successive phases are a significant fraction of a cycle, they might be limited to some maximum interval $\Delta\phi_{\max}$ as follows:
\begin{align}
w_i=&\min\left\{\phi_{i+1}-\phi_i, \, \Delta\phi_{\max} \right\}+ \nonumber \\
&+\min\left\{\phi_{i}-\phi_{i-1}, \, \Delta\phi_{\max} \right\}~~~~~\forall  i \in (2,n-1) \label{eq:w_gap_ib}\\
w_1=&\min\left\{\phi_{2}-\phi_1, \, \Delta\phi_{\max} \right\}+ \nonumber \\
&+\min\left\{\phi_{1}-\phi_{n}+2\pi, \, \Delta\phi_{\max} \right\} \label{eq:w_gap_1b}\\
w_n=&\min\left\{\phi_{n}-\phi_{n-1}, \, \Delta\phi_{\max} \right\}+ \nonumber \\
&+\min\left\{\phi_{1}-\phi_{n}+2\pi, \, \Delta\phi_{\max} \right\} \label{eq:w_gap_nb}
\end{align}
and, of course, then $W\leq 4\pi$.

\section{The {\em Hipparcos} periodic light curves}
\label{app:data}
\begin{figure*}
\includegraphics[width=2\columnwidth]{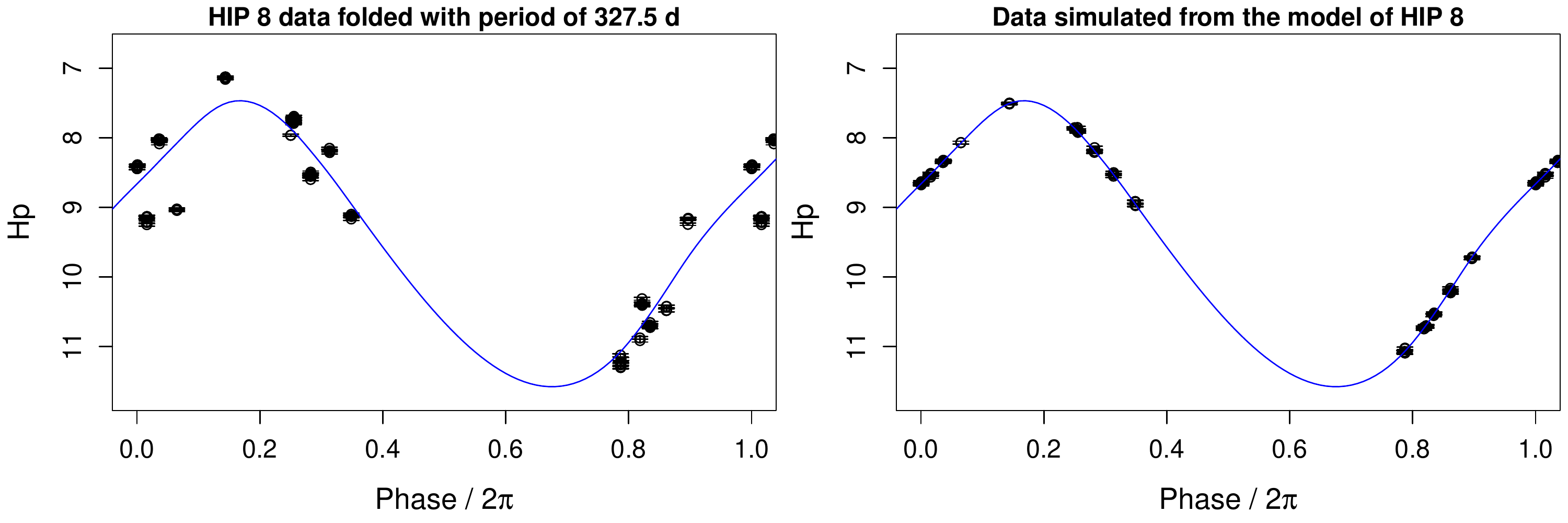}\\ 
\includegraphics[width=2\columnwidth]{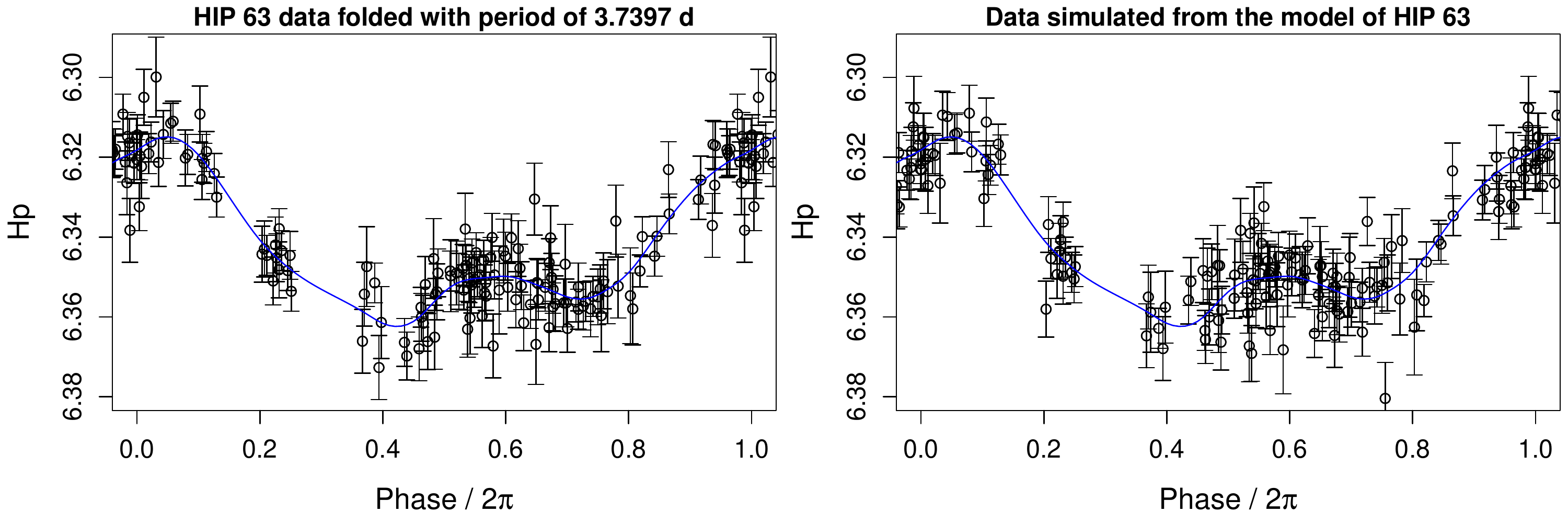}\\
\includegraphics[width=2\columnwidth]{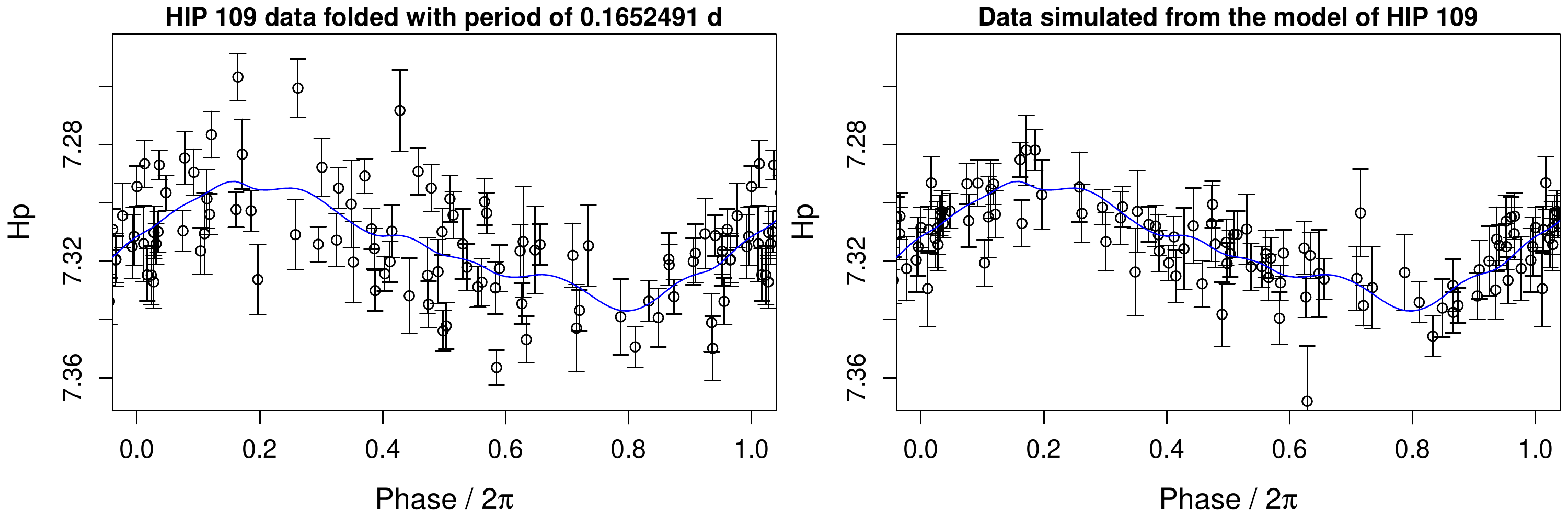}\\
\includegraphics[width=2\columnwidth]{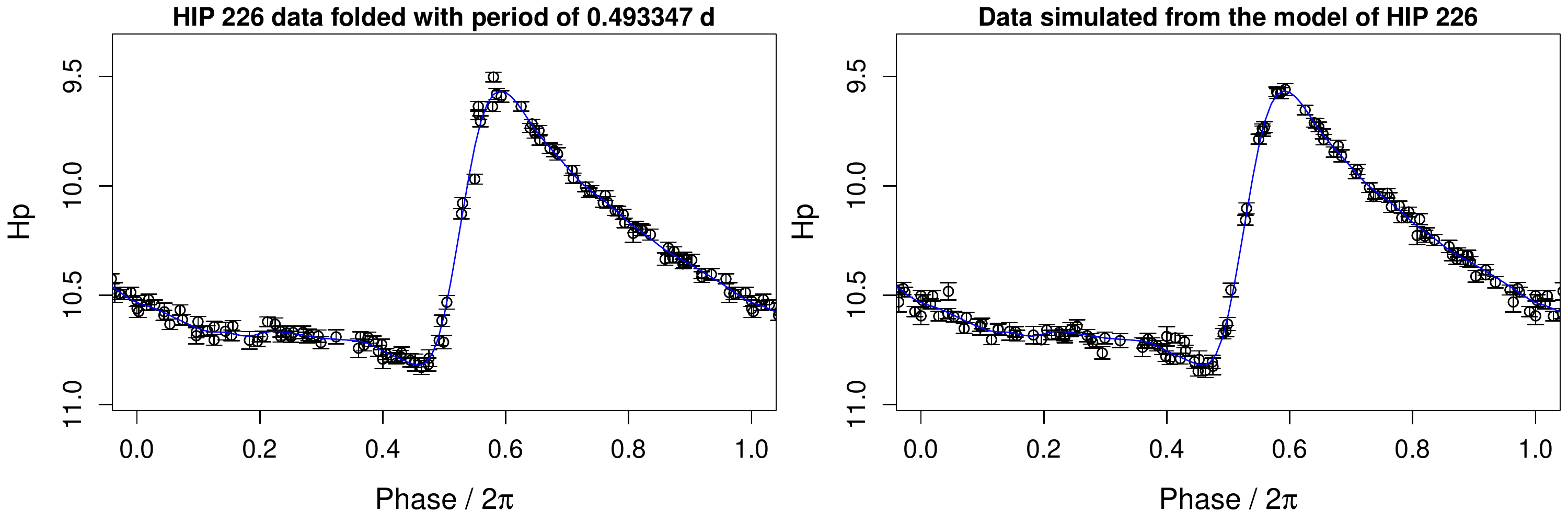}
\caption{Panels on the left-hand side present folded light curves from the {\it Hipparcos} periodic catalogue, together with the smoothing spline model employed to generate simulated data (shown in the panels on the right-hand side). 
Simulations served to assess the accuracy of statistical parameters with respect to the model (as reference), while classification experiments were performed on the original data. 
The illustration of the full set of light curves employed herein is available online (see supporting information).}
\label{fig:LC}
\end{figure*}

Light curves of the {\it Hipparcos} periodic variables are illustrated in Fig.~\ref{fig:LC}.
Panels on the left-hand side show the {\it Hipparcos} data folded with the catalogue period, while  the right-hand panels present data  simulated around the model according to the measurement uncertainties (assumed Gaussian) and the observed phases.
Simulated data were employed to assess the accuracy of statistical estimators (with respect to the model), while classification was performed on the original data.
Figure~\ref{fig:LC} provides sample light curves for 4 of the 2683 {\it Hipparcos} sources available online (see supporting information).\footnote{The 29 objects with no data associated with good quality flags (i.e., field HT4 $> 1$ only) were not included in Fig.~\ref{fig:LC}.
} 

\section{Statistical parameters}
\label{app:def}
The statistical estimators employed to test different weighting schemes are defined below.

\subsection{Variance, skewness and kurtosis}
Sample noise-unbiased weighted moments and cumulants are defined following  \citet{RimoldiniIntrinsic}.
Denoting the sample weighted central moments of order $r$  by
\begin{equation}
m_r=\frac{1}{W}\sum_{i=1}^{n}w_i (x_i-\bar{x})^r, \mbox{~~~where~~~}W=\sum_{i=1}^n w_i,
\end{equation}
and representing the Gaussian uncertainties by $\epsilon_i$, 
the noise-unbiased estimators (starred) are defined as follows:
\begin{align}
m^*_2 =&\, m_2-\frac{1}{W}\sum_{i=1}^{n}w_i  \epsilon_i^2\left(1-\frac{w_i}{W} \right) = k^*_2 \\
m^*_3 =&\,m_3-\frac{3}{W}\sum_{i=1}^{n}w_i \epsilon_i^2  \left(x_i-\bar{x}\right)\left(1-\frac{2 w_i}{W} \right)= k^*_3 \\
m^*_4  =&\,m_4- \frac{6}{W}\sum_{i=1}^{n}w_i  \epsilon_i^2 \left[\left(x_i-\bar{x}\right)^2\left(1-\frac{2w_i}{W} \right)+\right.\nonumber\\
&\left.-\frac{\epsilon_i^2}{2}\left(1-\frac{2w_i}{W}\right)^2+\frac{m^*_2 w_i}{W}\right] 
-\frac{3}{W^4}\left(\sum_{i=1}^{n} w_i^2\epsilon_i^2\right)^2\\
(m_2^2)^* =&\left(m^*_2\right)^2 -\frac{4}{W^2}\sum_{i=1}^{n} w_i^2\epsilon_i^2 \left[ \left(x_i-\bar{x}\right)^2 -\frac{\epsilon_i^2}{2}\left(1-\frac{2w_i}{W} \right)\right]+\nonumber\\
&+\frac{2}{W^4}\left(\sum_{i=1}^{n} w_i^2\epsilon_i^2\right)^2\\
k^*_4  = &\,m^*_4-3\,(m_2^2)^* \\
g^*_1=&\,k^*_3/(k^*_2)^{3/2} \\
g^*_2=&\,k^*_4/(k^*_2)^2.
\end{align}

\subsection{Percentiles}
Percentiles depend on the rank of sorted values, thus they are less sensitive to extreme values than moments and cumulants which involve powers of deviations from the mean and average over all elements.
The $m$-th percentile $P_m(\mathbf{x})$ is defined as the (interpolated) value such that $m$ per cent of the data are smaller than  $P_m(\mathbf{x})$.
Two common percentiles are the median $P_{50}(\mathbf{x})$ and  interquartile range $\mbox{IQR}=P_{75}(\mathbf{x})-P_{25}(\mathbf{x})$.

Denoting  the list of measurements $x_i$ sorted in increasing values by $\{x_{(1)},...,x_{(n)} \}$, associated with weights $\{w_{(1)},...,w_{(n)} \}$, respectively, the  $m$-th weighted percentile $P_m(\mathbf{x})$ is defined as follows:
\begin{equation}
P_m(\mathbf{x})=\left\{ 
\begin{array}{ll}
x_{(1)} &  \mbox{if }  0< m\leq p_1 \\
x_{(i)}+\frac{m-p_i}{p_{i+1}-p_i}\,\left(x_{(i+1)}-x_{(i)}\right)& \mbox{if } p_i\leq m \leq p_{i+1}  \\
x_{(n)} &  \mbox{if }  p_n \leq m < 100
\end{array}
\right. 
\end{equation}
where
\begin{equation}
 p_i=\frac{100}{W}\left(\sum_{j=1}^i w_{(j)} - \frac{w_{(i)}}{2} \right)~~~\mbox{and}~~x_{(k)}\leq x_{(k+1)}~~\forall k<n.
\end{equation}

\section{Scatter plots of related estimators}
\label{app:scatter}
\begin{figure*}
\begin{minipage}{\columnwidth}
\center
(a)\\
\vspace{-0.2cm}
\includegraphics[width=\columnwidth]{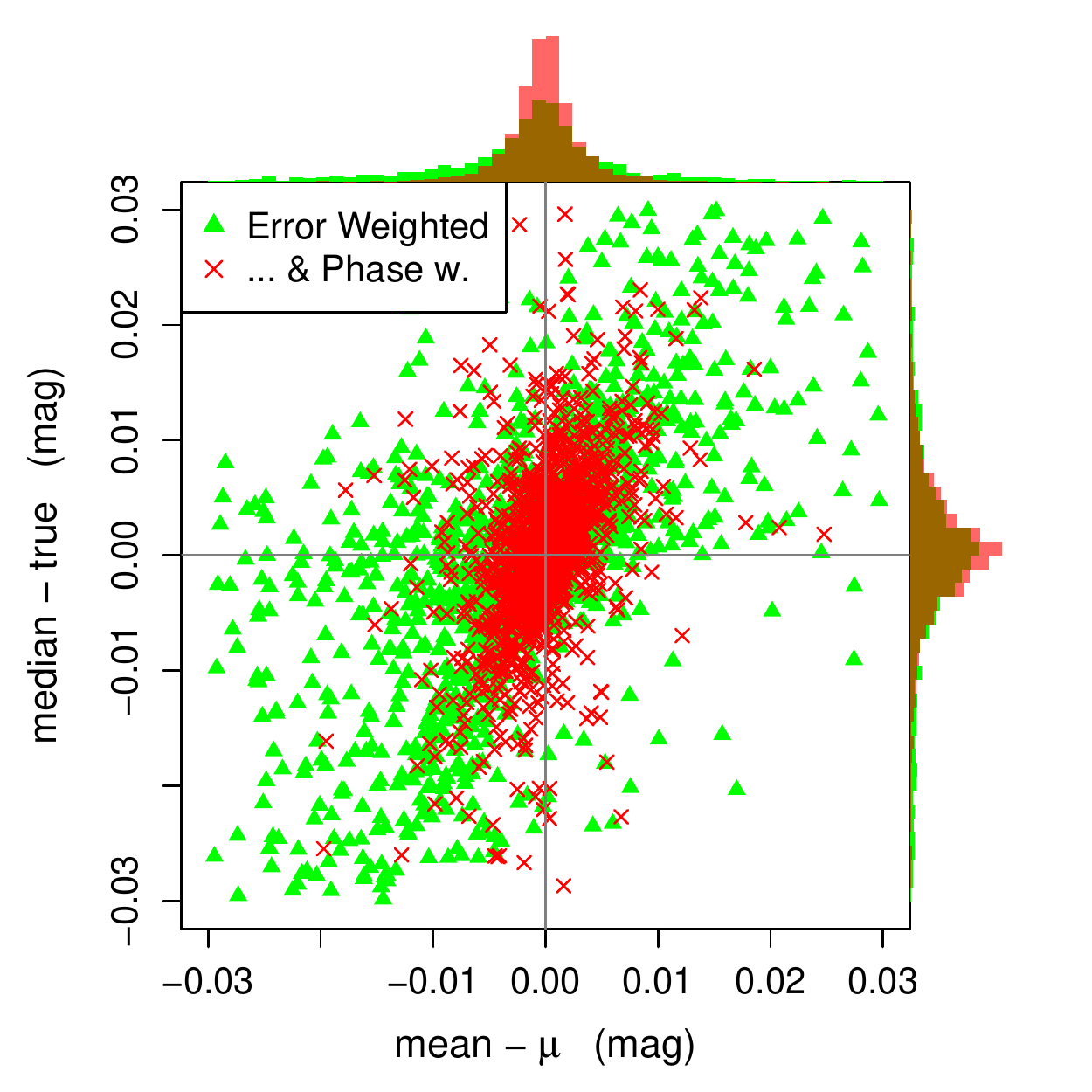}\\
(c)\\
\vspace{-0.2cm}
\includegraphics[width=\columnwidth]{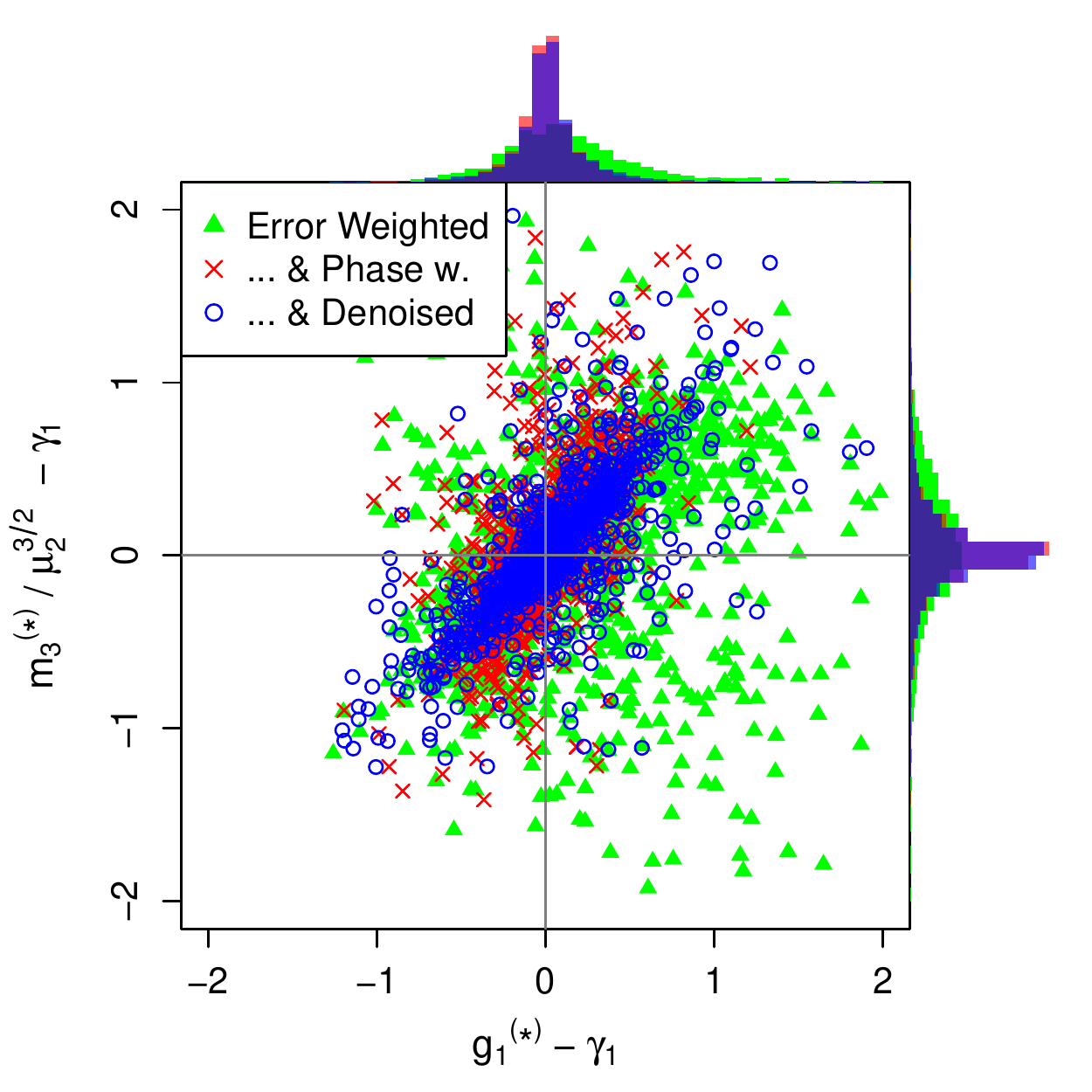}
\end{minipage}
\begin{minipage}{\columnwidth}
\center
(b)\\
\vspace{-0.2cm}
\includegraphics[width=\columnwidth]{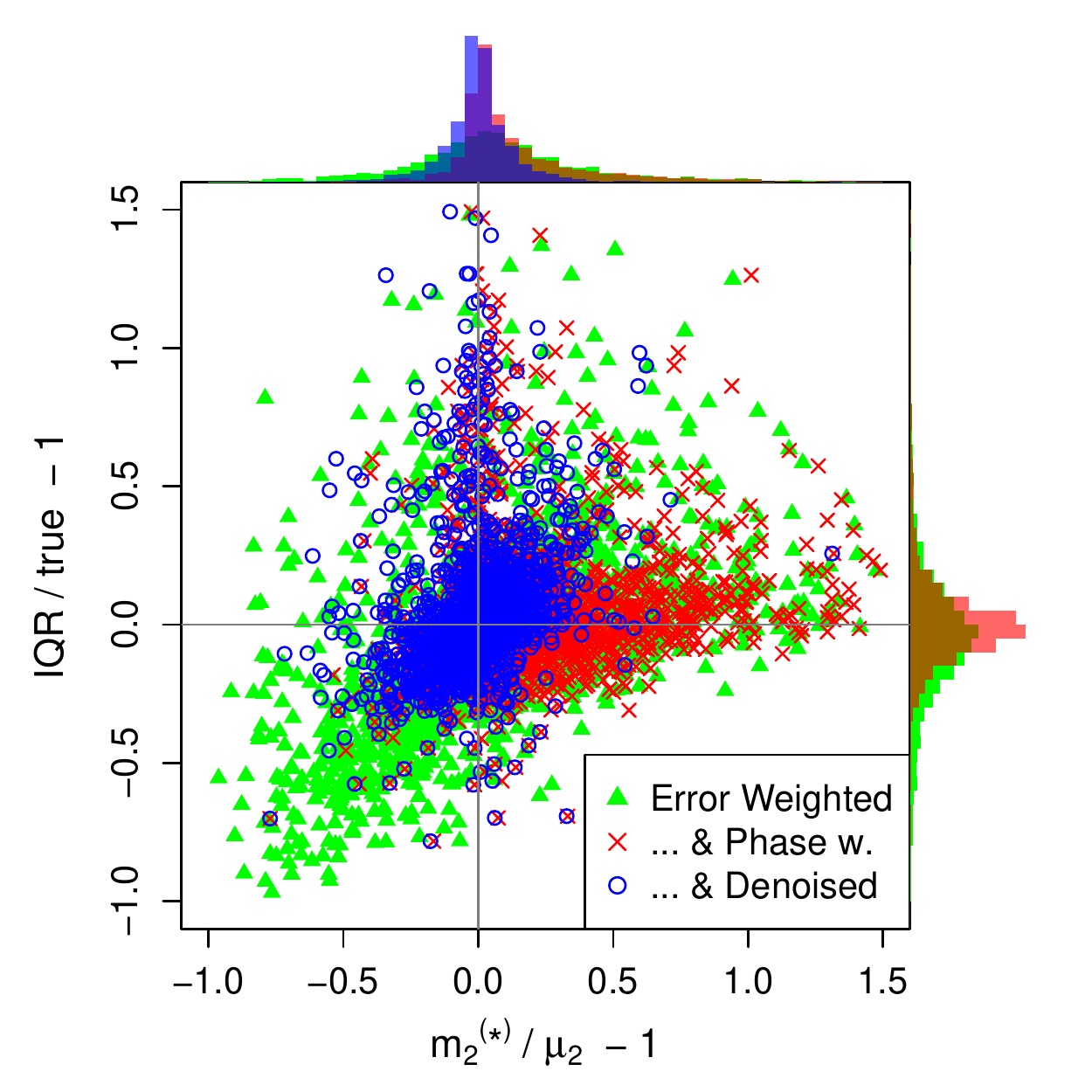}\\
(d)\\
\vspace{-0.2cm}
\includegraphics[width=\columnwidth]{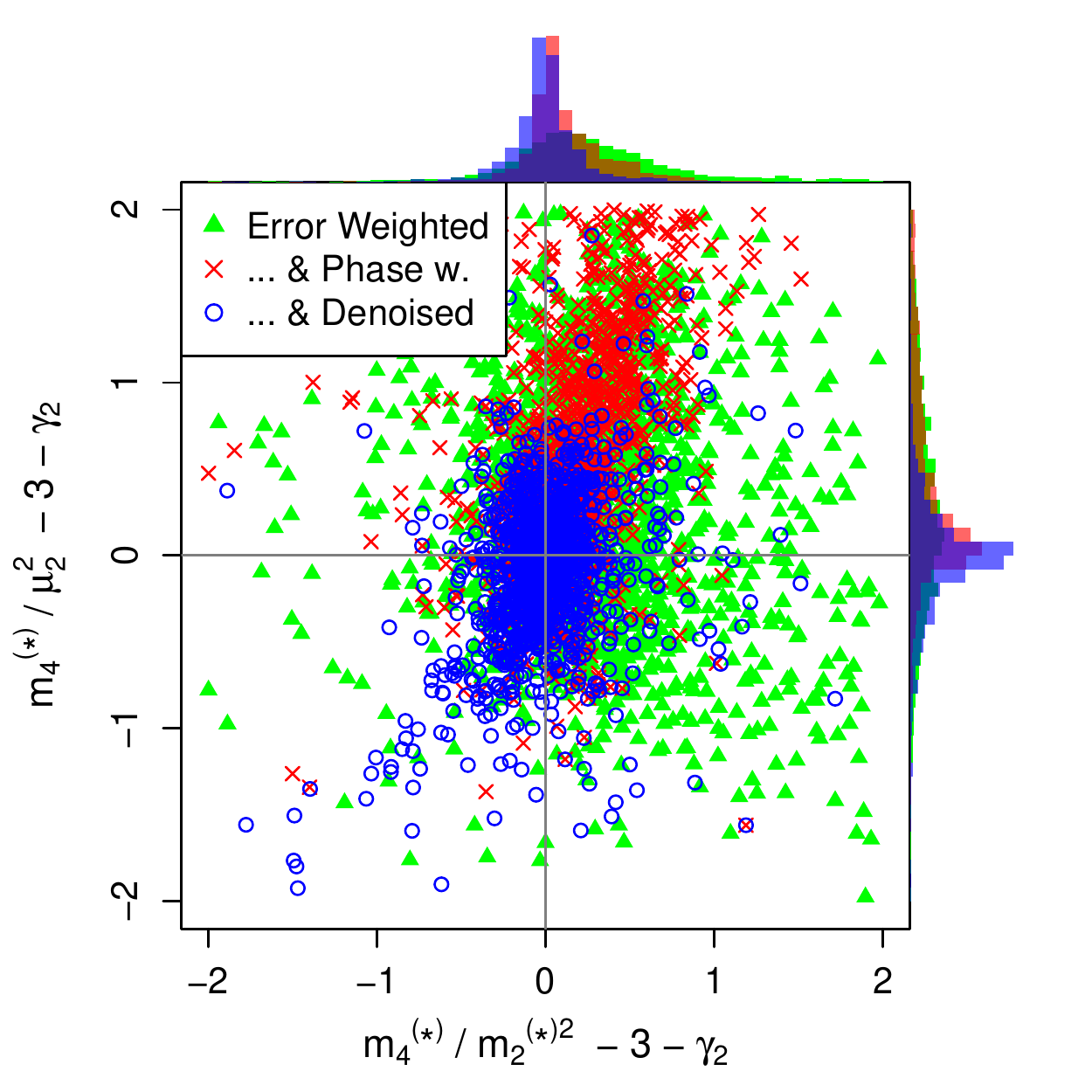}
\end{minipage}
\caption{Deviations from the true values of (a) mean and median, (b) variance and interquartile range,  (c) skewness and (d) kurtosis moments standardized by the estimated and true variances. The triangles and histograms in green denote error-weighted estimators, the crosses and histograms in red indicate error-phase weighted estimators, and the circles and histograms in blue represent noise-unbiased error-phase weighted estimators.}
\label{fig:scatter1}
\end{figure*}

\begin{figure}
\includegraphics[width=\columnwidth]{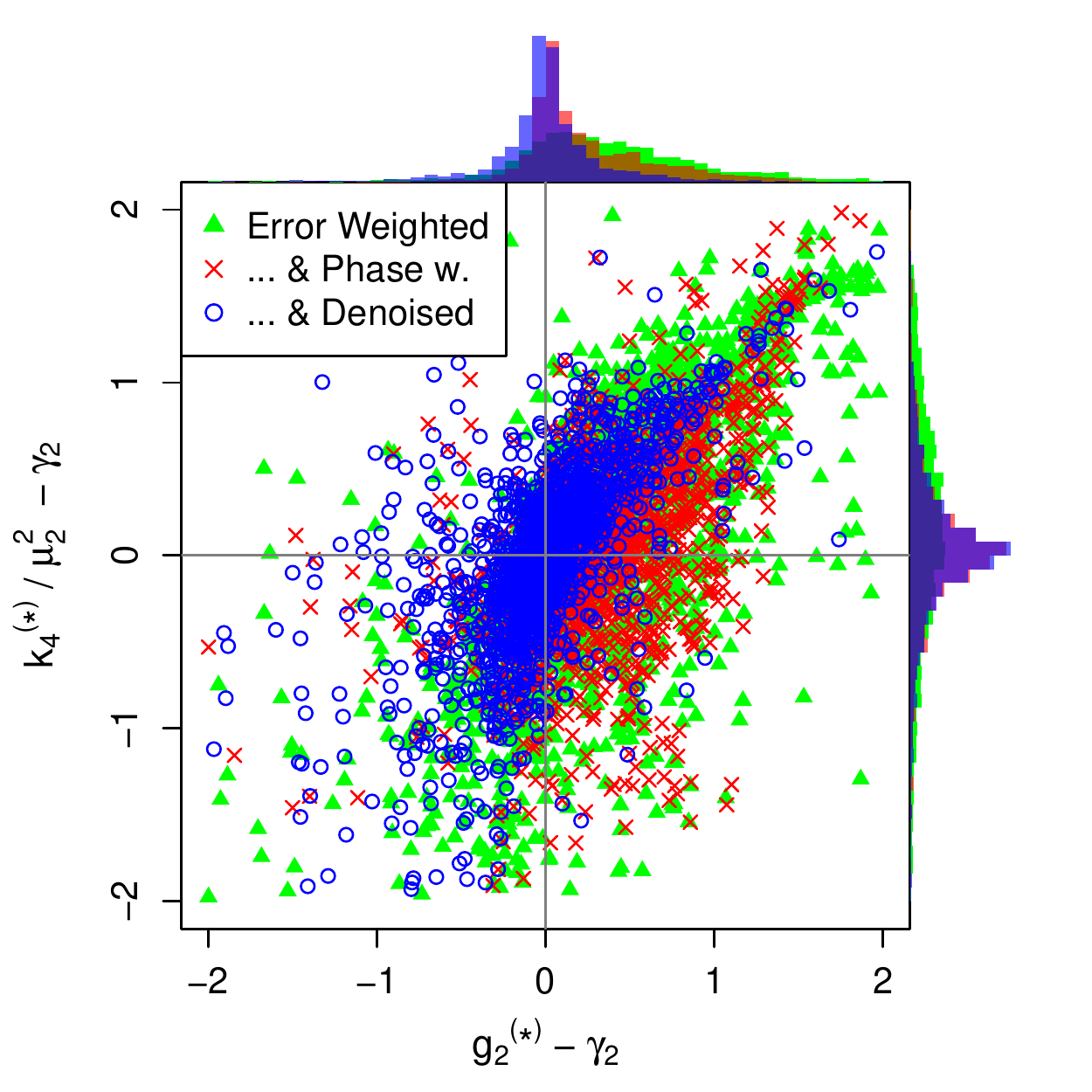}
\caption{Deviations from the true values of the kurtosis cumulants, standardized by the estimated and true variances.  The triangles and histograms in green denote error-weighted estimators, the crosses and histograms in red indicate error-phase weighted estimators, and the circles and histograms in blue represent noise-unbiased error-phase weighted estimators.}
\label{fig:scatter2}
\end{figure}

Figures~\ref{fig:scatter1}--\ref{fig:scatter2} compare statistical parameters which provide similar information, such as robust versus non-robust or normalized versus non-normalized estimators, as a function of the weighting scheme.
Error-phase weighted (and noise-unbiased, when applicable) estimators  correspond to the most strongly peaked distributions around the correct values.

\section{Unweighted estimators as a function of variability type}
\label{app:scatterUNWEIGHTED}
\begin{figure*}
\begin{minipage}{\columnwidth}
\center
~~~~~~~~~(a)\\
\vspace{-0.2cm}
\includegraphics[width=\columnwidth]{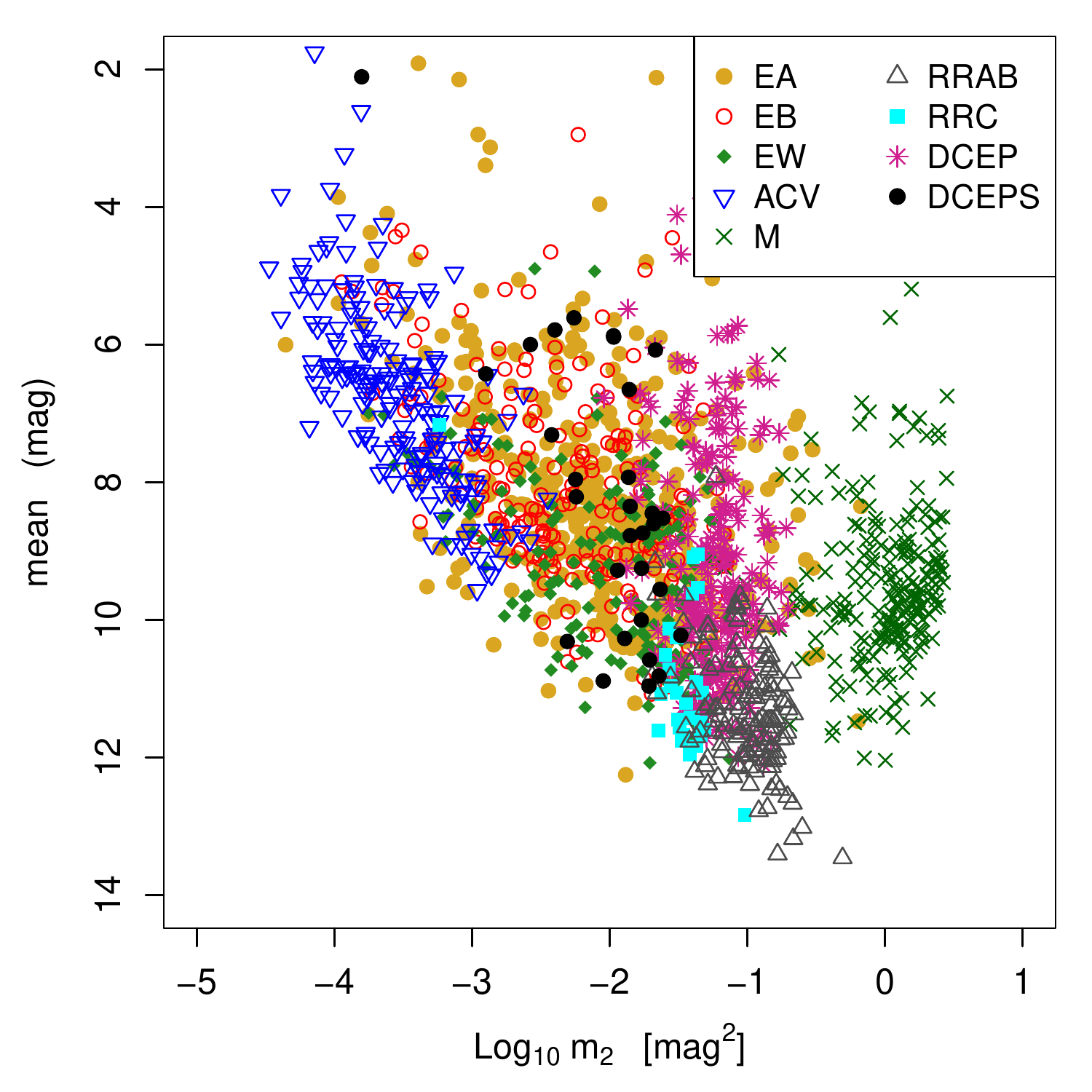}\\
~~~~~~~~~(c)\\
\vspace{-0.2cm}
\includegraphics[width=\columnwidth]{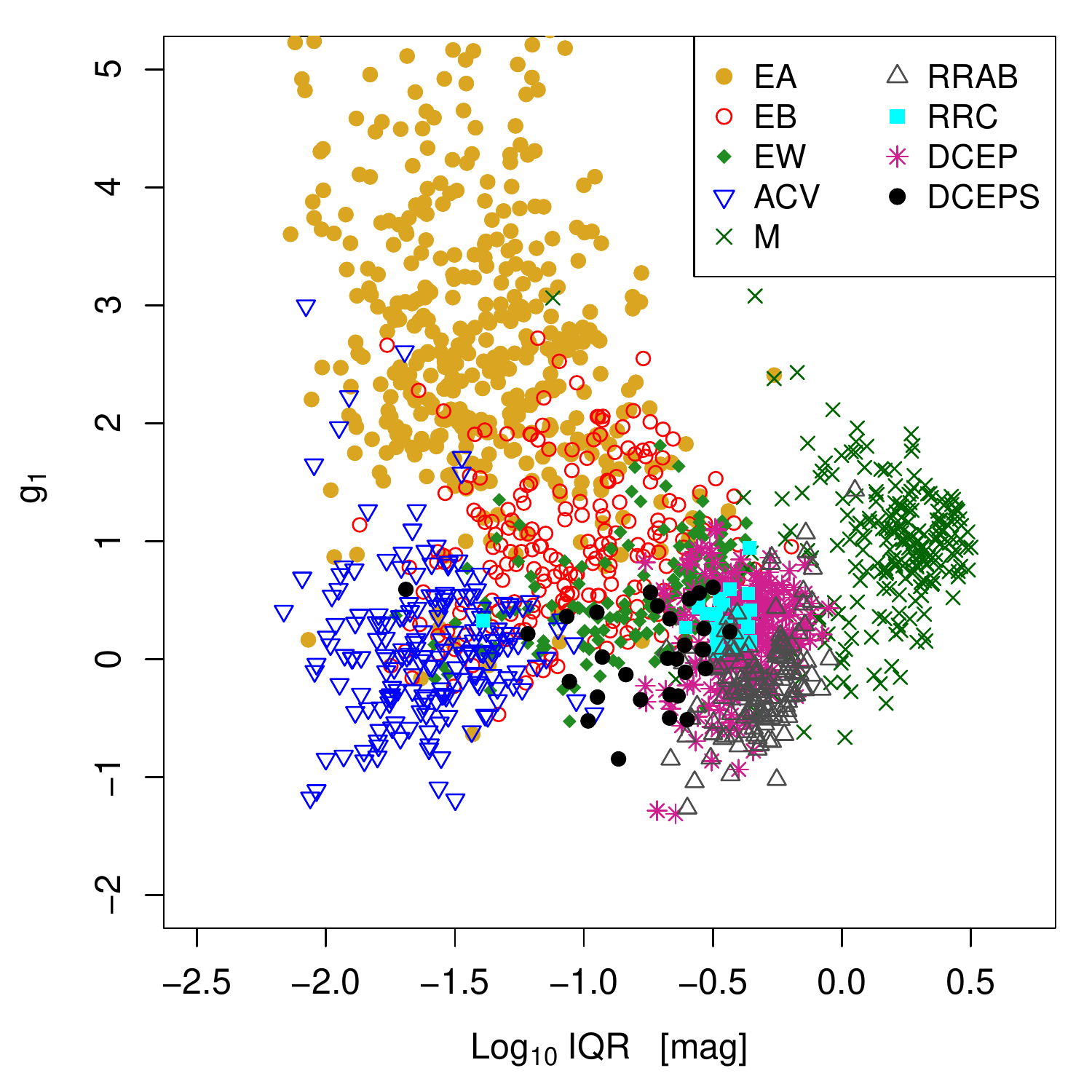}
\end{minipage}
\begin{minipage}{\columnwidth}
\center
~~~~~~~~~(b)\\
\vspace{-0.2cm}
\includegraphics[width=\columnwidth]{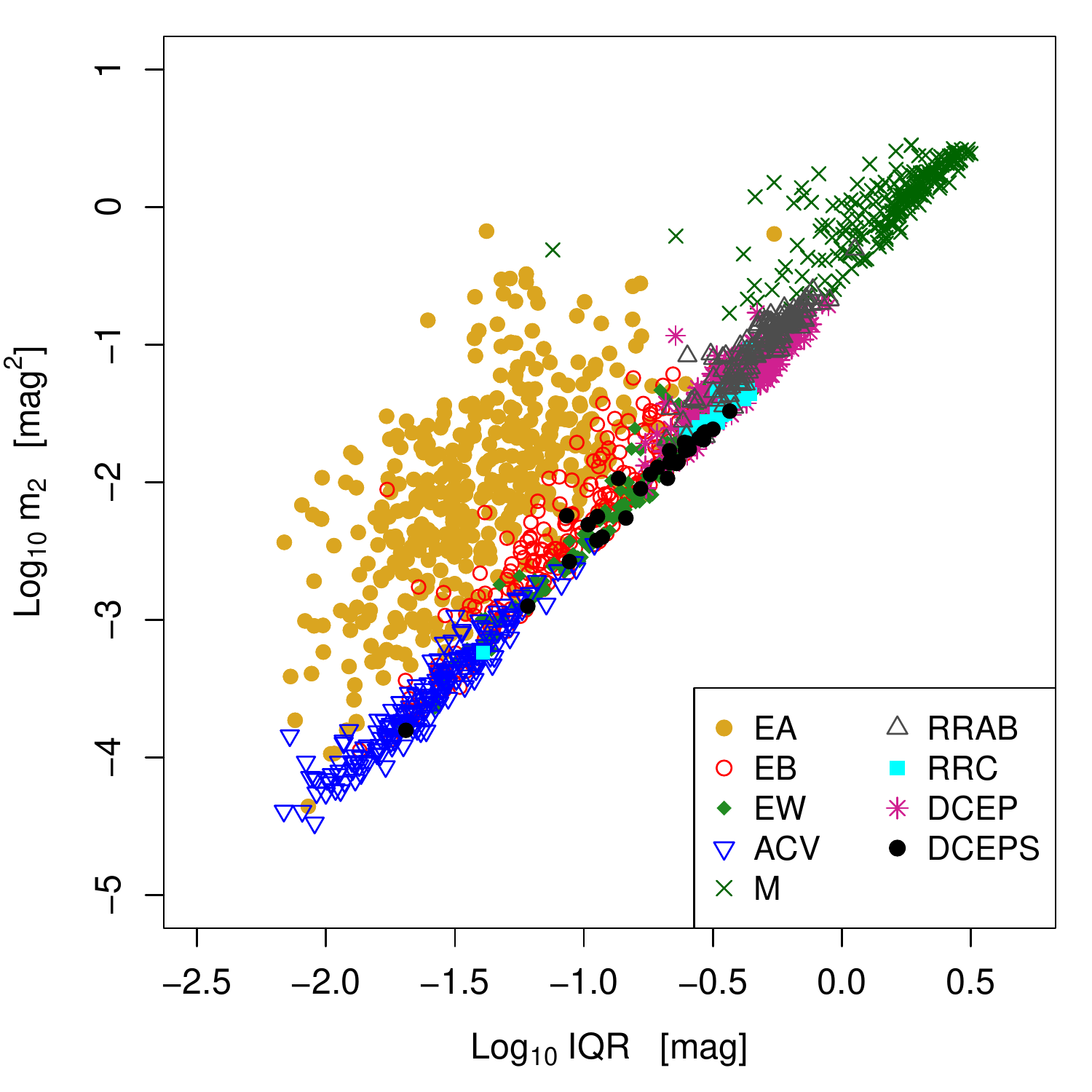}\\
~~~~~~~~~(d)\\
\vspace{-0.2cm}
\includegraphics[width=\columnwidth]{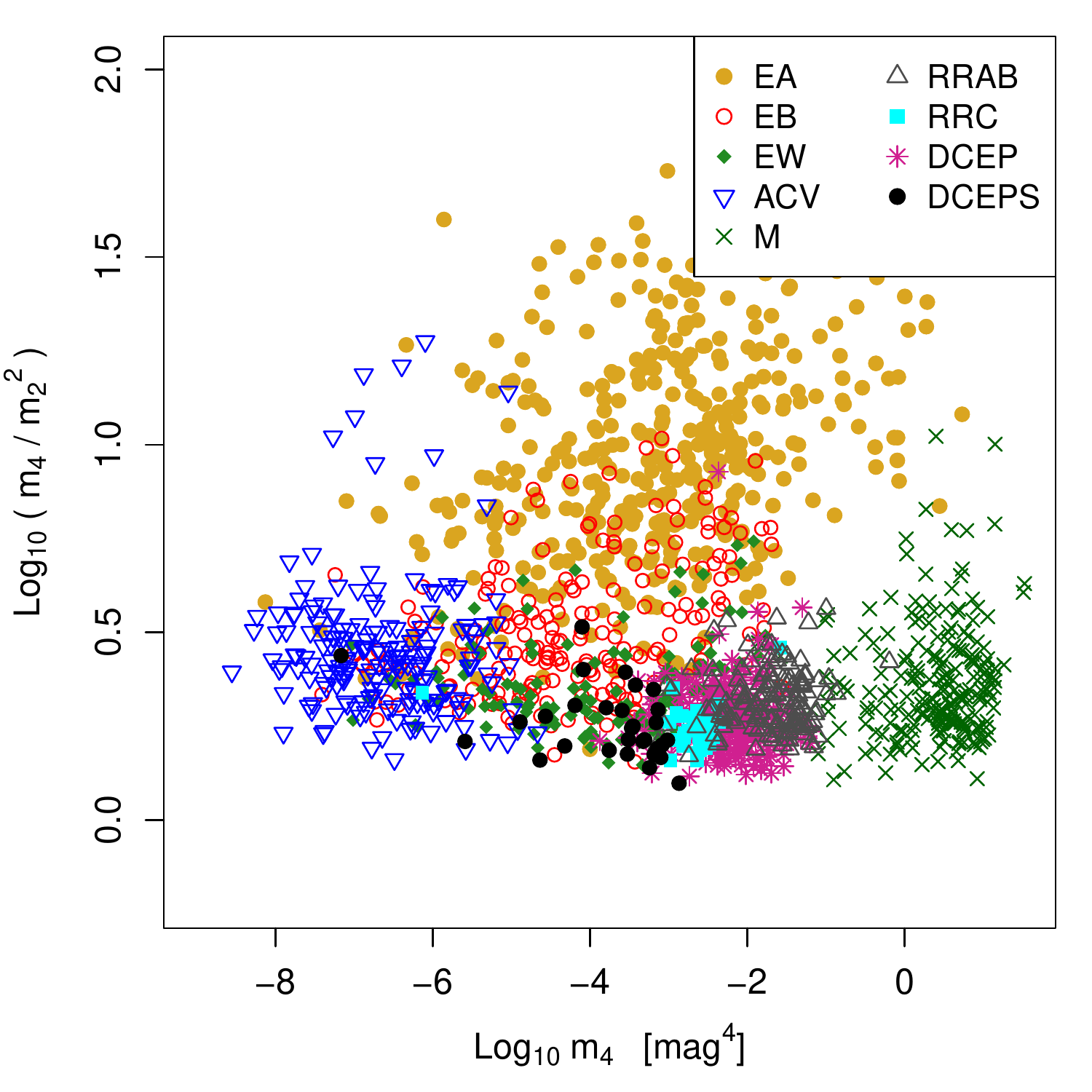}
\end{minipage}
\caption{A selection of unweighted estimators employed for classification is illustrated as a function of variability type.
Class labels are described in Table~\ref{tab:ts} and denoted by symbols as shown in the legend of each panel.}
\label{fig:scatterUNWEIGHTED}
\end{figure*}

Figure~\ref{fig:scatterUNWEIGHTED} presents a selection of unweighted estimators for stars of different variability types. 
This Figure is intended to be compared to Fig.~\ref{fig:scatter}, which illustrates the same information for phase-error weighted and noise-unbiased estimators.

\textcolor{black}{
\section*{Supporting Information}
Additional Supporting Information may be found in the online version of this paper: \\
{\bf Figure B1.} The folded light curves from the {\it Hipparcos} periodic catalogue and the smoothing spline models employed to generate simulated data. 
}

\bsp
\label{lastpage}


\begin{thebibliography}{99}
\bibitem[\protect\citeauthoryear{Auvergne et al.}{2009}]{corot}Auvergne M. et al., 2009, A\&A, 506, 411
\bibitem[\protect\citeauthoryear{Blomme et al.}{2010}]{BlommeA}Blomme J. et al., 2010, ApJ, 713, L204
\bibitem[\protect\citeauthoryear{Blomme et al.}{2011}]{BlommeB}Blomme J. et al., 2011, MNRAS, 418, 96
\bibitem[\protect\citeauthoryear{Borucki et al.}{2010}]{kepler}Borucki W.J. et al., 2010, Science, 327, 977
\bibitem[\protect\citeauthoryear{Breiman}{2001}]{Breiman}Breiman L., 2001, Machine Learning, 45, 5
\bibitem[\protect\citeauthoryear{Carbonell, Oliver \& Ballester}{1992}]{Carbonell}Carbonell M.,  Oliver R., Ballester J.L., 1992, A\&A, 264, 350
\bibitem[\protect\citeauthoryear{Debosscher et al.}{2007}]{DebosscherA}Debosscher J., Sarro L.M., Aerts C., Cuypers J., Vandenbussche B., Garrido R., Solano E., 2007, A\&A, 475, 1159
\bibitem[\protect\citeauthoryear{Debosscher et al.}{2009}]{DebosscherB}Debosscher J. et al., 2009, A\&A, 506, 519
\bibitem[\protect\citeauthoryear{Dubath et al.}{2011}]{Dubath}Dubath P. et al., 2011, MNRAS, 414, 2602
\bibitem[\protect\citeauthoryear{ESA}{1997}]{ESA} European Space Agency, 1997, The Hipparcos and Tycho Catalogues, ESA~SP-1200
\bibitem[\protect\citeauthoryear{Eyer}{1998}]{EyerThesis}Eyer L., 1998,  Les \'{e}toiles variables de la mission HIPPARCOS, PhD Thesis N\textsuperscript{o}~3002, Universit\'{e} de Gen\`{e}ve
\bibitem[\protect\citeauthoryear{Eyer \& Mignard}{2005}]{Eyer2005}Eyer L.,  Mignard F., 2005, MNRAS, 361, 1136
\bibitem[\protect\citeauthoryear{Eyer et al.}{1994}]{Eyer1994}Eyer L., Grenon M., Falin J.-L., Froeschle M., Mignard F., 1994, Solar Physics, 152, 91
\bibitem[\protect\citeauthoryear{Ivezi\'c et al.}{1994}]{Ivezic}\textcolor{black}{Ivezi\'c \v{Z}. et al., 2011, preprint (arXiv:0805.2366)}
\bibitem[\protect\citeauthoryear{Kaiser et al.}{2002}]{Kaiser}\textcolor{black}{Kaiser N. et al., 2002, in Tyson J. A., Wolff S., eds, Proc. SPIE Vol. 4836, Survey and Other Telescope Technologies and Discoveries. SPIE, Bellingham, p. 154}
\bibitem[\protect\citeauthoryear{Koen}{2005}]{Koen}Koen C., 2005, MNRAS, 361, 887
\bibitem[\protect\citeauthoryear{Liaw \&  Wiener}{2002}]{Rrf}Liaw A., Wiener M., 2002, R News, 2 (3), 18
\bibitem[\protect\citeauthoryear{Perryman et al.}{1997}]{Perryman}Perryman M.A.C. et al., 1997, A\&A, 323, L49
\bibitem[\protect\citeauthoryear{Perryman et al.}{2001}]{PerrymanGaia}Perryman M.A.C. et al., 2001, A\&A, 369, 339
\bibitem[\protect\citeauthoryear{R Development Core Team}{2013}]{R}R Development Core Team, 2013, R: A Language and Environment for Statistical Computing, R Foundation for Statistical Computing, Vienna, Austria
\bibitem[\protect\citeauthoryear{Richards et al.}{2011}]{Richards}Richards J.W. et al., 2011, ApJ, 733, 10
\bibitem[\protect\citeauthoryear{Rimoldini et al.}{2012}]{RimoldiniUnsolved}Rimoldini L. et al., 2012, MNRAS,  427, 2917
\bibitem[\protect\citeauthoryear{Rimoldini}{2013a}]{RimoldiniUnbiased}Rimoldini L., 2013a, preprint (\href{http://xxx.lanl.gov/abs/1304.6564}{arXiv:1304.6564})
\bibitem[\protect\citeauthoryear{Rimoldini}{2013b}]{RimoldiniIntrinsic}Rimoldini L., 2013b, preprint (\href{http://xxx.lanl.gov/abs/1304.6715}{arXiv:1304.6715})
\bibitem[\protect\citeauthoryear{Rybicki \& Press}{1992}]{Rybicki}\textcolor{black}{Rybicki G.B., Press W.H., 1992, ApJ, 398, 169}
\bibitem[\protect\citeauthoryear{Ruppert, Wand, \& Carroll}{2003}]{GCV}Ruppert D., Wand M.P., Carroll R.J., 2003, Nonparametric Regression, Cambridge University Press
\bibitem[\protect\citeauthoryear{Sarro et al.}{2009}]{Sarro}Sarro L.M., Debosscher J., Aerts C., L\'{o}pez M., 2009, A\&A, 506, 535
\bibitem[\protect\citeauthoryear{Scargle}{1989}]{Scargle1989}\textcolor{black}{Scargle J.D., 1989, ApJ, 343, 874}
\bibitem[\protect\citeauthoryear{Scargle}{1990}]{Scargle1990}\textcolor{black}{Scargle J.D., 1990, ApJ, 359, 469}
\bibitem[\protect\citeauthoryear{Stuart \& Ord}{1969}]{Kendall}Stuart A., Ord J., 1969, Kendall's Advanced Theory of Statistics, Charles Griffin \& Co.~Ltd, London
\bibitem[\protect\citeauthoryear{van Leeuwen}{1997}]{VanLeeuwen} van Leeuwen F., 1997, Sp. Sci. Rev., 81, 201
\bibitem[\protect\citeauthoryear{Vio, Strohmer \& Wamsteker}{2000}]{Vio}Vio R., Strohmer T., Wamsteker W., 2000, PASP, 112, 74
\bibitem[\protect\citeauthoryear{Watson, Henden \&  Price}{Watson et al.}{2012}]{Watson} Watson C., Henden A.A., Price A., 2012, VizieR Online Data Catalogue, AAVSO International Variable Star Index, Version 2012-09-23
\end{thebibliography}
\end{document}